\documentclass[epj,nopacs]{svjour}
\usepackage{cite,mcite}
\usepackage[pdftex]{graphicx}
\usepackage{subfigure}
\usepackage{atlasphysics}
\usepackage{amsmath}
\usepackage{amssymb}
\usepackage{mathptmx}
\usepackage{helvet}
\newcommand{\pho}{\phantom{0}}

\begin{document}
\hugehead

\date{\today}
% \vspace*{-1cm}
\title{The ATLAS Inner Detector commissioning and calibration}
%\author{\ }\institute{\ }
% ATLAS Collaboration author list for 03-MAR-2010
% Data extracted on 02-Jun-2010 for paperid 23
\author{
G.~Aad$^{\rm 48}$,
B.~Abbott$^{\rm 111}$,
J.~Abdallah$^{\rm 11}$,
A.A.~Abdelalim$^{\rm 49}$,
A.~Abdesselam$^{\rm 118}$,
O.~Abdinov$^{\rm 10}$,
B.~Abi$^{\rm 112}$,
M.~Abolins$^{\rm 88}$,
H.~Abramowicz$^{\rm 152}$,
H.~Abreu$^{\rm 115}$
% ...
}
\institute{
$^{1}$ University at Albany, 1400 Washington Ave, Albany, NY 12222, United States of America\\
$^{2}$ University of Alberta, Department of Physics, Centre for Particle Physics, Edmonton, AB T6G 2G7, Canada\\
$^{3}$ Ankara University$^{(a)}$, Faculty of Sciences, Department of Physics, TR 061000 Tandogan, Ankara; Dumlupinar University$^{(b)}$, Faculty of Arts and Sciences, Department of Physics, Kutahya; Gazi University$^{(c)}$, Faculty of Arts and Sciences, Department of Physics, 06500, Teknikokullar, Ankara; TOBB University of Economics and Technology$^{(d)}$, Faculty of Arts and Sciences, Division of Physics, 06560, Sogutozu, Ankara; Turkish Atomic Energy Authority$^{(e)}$, 06530, Lodumlu, Ankara, Turkey\\
$^{4}$ LAPP, Universit\'e de Savoie, CNRS/IN2P3, Annecy-le-Vieux, France\\
$^{5}$ Argonne National Laboratory, High Energy Physics Division, 9700 S. Cass Avenue, Argonne IL 60439, United States of America\\
$^{6}$ University of Arizona, Department of Physics, Tucson, AZ 85721, United States of America\\
$^{7}$ The University of Texas at Arlington, Department of Physics, Box 19059, Arlington, TX 76019, United States of America\\
$^{8}$ University of Athens, Nuclear \& Particle Physics, Department of Physics, Panepistimiopouli, Zografou, GR 15771 Athens, Greece\\
$^{9}$ National Technical University of Athens, Physics Department, 9-Iroon Polytechniou, GR 15780 Zografou, Greece\\
$^{10}$ Institute of Physics, Azerbaijan Academy of Sciences, H. Javid Avenue 33, AZ 143 Baku, Azerbaijan\\
$^{11}$ Institut de F\'isica d'Altes Energies, IFAE, Edifici Cn, Universitat Aut\`onoma  de Barcelona,  ES - 08193 Bellaterra (Barcelona), Spain\\
$^{12}$ University of Belgrade$^{(a)}$, Institute of Physics, P.O. Box 57, 11001 Belgrade; Vinca Institute of Nuclear Sciences$^{(b)}$, Mihajla Petrovica Alasa 12-14, 11001 Belgrade, Serbia\\
$^{13}$ University of Bergen, Department for Physics and Technology, Allegaten 55, NO - 5007 Bergen, Norway\\
$^{14}$ Lawrence Berkeley National Laboratory and University of California, Physics Division, MS50B-6227, 1 Cyclotron Road, Berkeley, CA 94720, United States of America\\
$^{15}$ Humboldt University, Institute of Physics, Berlin, Newtonstr. 15, D-12489 Berlin, Germany\\
$^{16}$ University of Bern,
Albert Einstein Center for Fundamental Physics,
Laboratory for High Energy Physics, Sidlerstrasse 5, CH - 3012 Bern, Switzerland\\
$^{17}$ University of Birmingham, School of Physics and Astronomy, Edgbaston, Birmingham B15 2TT, United Kingdom\\
$^{18}$ Bogazici University$^{(a)}$, Faculty of Sciences, Department of Physics, TR - 80815 Bebek-Istanbul; Dogus University$^{(b)}$, Faculty of Arts and Sciences, Department of Physics, 34722, Kadikoy, Istanbul; $^{(c)}$Gaziantep University, Faculty of Engineering, Department of Physics Engineering, 27310, Sehitkamil, Gaziantep, Turkey; Istanbul Technical University$^{(d)}$, Faculty of Arts and Sciences, Department of Physics, 34469, Maslak, Istanbul, Turkey\\
$^{19}$ INFN Sezione di Bologna$^{(a)}$; Universit\`a  di Bologna, Dipartimento di Fisica$^{(b)}$, viale C. Berti Pichat, 6/2, IT - 40127 Bologna, Italy\\
$^{20}$ University of Bonn, Physikalisches Institut, Nussallee 12, D - 53115 Bonn, Germany\\
$^{21}$ Boston University, Department of Physics,  590 Commonwealth Avenue, Boston, MA 02215, United States of America\\
$^{22}$ Brandeis University, Department of Physics, MS057, 415 South Street, Waltham, MA 02454, United States of America\\
$^{23}$ Universidade Federal do Rio De Janeiro, COPPE/EE/IF $^{(a)}$, Caixa Postal 68528, Ilha do Fundao, BR - 21945-970 Rio de Janeiro; $^{(b)}$Universidade de Sao Paulo, Instituto de Fisica, R.do Matao Trav. R.187, Sao Paulo - SP, 05508 - 900, Brazil\\
$^{24}$ Brookhaven National Laboratory, Physics Department, Bldg. 510A, Upton, NY 11973, United States of America\\
$^{25}$ National Institute of Physics and Nuclear Engineering$^{(a)}$, Bucharest-Magurele, Str. Atomistilor 407,  P.O. Box MG-6, R-077125, Romania; University Politehnica Bucharest$^{(b)}$, Rectorat - AN 001, 313 Splaiul Independentei, sector 6, 060042 Bucuresti; West University$^{(c)}$ in Timisoara, Bd. Vasile Parvan 4, Timisoara, Romania\\
$^{26}$ Universidad de Buenos Aires, FCEyN, Dto. Fisica, Pab I - C. Universitaria, 1428 Buenos Aires, Argentina\\
$^{27}$ University of Cambridge, Cavendish Laboratory, J J Thomson Avenue, Cambridge CB3 0HE, United Kingdom\\
$^{28}$ Carleton University, Department of Physics, 1125 Colonel By Drive,  Ottawa ON  K1S 5B6, Canada\\
$^{29}$ CERN, CH - 1211 Geneva 23, Switzerland\\
$^{30}$ University of Chicago, Enrico Fermi Institute, 5640 S. Ellis Avenue, Chicago, IL 60637, United States of America\\
$^{31}$ Pontificia Universidad Cat\'olica de Chile, Facultad de Fisica, Departamento de Fisica$^{(a)}$, Avda. Vicuna Mackenna 4860, San Joaquin, Santiago; Universidad T\'ecnica Federico Santa Mar\'ia, Departamento de F\'isica$^{(b)}$, Avda. Esp\~ana 1680, Casilla 110-V,  Valpara\'iso, Chile\\
$^{32}$ Institute of High Energy Physics, Chinese Academy of Sciences$^{(a)}$, P.O. Box 918, 19 Yuquan Road, Shijing Shan District, CN - Beijing 100049; University of Science \& Technology of China (USTC), Department of Modern Physics$^{(b)}$, Hefei, CN - Anhui 230026; Nanjing University, Department of Physics$^{(c)}$, 22 Hankou Road, Nanjing, 210093; Shandong University, High Energy Physics Group$^{(d)}$, Jinan, CN - Shandong 250100, China\\
$^{33}$ Laboratoire de Physique Corpusculaire, Clermont Universit\'e, Universit\'e Blaise Pascal, CNRS/IN2P3, FR - 63177 Aubiere Cedex, France\\
$^{34}$ Columbia University, Nevis Laboratory, 136 So. Broadway, Irvington, NY 10533, United States of America\\
$^{35}$ University of Copenhagen, Niels Bohr Institute, Blegdamsvej 17, DK - 2100 Kobenhavn 0, Denmark\\
$^{36}$ INFN Gruppo Collegato di Cosenza$^{(a)}$; Universit\`a della Calabria, Dipartimento di Fisica$^{(b)}$, IT-87036 Arcavacata di Rende, Italy\\
$^{37}$ Faculty of Physics and Applied Computer Science of the AGH-University of Science and Technology, (FPACS, AGH-UST), al. Mickiewicza 30, PL-30059 Cracow, Poland\\
$^{38}$ The Henryk Niewodniczanski Institute of Nuclear Physics, Polish Academy of Sciences, ul. Radzikowskiego 152, PL - 31342 Krakow, Poland\\
$^{39}$ Southern Methodist University, Physics Department, 106 Fondren Science Building, Dallas, TX 75275-0175, United States of America\\
$^{40}$ University of Texas at Dallas, 800 West Campbell Road, Richardson, TX 75080-3021, United States of America\\
$^{41}$ DESY, Notkestr. 85, D-22603 Hamburg and Platanenallee 6, D-15738 Zeuthen, Germany\\
$^{42}$ TU Dortmund, Experimentelle Physik IV, DE - 44221 Dortmund, Germany\\
$^{43}$ Technical University Dresden, Institut f\"{u}r Kern- und Teilchenphysik, Zellescher Weg 19, D-01069 Dresden, Germany\\
$^{44}$ Duke University, Department of Physics, Durham, NC 27708, United States of America\\
$^{45}$ University of Edinburgh, School of Physics \& Astronomy, James Clerk Maxwell Building, The Kings Buildings, Mayfield Road, Edinburgh EH9 3JZ, United Kingdom\\
$^{46}$ Fachhochschule Wiener Neustadt; Johannes Gutenbergstrasse 3 AT - 2700 Wiener Neustadt, Austria\\
$^{47}$ INFN Laboratori Nazionali di Frascati, via Enrico Fermi 40, IT-00044 Frascati, Italy\\
$^{48}$ Albert-Ludwigs-Universit\"{a}t, Fakult\"{a}t f\"{u}r Mathematik und Physik, Hermann-Herder Str. 3, D - 79104 Freiburg i.Br., Germany\\
$^{49}$ Universit\'e de Gen\`eve, Section de Physique, 24 rue Ernest Ansermet, CH - 1211 Geneve 4, Switzerland\\
$^{50}$ INFN Sezione di Genova$^{(a)}$; Universit\`a  di Genova, Dipartimento di Fisica$^{(b)}$, via Dodecaneso 33, IT - 16146 Genova, Italy\\
$^{51}$ Institute of Physics of the Georgian Academy of Sciences, 6 Tamarashvili St., GE - 380077 Tbilisi; Tbilisi State University, HEP Institute, University St. 9, GE - 380086 Tbilisi, Georgia\\
$^{52}$ Justus-Liebig-Universit\"{a}t Giessen, II Physikalisches Institut, Heinrich-Buff Ring 16,  D-35392 Giessen, Germany\\
$^{53}$ University of Glasgow, Department of Physics and Astronomy, Glasgow G12 8QQ, United Kingdom\\
$^{54}$ Georg-August-Universit\"{a}t, II. Physikalisches Institut, Friedrich-Hund Platz 1, D-37077 G\"{o}ttingen, Germany\\
$^{55}$ Laboratoire de Physique Subatomique et de Cosmologie, CNRS/IN2P3, Universit\'e Joseph Fourier, INPG, 53 avenue des Martyrs, FR - 38026 Grenoble Cedex, France\\
$^{56}$ Hampton University, Department of Physics, Hampton, VA 23668, United States of America\\
$^{57}$ Harvard University, Laboratory for Particle Physics and Cosmology, 18 Hammond Street, Cambridge, MA 02138, United States of America\\
$^{58}$ Ruprecht-Karls-Universit\"{a}t Heidelberg: Kirchhoff-Institut f\"{u}r Physik$^{(a)}$, Im Neuenheimer Feld 227, D-69120 Heidelberg; Physikalisches Institut$^{(b)}$, Philosophenweg 12, D-69120 Heidelberg; ZITI Ruprecht-Karls-University Heidelberg$^{(c)}$, Lehrstuhl f\"{u}r Informatik V, B6, 23-29, DE - 68131 Mannheim, Germany\\
$^{59}$ Hiroshima University, Faculty of Science, 1-3-1 Kagamiyama, Higashihiroshima-shi, JP - Hiroshima 739-8526, Japan\\
$^{60}$ Hiroshima Institute of Technology, Faculty of Applied Information Science, 2-1-1 Miyake Saeki-ku, Hiroshima-shi, JP - Hiroshima 731-5193, Japan\\
$^{61}$ Indiana University, Department of Physics,  Swain Hall West 117, Bloomington, IN 47405-7105, United States of America\\
$^{62}$ Institut f\"{u}r Astro- und Teilchenphysik, Technikerstrasse 25, A - 6020 Innsbruck, Austria\\
$^{63}$ University of Iowa, 203 Van Allen Hall, Iowa City, IA 52242-1479, United States of America\\
$^{64}$ Iowa State University, Department of Physics and Astronomy, Ames High Energy Physics Group,  Ames, IA 50011-3160, United States of America\\
$^{65}$ Joint Institute for Nuclear Research, JINR Dubna, RU - 141 980 Moscow Region, Russia\\
$^{66}$ KEK, High Energy Accelerator Research Organization, 1-1 Oho, Tsukuba-shi, Ibaraki-ken 305-0801, Japan\\
$^{67}$ Kobe University, Graduate School of Science, 1-1 Rokkodai-cho, Nada-ku, JP Kobe 657-8501, Japan\\
$^{68}$ Kyoto University, Faculty of Science, Oiwake-cho, Kitashirakawa, Sakyou-ku, Kyoto-shi, JP - Kyoto 606-8502, Japan\\
$^{69}$ Kyoto University of Education, 1 Fukakusa, Fujimori, fushimi-ku, Kyoto-shi, JP - Kyoto 612-8522, Japan\\
$^{70}$ Universidad Nacional de La Plata, FCE, Departamento de F\'{i}sica, IFLP (CONICET-UNLP),   C.C. 67,  1900 La Plata, Argentina\\
$^{71}$ Lancaster University, Physics Department, Lancaster LA1 4YB, United Kingdom\\
$^{72}$ INFN Sezione di Lecce$^{(a)}$; Universit\`a  del Salento, Dipartimento di Fisica$^{(b)}$Via Arnesano IT - 73100 Lecce, Italy\\
$^{73}$ University of Liverpool, Oliver Lodge Laboratory, P.O. Box 147, Oxford Street,  Liverpool L69 3BX, United Kingdom\\
$^{74}$ Jo\v{z}ef Stefan Institute and University of Ljubljana, Department  of Physics, SI-1000 Ljubljana, Slovenia\\
$^{75}$ Queen Mary University of London, Department of Physics, Mile End Road, London E1 4NS, United Kingdom\\
$^{76}$ Royal Holloway, University of London, Department of Physics, Egham Hill, Egham, Surrey TW20 0EX, United Kingdom\\
$^{77}$ University College London, Department of Physics and Astronomy, Gower Street, London WC1E 6BT, United Kingdom\\
$^{78}$ Laboratoire de Physique Nucl\'eaire et de Hautes Energies, Universit\'e Pierre et Marie Curie (Paris 6), Universit\'e Denis Diderot (Paris-7), CNRS/IN2P3, Tour 33, 4 place Jussieu, FR - 75252 Paris Cedex 05, France\\
$^{79}$ Lunds universitet, Naturvetenskapliga fakulteten, Fysiska institutionen, Box 118, SE - 221 00 Lund, Sweden\\
$^{80}$ Universidad Autonoma de Madrid, Facultad de Ciencias, Departamento de Fisica Teorica, ES - 28049 Madrid, Spain\\
$^{81}$ Universit\"{a}t Mainz, Institut f\"{u}r Physik, Staudinger Weg 7, DE - 55099 Mainz, Germany\\
$^{82}$ University of Manchester, School of Physics and Astronomy, Manchester M13 9PL, United Kingdom\\
$^{83}$ CPPM, Aix-Marseille Universit\'e, CNRS/IN2P3, Marseille, France\\
$^{84}$ University of Massachusetts, Department of Physics, 710 North Pleasant Street, Amherst, MA 01003, United States of America\\
$^{85}$ McGill University, High Energy Physics Group, 3600 University Street, Montreal, Quebec H3A 2T8, Canada\\
$^{86}$ University of Melbourne, School of Physics, AU - Parkville, Victoria 3010, Australia\\
$^{87}$ The University of Michigan, Department of Physics, 2477 Randall Laboratory, 500 East University, Ann Arbor, MI 48109-1120, United States of America\\
$^{88}$ Michigan State University, Department of Physics and Astronomy, High Energy Physics Group, East Lansing, MI 48824-2320, United States of America\\
$^{89}$ INFN Sezione di Milano$^{(a)}$; Universit\`a  di Milano, Dipartimento di Fisica$^{(b)}$, via Celoria 16, IT - 20133 Milano, Italy\\
$^{90}$ B.I. Stepanov Institute of Physics, National Academy of Sciences of Belarus, Independence Avenue 68, Minsk 220072, Republic of Belarus\\
$^{91}$ National Scientific \& Educational Centre for Particle \& High Energy Physics, NC PHEP BSU, M. Bogdanovich St. 153, Minsk 220040, Republic of Belarus\\
$^{92}$ Massachusetts Institute of Technology, Department of Physics, Room 24-516, Cambridge, MA 02139, United States of America\\
$^{93}$ University of Montreal, Group of Particle Physics, C.P. 6128, Succursale Centre-Ville, Montreal, Quebec, H3C 3J7  , Canada\\
$^{94}$ P.N. Lebedev Institute of Physics, Academy of Sciences, Leninsky pr. 53, RU - 117 924 Moscow, Russia\\
$^{95}$ Institute for Theoretical and Experimental Physics (ITEP), B. Cheremushkinskaya ul. 25, RU 117 218 Moscow, Russia\\
$^{96}$ Moscow Engineering \& Physics Institute (MEPhI), Kashirskoe Shosse 31, RU - 115409 Moscow, Russia\\
$^{97}$ Lomonosov Moscow State University Skobeltsyn Institute of Nuclear Physics (MSU SINP), 1(2), Leninskie gory, GSP-1, Moscow 119991 Russian Federation, Russia\\
$^{98}$ Ludwig-Maximilians-Universit\"at M\"unchen, Fakult\"at f\"ur Physik, Am Coulombwall 1,  DE - 85748 Garching, Germany\\
$^{99}$ Max-Planck-Institut f\"ur Physik, (Werner-Heisenberg-Institut), F\"ohringer Ring 6, 80805 M\"unchen, Germany\\
$^{100}$ Nagasaki Institute of Applied Science, 536 Aba-machi, JP Nagasaki 851-0193, Japan\\
$^{101}$ Nagoya University, Graduate School of Science, Furo-Cho, Chikusa-ku, Nagoya, 464-8602, Japan\\
$^{102}$ INFN Sezione di Napoli$^{(a)}$; Universit\`a  di Napoli, Dipartimento di Scienze Fisiche$^{(b)}$, Complesso Universitario di Monte Sant'Angelo, via Cinthia, IT - 80126 Napoli, Italy\\
$^{103}$  University of New Mexico, Department of Physics and Astronomy, MSC07 4220, Albuquerque, NM 87131 USA, United States of America\\
$^{104}$ Radboud University Nijmegen/NIKHEF, Department of Experimental High Energy Physics, Heyendaalseweg 135, NL-6525 AJ, Nijmegen, Netherlands\\
$^{105}$ Nikhef National Institute for Subatomic Physics, and University of Amsterdam, Science Park 105, 1098 XG Amsterdam, Netherlands\\
$^{106}$ Department of Physics, Northern Illinois University, LaTourette Hall
Normal Road, DeKalb, IL 60115, United States of America\\
$^{107}$ Budker Institute of Nuclear Physics (BINP), RU - Novosibirsk 630 090, Russia\\
$^{108}$ New York University, Department of Physics, 4 Washington Place, New York NY 10003, USA, United States of America\\
$^{109}$ Ohio State University, 191 West Woodruff Ave, Columbus, OH 43210-1117, United States of America\\
$^{110}$ Okayama University, Faculty of Science, Tsushimanaka 3-1-1, Okayama 700-8530, Japan\\
$^{111}$ University of Oklahoma, Homer L. Dodge Department of Physics and Astronomy, 440 West Brooks, Room 100, Norman, OK 73019-0225, United States of America\\
$^{112}$ Oklahoma State University, Department of Physics, 145 Physical Sciences Building, Stillwater, OK 74078-3072, United States of America\\
$^{113}$ Palack\'y University, 17.listopadu 50a,  772 07  Olomouc, Czech Republic\\
$^{114}$ University of Oregon, Center for High Energy Physics, Eugene, OR 97403-1274, United States of America\\
$^{115}$ LAL, Univ. Paris-Sud, IN2P3/CNRS, Orsay, France\\
$^{116}$ Osaka University, Graduate School of Science, Machikaneyama-machi 1-1, Toyonaka, Osaka 560-0043, Japan\\
$^{117}$ University of Oslo, Department of Physics, P.O. Box 1048,  Blindern, NO - 0316 Oslo 3, Norway\\
$^{118}$ Oxford University, Department of Physics, Denys Wilkinson Building, Keble Road, Oxford OX1 3RH, United Kingdom\\
$^{119}$ INFN Sezione di Pavia$^{(a)}$; Universit\`a  di Pavia, Dipartimento di Fisica Nucleare e Teorica$^{(b)}$, Via Bassi 6, IT-27100 Pavia, Italy\\
$^{120}$ University of Pennsylvania, Department of Physics, High Energy Physics Group, 209 S. 33rd Street, Philadelphia, PA 19104, United States of America\\
$^{121}$ Petersburg Nuclear Physics Institute, RU - 188 300 Gatchina, Russia\\
$^{122}$ INFN Sezione di Pisa$^{(a)}$; Universit\`a   di Pisa, Dipartimento di Fisica E. Fermi$^{(b)}$, Largo B. Pontecorvo 3, IT - 56127 Pisa, Italy\\
$^{123}$ University of Pittsburgh, Department of Physics and Astronomy, 3941 O'Hara Street, Pittsburgh, PA 15260, United States of America\\
$^{124}$ Laboratorio de Instrumentacao e Fisica Experimental de Particulas - LIP$^{(a)}$, Avenida Elias Garcia 14-1, PT - 1000-149 Lisboa, Portugal; Universidad de Granada, Departamento de Fisica Teorica y del Cosmos and CAFPE$^{(b)}$, E-18071 Granada, Spain\\
$^{125}$ Institute of Physics, Academy of Sciences of the Czech Republic, Na Slovance 2, CZ - 18221 Praha 8, Czech Republic\\
$^{126}$ Charles University in Prague, Faculty of Mathematics and Physics, Institute of Particle and Nuclear Physics, V Holesovickach 2, CZ - 18000 Praha 8, Czech Republic\\
$^{127}$ Czech Technical University in Prague, Zikova 4, CZ - 166 35 Praha 6, Czech Republic\\
$^{128}$ State Research Center Institute for High Energy Physics, Moscow Region, 142281, Protvino, Pobeda street, 1, Russia\\
$^{129}$ Rutherford Appleton Laboratory, Science and Technology Facilities Council, Harwell Science and Innovation Campus, Didcot OX11 0QX, United Kingdom\\
$^{130}$ University of Regina, Physics Department, Canada\\
$^{131}$ Ritsumeikan University, Noji Higashi 1 chome 1-1, JP - Kusatsu, Shiga 525-8577, Japan\\
$^{132}$ INFN Sezione di Roma I$^{(a)}$; Universit\`a  La Sapienza, Dipartimento di Fisica$^{(b)}$, Piazzale A. Moro 2, IT- 00185 Roma, Italy\\
$^{133}$ INFN Sezione di Roma Tor Vergata$^{(a)}$; Universit\`a di Roma Tor Vergata, Dipartimento di Fisica$^{(b)}$ , via della Ricerca Scientifica, IT-00133 Roma, Italy\\
$^{134}$ INFN Sezione di  Roma Tre$^{(a)}$; Universit\`a Roma Tre, Dipartimento di Fisica$^{(b)}$, via della Vasca Navale 84, IT-00146  Roma, Italy\\
$^{135}$ R\'eseau Universitaire de Physique des Hautes Energies (RUPHE): Universit\'e Hassan II, Facult\'e des Sciences Ain Chock$^{(a)}$, B.P. 5366, MA - Casablanca; Centre National de l'Energie des Sciences Techniques Nucleaires (CNESTEN)$^{(b)}$, B.P. 1382 R.P. 10001 Rabat 10001; Universit\'e Mohamed Premier$^{(c)}$, LPTPM, Facult\'e des Sciences, B.P.717. Bd. Mohamed VI, 60000, Oujda ; Universit\'e Mohammed V, Facult\'e des Sciences$^{(d)}$4 Avenue Ibn Battouta, BP 1014 RP, 10000 Rabat, Morocco\\
$^{136}$ CEA, DSM/IRFU, Centre d'Etudes de Saclay, FR - 91191 Gif-sur-Yvette, France\\
$^{137}$ University of California Santa Cruz, Santa Cruz Institute for Particle Physics (SCIPP), Santa Cruz, CA 95064, United States of America\\
$^{138}$ University of Washington, Seattle, Department of Physics, Box 351560, Seattle, WA 98195-1560, United States of America\\
$^{139}$ University of Sheffield, Department of Physics \& Astronomy, Hounsfield Road, Sheffield S3 7RH, United Kingdom\\
$^{140}$ Shinshu University, Department of Physics, Faculty of Science, 3-1-1 Asahi, Matsumoto-shi, JP - Nagano 390-8621, Japan\\
$^{141}$ Universit\"{a}t Siegen, Fachbereich Physik, D 57068 Siegen, Germany\\
$^{142}$ Simon Fraser University, Department of Physics, 8888 University Drive, CA - Burnaby, BC V5A 1S6, Canada\\
$^{143}$ SLAC National Accelerator Laboratory, Stanford, California 94309, United States of America\\
$^{144}$ Comenius University, Faculty of Mathematics, Physics \& Informatics$^{(a)}$, Mlynska dolina F2, SK - 84248 Bratislava; Institute of Experimental Physics of the Slovak Academy of Sciences, Dept. of Subnuclear Physics$^{(b)}$, Watsonova 47, SK - 04353 Kosice, Slovak Republic\\
$^{145}$ Stockholm University: Department of Physics$^{(a)}$; The Oskar Klein Centre$^{(b)}$, AlbaNova, SE - 106 91 Stockholm, Sweden\\
$^{146}$ Royal Institute of Technology (KTH), Physics Department, SE - 106 91 Stockholm, Sweden\\
$^{147}$ Stony Brook University, Department of Physics and Astronomy, Nicolls Road, Stony Brook, NY 11794-3800, United States of America\\
$^{148}$ University of Sussex, Department of Physics and Astronomy
Pevensey 2 Building, Falmer, Brighton BN1 9QH, United Kingdom\\
$^{149}$ University of Sydney, School of Physics, AU - Sydney NSW 2006, Australia\\
$^{150}$ Insitute of Physics, Academia Sinica, TW - Taipei 11529, Taiwan\\
$^{151}$ Technion, Israel Inst. of Technology, Department of Physics, Technion City, IL - Haifa 32000, Israel\\
$^{152}$ Tel Aviv University, Raymond and Beverly Sackler School of Physics and Astronomy, Ramat Aviv, IL - Tel Aviv 69978, Israel\\
$^{153}$ Aristotle University of Thessaloniki, Faculty of Science, Department of Physics, Division of Nuclear \& Particle Physics, University Campus, GR - 54124, Thessaloniki, Greece\\
$^{154}$ The University of Tokyo, International Center for Elementary Particle Physics and Department of Physics, 7-3-1 Hongo, Bunkyo-ku, JP - Tokyo 113-0033, Japan\\
$^{155}$ Tokyo Metropolitan University, Graduate School of Science and Technology, 1-1 Minami-Osawa, Hachioji, Tokyo 192-0397, Japan\\
$^{156}$ Tokyo Institute of Technology, 2-12-1-H-34 O-Okayama, Meguro, Tokyo 152-8551, Japan\\
$^{157}$ University of Toronto, Department of Physics, 60 Saint George Street, Toronto M5S 1A7, Ontario, Canada\\
$^{158}$ TRIUMF$^{(a)}$, 4004 Wesbrook Mall, Vancouver, B.C. V6T 2A3; $^{(b)}$York University, Department of Physics and Astronomy, 4700 Keele St., Toronto, Ontario, M3J 1P3, Canada\\
$^{159}$ University of Tsukuba, Institute of Pure and Applied Sciences, 1-1-1 Tennoudai, Tsukuba-shi, JP - Ibaraki 305-8571, Japan\\
$^{160}$ Tufts University, Science \& Technology Center, 4 Colby Street, Medford, MA 02155, United States of America\\
$^{161}$ Universidad Antonio Narino, Centro de Investigaciones, Cra 3 Este No.47A-15, Bogota, Colombia\\
$^{162}$ University of California, Irvine, Department of Physics \& Astronomy, CA 92697-4575, United States of America\\
$^{163}$ INFN Gruppo Collegato di Udine$^{(a)}$; ICTP$^{(b)}$, Strada Costiera 11, IT-34014, Trieste; Universit\`a  di Udine, Dipartimento di Fisica$^{(c)}$, via delle Scienze 208, IT - 33100 Udine, Italy\\
$^{164}$ University of Illinois, Department of Physics, 1110 West Green Street, Urbana, Illinois 61801, United States of America\\
$^{165}$ University of Uppsala, Department of Physics and Astronomy, P.O. Box 516, SE -751 20 Uppsala, Sweden\\
$^{166}$ Instituto de F\'isica Corpuscular (IFIC) Centro Mixto UVEG-CSIC, Apdo. 22085  ES-46071 Valencia, Dept. F\'isica At. Mol. y Nuclear; Univ. of Valencia, and Instituto de Microelectr\'onica de Barcelona (IMB-CNM-CSIC) 08193 Bellaterra Barcelona, Spain\\
$^{167}$ University of British Columbia, Department of Physics, 6224 Agricultural Road, CA - Vancouver, B.C. V6T 1Z1, Canada\\
$^{168}$ University of Victoria, Department of Physics and Astronomy, P.O. Box 3055, Victoria B.C., V8W 3P6, Canada\\
$^{169}$ Waseda University, WISE, 3-4-1 Okubo, Shinjuku-ku, Tokyo, 169-8555, Japan\\
$^{170}$ The Weizmann Institute of Science, Department of Particle Physics, P.O. Box 26, IL - 76100 Rehovot, Israel\\
$^{171}$ University of Wisconsin, Department of Physics, 1150 University Avenue, WI 53706 Madison, Wisconsin, United States of America\\
$^{172}$ Julius-Maximilians-University of W\"urzburg, Physikalisches Institute, Am Hubland, 97074 W\"urzburg, Germany\\
$^{173}$ Bergische Universit\"{a}t, Fachbereich C, Physik, Postfach 100127, Gauss-Strasse 20, D- 42097 Wuppertal, Germany\\
$^{174}$ Yale University, Department of Physics, PO Box 208121, New Haven CT, 06520-8121, United States of America\\
$^{175}$ Yerevan Physics Institute, Alikhanian Brothers Street 2, AM - 375036 Yerevan, Armenia\\
$^{176}$ ATLAS-Canada Tier-1 Data Centre, TRIUMF, 4004 Wesbrook Mall, Vancouver, BC, V6T 2A3, Canada\\
$^{177}$ GridKA Tier-1 FZK, Forschungszentrum Karlsruhe GmbH, Steinbuch Centre for Computing (SCC), Hermann-von-Helmholtz-Platz 1, 76344 Eggenstein-Leopoldshafen, Germany\\
$^{178}$ Port d'Informacio Cientifica (PIC), Universitat Autonoma de Barcelona (UAB), Edifici D, E-08193 Bellaterra, Spain\\
$^{179}$ Centre de Calcul CNRS/IN2P3, Domaine scientifique de la Doua, 27 bd du 11 Novembre 1918, 69622 Villeurbanne Cedex, France\\
$^{180}$ INFN-CNAF, Viale Berti Pichat 6/2, 40127 Bologna, Italy\\
$^{181}$ Nordic Data Grid Facility, NORDUnet A/S, Kastruplundgade 22, 1, DK-2770 Kastrup, Denmark\\
$^{182}$ SARA Reken- en Netwerkdiensten, Science Park 121, 1098 XG Amsterdam, Netherlands\\
$^{183}$ Academia Sinica Grid Computing, Institute of Physics, Academia Sinica, No.128, Sec. 2, Academia Rd.,   Nankang, Taipei, Taiwan 11529, Taiwan\\
$^{184}$ UK-T1-RAL Tier-1, Rutherford Appleton Laboratory, Science and Technology Facilities Council, Harwell Science and Innovation Campus, Didcot OX11 0QX, United Kingdom\\
$^{185}$ RHIC and ATLAS Computing Facility, Physics Department, Building 510, Brookhaven National Laboratory, Upton, New York 11973, United States of America\\
$^{a}$ Also at CPPM, Marseille, France.\\
$^{b}$ Also at TRIUMF,  Vancouver,  Canada\\
$^{c}$ Also at FPACS, AGH-UST,  Cracow, Poland\\
$^{d}$ Also at TRIUMF, Vancouver, Canada\\
$^{e}$ Now at CERN\\
$^{f}$ Also at  Universit\`a di Napoli  Parthenope, Napoli, Italy\\
$^{g}$ Also at Institute of Particle Physics (IPP), Canada\\
$^{h}$ Also at  Universit\`a di Napoli  Parthenope, via A. Acton 38, IT - 80133 Napoli, Italy\\
$^{i}$ Louisiana Tech University, 305 Wisteria Street, P.O. Box 3178, Ruston, LA 71272, United States of America   \\
$^{j}$ At California State University, Fresno, USA\\
$^{k}$ Also at TRIUMF, 4004 Wesbrook Mall, Vancouver, B.C. V6T 2A3, Canada\\
$^{l}$ Currently at Istituto Universitario di Studi Superiori IUSS, Pavia, Italy\\
$^{m}$ Also at FPACS, AGH-UST, Cracow, Poland\\
$^{n}$ Also at California Institute of Technology,  Pasadena, USA \\
$^{o}$ Louisiana Tech University, Ruston, USA  \\
$^{p}$ Also at University of Montreal, Montreal, Canada\\
$^{q}$ Also at Institut f\"ur Experimentalphysik, Universit\"at Hamburg,  Hamburg, Germany\\
$^{r}$ Also at Petersburg Nuclear Physics Institute, Gatchina, Russia\\
$^{s}$ Also at Institut f\"ur Experimentalphysik, Universit\"at Hamburg,  Luruper Chaussee 149, 22761 Hamburg, Germany\\
$^{t}$ Also at School of Physics and Engineering, Sun Yat-sen University, China\\
$^{u}$ Also at School of Physics, Shandong University, Jinan, China\\
$^{v}$ Also at California Institute of Technology, Pasadena, USA\\
$^{w}$ Also at Rutherford Appleton Laboratory, Didcot, UK \\
$^{x}$ Also at school of physics, Shandong University, Jinan\\
$^{y}$ Also at Rutherford Appleton Laboratory, Didcot , UK\\
$^{z}$ Now at KEK\\
$^{aa}$ University of South Carolina, Columbia, USA \\
$^{ab}$ Also at KFKI Research Institute for Particle and Nuclear Physics, Budapest, Hungary\\
$^{ac}$ University of South Carolina, Dept. of Physics and Astronomy, 700 S. Main St, Columbia, SC 29208, United States of America\\
$^{ad}$ Also at Institute of Physics, Jagiellonian University, Cracow, Poland\\
$^{ae}$ Louisiana Tech University, Ruston, USA\\
$^{af}$ Also at School of Physics and Engineering, Sun Yat-sen University, Taiwan\\
$^{ag}$ University of South Carolina, Columbia, USA\\
$^{ah}$ Transfer to LHCb 31.01.2010\\
$^{ai}$ Also at Nanjing University, China\\
$^{*}$ Deceased
}

%\input{foreword}

%
% The abstract
%
\abstract{The ATLAS Inner Detector is a composite tracking system consisting of 
silicon pixels, silicon strips and straw tubes in a 2~T magnetic field. 
Its installation was 
completed in August 2008 and the detector took part in data-taking with 
single LHC beams and cosmic rays. The initial detector operation, 
hardware commissioning and in-situ calibrations are described.
Tracking performance has been measured with 7.6~million cosmic-ray events, 
collected using a tracking trigger and reconstructed with modular 
pattern-recognition and fitting software.
The intrinsic hit efficiency and tracking trigger efficiencies are 
close to 100\%. Lorentz angle measurements for both electrons and holes,
specific energy-loss calibration and transition radiation turn-on 
measurements 
have been performed.
Different alignment techniques have been used to 
reconstruct the detector geometry. After the initial 
alignment, a transverse impact parameter 
resolution of
$22.1\pm 0.9$~$\mu$m  
and a relative momentum resolution  
$\sigma_p/p=(4.83\pm 0.16)\times 10^{-4}\ \mathrm{GeV}^{-1} \times\pT$
have been measured for high momentum tracks.
%The excellent resolution observed in this early commissioning run will 
%ensure that the design performance will be reached shortly after the first LHC 
%collisions.
}
\authorrunning{G. Aad et al.}
\titlerunning{The ATLAS Inner Detector commissioning}
\maketitle
%
%%%%%%%%%%%%%%%%%%%%%%%%%%%%%%%%%%%%%%%%%%%%%%%%%%%%%%%%%%%%%%%%%%%%%%%%%%%%%%%
% Introduction
%%%%%%%%%%%%%%%%%%%%%%%%%%%%%%%%%%%%%%%%%%%%%%%%%%%%%%%%%%%%%%%%%%%%%%%%%%%%%%%
%
%\linenumbers

\section{Introduction}

The ATLAS detector \cite{bib:ATLASDetectorPaper} is one of two large general-purpose detectors 
designed to probe new physics at the unprecedented energies and luminosities available at the 
Large Hadron Collider at CERN~\cite{bib:LHCPaper}. ATLAS is divided into three major regions: 
a large toroidal-field high-precision muon spectrometer surrounding a set of high-granularity 
calorimeters which, in turn, surround an optimized, multi-technology tracker situated in a 
2~T magnetic field provided by a solenoid. 

This central tracking detector is referred to as the Inner Detector (ID). 
This paper describes the commissioning and calibration of the Inner Detector from its final 
installation in August of 2008 through cosmic-ray data-taking until the end of the year. In 
this period the full tracking system operated for the first time. The aim of this commissioning 
phase was to prepare the detector for LHC collisions which took place in 2009. The necessary steps were:
\begin{itemize}
\item to operate all the services and controls,
\item to perform an in-situ calibration of the detector,
\item to synchronise all sub-detectors,
\item to measure efficiency and noise occupancy for each sub-detector in combined operation,
\item to test the reconstruction software and the tracking triggers on real data,
\item to perform an initial alignment of the detector. 
\end{itemize}

A significant component of the commissioning involved setting up the hardware and 
software infrastructure needed to operate the detector. This included the calibration
procedures, which will be repeated regularly during proton-proton data-taking periods. The most 
relevant aspects are therefore described here.

Cosmic-ray events were used to perform a preliminary alignment and 
to commission the track reconstruction. They
mostly consist of a single muon traversing the whole detector,
and have a hard momentum spectrum.
Their kinematics makes them particularly suitable for some specific measurements,
for example intrinsic detector efficiency, track resolution and 
study of detector response to ionisation as a function of momentum and 
incident angle.

The layout of the paper is as follows. 
The main components of the ID are briefly described in Section~\ref{sec:ID}. The operating modes 
and conditions during the different data-taking periods, the reconstruction software and
the tracking triggers are described in Section~\ref{sec:data-operation}. 
The synchronisation of the sub-detectors is presented in Section~\ref{sec:timing} and the calibration procedures and results
in Section~\ref{sec:commissioning}.
Section~\ref{sec:alignment} describes the alignment, while Section~\ref{sec:performance} presents
measurements of the detector performance: intrinsic efficiency, 
the Lorentz angle in silicon for both electrons and holes, resolution of tracking parameters,
the specific energy loss for particle 
identification at low momentum and the observation of transition radiation turn-on.

In the following, the ATLAS coordinate system will be used. The nominal interaction point is defined as the origin of a right-handed coordinate system. The beam direction defines the $z$-axis and the $x$-$y$ plane is transverse to it. The positive $x$-axis is defined as pointing from the interaction point to the centre of the
LHC ring and the positive $y$-axis points upwards. Cylindrical coordinates $R$ and $\phi$ are often used in the transverse plane. 
The pseudorapidity $\eta$ is defined in terms of the polar angle $\theta$:
$\eta=-\ln\tan(\theta/2)$.

Tracks are described using the parameters of a helical trajectory at the point of closest approach 
to the $z$-axis: the transverse impact parameter, $d_0$, the $z$ coordinate, $z_0$, the angles of 
the momentum direction, $\phi_0$ and $\theta$, and the inverse of the particle momentum multiplied 
by the charge, $q/p$.

%
%%%%%%%%%%%%%%%%%%%%%%%%%%%%%%%%%%%%%%%%%%%%%%%%%%%%%%%%%%%%%%%%%%%%%%%%%%%%%%%
% Detector
%%%%%%%%%%%%%%%%%%%%%%%%%%%%%%%%%%%%%%%%%%%%%%%%%%%%%%%%%%%%%%%%%%%%%%%%%%%%%%%
%
\section{The ATLAS Inner Detector}
\label{sec:ID}
\begin{figure*}
\includegraphics[width=2\columnwidth]{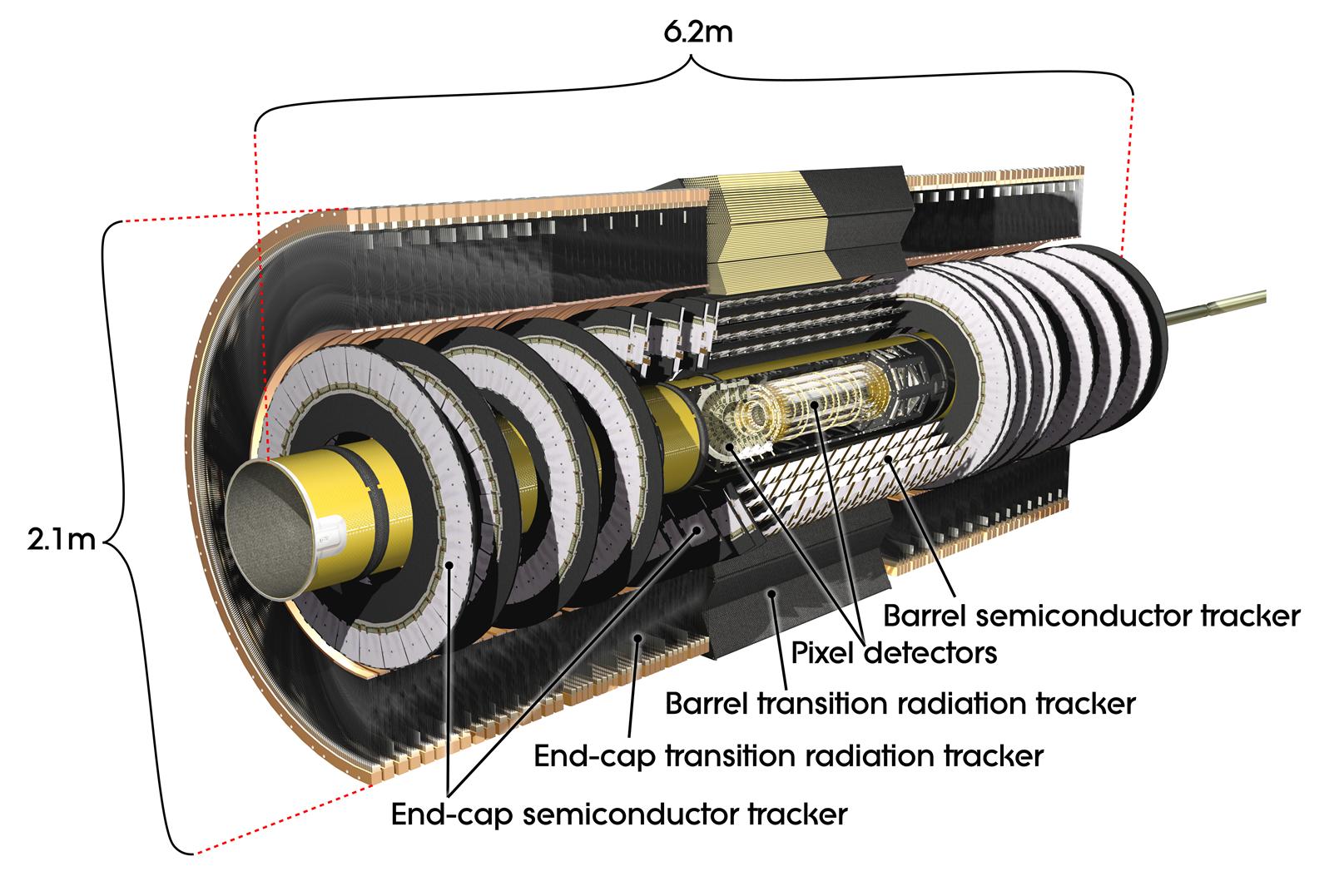}
\caption{Cut-away image of the ATLAS Inner Detector}
\label{fig:InnerDetector}
\end{figure*}

%%%%%%%%%%%%%%%%%%%%%%%
% Intro
%%%%%%%%%%%%%%%%%%%%%%%
The layout of the Inner Detector is shown in Fig.~\ref{fig:InnerDetector}. The 
acceptance in pseudorapidity is $|\eta|<2.5$ for particles coming from the LHC 
beam-interaction region, with full coverage in $\phi$. The detector has been designed 
to provide a transverse momentum resolution, in the plane perpendicular to the beam axis, of 
%$\sigma_{p_{\mathrm T}}/p_{\mathrm T} = 0.05\% p_{\mathrm T} \mbox{ [GeV]}\oplus1 \%$ 
$\sigma_{\pT}/\pT = 0.05\% \pT \GeV \oplus1 \%$ 
and a transverse impact parameter resolution of 10~$\mu$m for high momentum particles 
in the central $\eta$ region~\cite{bib:ATLASDetectorPaper}. The Inner Detector comprises
three complementary sub-detectors: the Pixel Detector, the SemiConductor Tracker and the 
Transition Radiation Tracker.
Relevant features are described briefly below; full details can be found 
in~\cite{bib:ATLASDetectorPaper}. 

%%%%%%%%%%%%%%%%%%%%%%%
% Pixel
%%%%%%%%%%%%%%%%%%%%%%%
{\bf The Pixel Detector} sensitive elements cover radial distances between 50.5~mm and 150~mm. 
The detector consists of 1744 silicon pixel modules~\cite{bib:PixelDetectorPaper} 
arranged in three concentric barrel layers and two endcaps of three disks each.
It provides typically three measurement points for particles originating in the 
beam-interaction region.
Each module covers an active area of 16.4~mm$\times$60.8~mm and contains 47\,232 pixels,
most of size 50~$\mu$m~$\times$~400~$\mu$m. The direction of the shorter 
pitch defines the local $x$-coordinate on the module and corresponds to the 
high-precision position measurement in the $R\phi$ plane. 
The longer pitch, corresponding to the local $y$-coordinate,
is oriented approximately along the $z$ direction in the 
barrel and along $R$ in the endcaps.  A module is read out by 16 radiation-hard 
front-end chips~\cite{bib:FEI3chip} bump-bonded to the sensor; the total number
of readout channels is $\sim$80.4 million. 
Hits in a pixel are read out if the signal 
exceeds a tunable threshold. 
The pulse height is measured 
%with 8-bit dynamic range 
using the Time-over-Threshold (ToT) technique. 

%%%%%%%%%%%%%%%%%%%%%%%
% SCT
%%%%%%%%%%%%%%%%%%%%%%%
{\bf The SemiConductor Tracker} (SCT) sensitive elements span radial distances from 
299~mm to 560~mm. The detector consists of 4088 modules of silicon-strip 
detectors arranged in four concentric barrels 
%with radii between 255~mm and 549~mm, half-length 805~mm, 
and two endcaps of nine disks each. It provides typically eight strip measurements
(four space-points) for particles originating in the beam-interaction region.
The strips in the barrel are approximately
parallel to the solenoid field and beam axis, and have a constant pitch of
80~$\mu$m, while in the endcaps the strip direction is radial and
of variable pitch. Most modules~\cite{bib:SCTBarrelModules,bib:SCTEndcapModules}
consist of four silicon-strip sensors~\cite{bib:SCTsensors}; two sensors on each 
side are daisy-chained together to give 768 strips of approximately 12~cm in length.
A second pair of identical sensors is glued back-to-back with the first pair at a 
stereo angle of 40~mrad to provide space points. The strips are read out by 
radiation-hard front-end readout chips~\cite{bib:SCTABCD3TA}, each chip 
reading out 128 channels; the total number of readout channels is $\sim$6.3 million.
The hit information is binary: a hit is registered if the
pulse height in a channel exceeds a preset threshold, normally corresponding 
to a charge of 1~fC. 
%The modules are 
%calibrated by injecting known test charges into the front-end chips and measuring the
%occupancy (fraction of triggered events for which the pulse height exceeds the
%threshold) as a function of threshold. The resulting threshold scan data are
%analysed online to generate optimised calibrations for physics 
%data-taking~\cite{bib:SCTDAQ}. 
%For the cosmic-ray data presented here, hit information
%was recorded in the triggered bunch-crossing cycle, and in the cycles immediately
%before and immediately after the trigger; the resulting hit patterns allow the
%timing of the SCT with respect to the cosmic-ray trigger to be checked.

Measurements in the silicon detectors often perform a selection on the angle of a track incident 
on a module. 
The angle between a track and the normal to the plane of a sensor is called~$\alpha$.
The angle between a track and the normal to the sensor in the plane defined by the normal 
to the sensor and the local $x$-axis (i.e. the axis in the plane of the sensor corresponding
to the high-precision measurement in the Pixel Detector or perpendicular to the strip direction 
in the SCT) is termed $\phi_{\rm local}$.

%%%%%%%%%%%%%%%%%%%%%%%
% TRT
%%%%%%%%%%%%%%%%%%%%%%%
{\bf The Transition Radiation Tracker} (TRT) sensitive volume covers radial distances from
563~mm to 1066~mm.
%is the outermost sub-detector of the Inner 
%Detector, filling the solenoidal magnet volume from about 50~cm to 100~cm in $R$ and from 
%--2.7~m to +2.7~m in $z$. The TRT 
The detector consists of 298\,304 proportional drift tubes (straws), 4~mm in diameter, read out 
by 350\,848 channels of electronics. 
The straws in the barrel region are arranged in three cylindrical layers and 32 $\phi$ 
sectors; they have split anodes and are read out from each side~\cite{bib:TRTBarrel}. 
The straws in the endcap regions are radially oriented and arranged in 
80~wheel-like modular structures~\cite{bib:TRTEndcap}.
%In the central pseudorapidity region from roughly --1 to +1, 52544 
%axially oriented straw tubes with split anodes are arranged in 96 modules 
%(three cylindrical layers and 32 $\phi$ sectors). They are read out from each side 
%in the TRT barrel~\cite{bib:TRTBarrel}.
%In each of the TRT endcaps~\cite{bib:TRTEndcap}, extending from a rapidity of 1 to about 2
%on each side of the barrel, there are 122880 radially oriented straws organized in
%eighteen wheel-like modular structures. 
The TRT straw layout is designed so that charged 
particles with transverse momentum $\pT > 0.5 \GeV$ and with pseudorapidity 
$|\eta| < 2.0$ cross typically more than 30 straws.  
The TRT provides electron identification via transition radiation from polypropylene fibres
(barrel) or foils (endcaps) interleaved between the straws. The much higher energy of the
transition radiation photons ($\sim 6$~keV compared with the few hundred eV deposited by 
an ionising particle in the Xe, CO$_2$, O$_2$ gas) is detected by a second, high-threshold, discriminator in the 
radiation-hard front-end electronics~\cite{bib:TRTElectronics}. 

%%%%%%%%%%%%%%%%%%%%%%%
% BCM
%%%%%%%%%%%%%%%%%%%%%%%
{\bf The Beam Conditions Monitor}~(BCM)~\cite{bib:BCMPaper} is designed to
monitor the rate of background particles and to protect
the silicon trackers from instantaneous
high radiation doses caused by LHC beam incidents. 
The BCM consists of two stations, forward and backward, each with four modules located at a 
radius of 5.5~cm and at a distance of $\pm 1.84$~m from
the interaction point. Each module has two 
pCVD diamond sensors of $1\times 1~\mathrm{cm}^2$ surface area and 500 $\mu$m thickness mounted back-to-back.
The 1~ns signal rise-time allows the discrimination of particle hits due to collisions (in-time)
from background (out-of-time). The BCM signal provides both trigger information 
and an instantaneous hit-rate used as input to a beam-abort signal.

%%%%%%%%%%%%%%%%%%%%%%%
% Readout
%%%%%%%%%%%%%%%%%%%%%%%
{\bf Readout systems.} The Pixel and SCT detectors' readout systems use optical 
transmission for the outgoing module data and the incoming timing, trigger and control 
data. The transmission is based on VCSELs operating at a wavelength of 
850~nm and radiation-hard fibres\cite{bib:SCTOpticalLinks,bib:PixelOpticalLinks}. 
For each SCT module, there are two optical links operating at 40~Mbits/s
for the data readout. 
%and one optical link for the 40~MHz bunch crossing clock and the 40~Mbits/s trigger and 
%control data streams. 
Redundancy is implemented to allow for the loss of one optical link,
without significant loss of data. 
%The Pixel Detectors links operate at the same speed as 
%those for the SCT, except for the innermost layer which has data links running at 
%80~Mbits/s.
For the cosmic-ray data-taking, the Pixel Detector links also operated at 40~MBits/s. 
%For the SCT the optoelectronics is mounted either on the front end modules or on flex 
%circuits connected to the modules. The Pixel Detector readout uses twisted pair cables 
%to transmit the data from the modules to opto-boards at a patch panel where the signals 
%are converted to optical ones. 
The TRT uses shielded twisted-pair lines to transfer data to a patch panel inside 
the muon spectrometer, where up to 31 lines are multiplexed~\cite{bib:GOL} into one 
1.6 Gbits/s optical link.  

The off-detector readout electronics is based on custom-made Read-Out Driver (ROD) 
modules~\cite{bib:SirOD,bib:TRTROD}.
%The end of the optical links for the Pixels and SCT are in the Back of Crate (BOC) cards in the ROD. 
%The optoelectronics for the BOCs for Pixels and SCT is based on arrays of VCSELs and {\it p-i-n} silicon diodes. 
The RODs gather the data belonging to a single trigger into one packet (and in the case of the TRT perform 
data compression) and transmit the data to the ATLAS readout system using optical links operating at 
1.6~Gbits/s\cite{bib:GOL}. The RODs also perform monitoring and calibration tasks\cite{bib:SCTDAQ}.

%%%%%%%%%%%%%%%%%%%%%%%
% Cooling
%%%%%%%%%%%%%%%%%%%%%%%
%{\bf Cooling.} The silicon detectors of the Inner Detector are cooled with a bi-phase 
%system which uses the large latent heat of vaporization to achieve the required cooling 
%power with very low mass flow~\cite{bib:CoolingPaper}. 
{\bf Cooling.} The silicon detectors are cooled with a bi-phase 
evaporative system~\cite{bib:CoolingPaper} which is designed to deliver C$_3$F$_8$ 
fluid at $-25$~$^\circ$C in the low-mass cooling structures on the detector.
%allows the removal of a maximum of 16 kW and 46 kW power dissipated in the 
%Pixel Detector and SCT respectively. 
%The required cooling power is achieved with very low mass 
%flow by using a liquid with large latent heat of vaporization. 
The target temperature for the silicon sensors after irradiation is 0~$^\circ$C for the 
Pixel Detector and $-7$~$^\circ$C for the SCT; these values were chosen to mitigate the 
effects of radiation damage. 
In the commissioning phase in 2008 both detectors limited the 
coolant temperature to $-10$~$^\circ$C in the circuits cooling their sensors. The 
resulting sensor temperatures were in the range $-7$~$^\circ$C to $+5$~$^\circ$C, 
depending on layer and module type. In 2009 the coolant temperature was reduced.
Sensor temperatures were in the range $-17$~$^\circ$C to $-7$~$^\circ$C for the
Pixel Detector and $-7$~$^\circ$C to $-2$~$^\circ$C for the SCT.
%2009: SCT B3-5 set point -20 -> sensor -2.5
%          B6             -10 ->        +5
%          EC             -14 ->        -7 
%The refrigerant is distributed across the detector through 204 parallel 
%circuits which can be opened and closed individually. 
%To avoid condensation the SCT and Pixel Detector volumes 
%are flushed with N$_2$.

In contrast to the silicon detectors, the TRT operates at room temperature. The 
electronics is cooled by a monophase-liquid cooling loop separate from the Pixel and SCT 
bi-phase system.

\section{Data samples and operation conditions}
\subsection{Data-taking periods}
\label{sec:data-operation}
\label{sec:data-taking}

In 2008 the Inner Detector participated in three main data-taking periods:
\begin{itemize}
 \item 
  Single-beam LHC running. Particularly relevant were the so called 
  {\em beam-splash} events, where the LHC beams were directed
  into the tertiary 
  collimators located 150~m from the interaction point, in order to provide 
  secondary particles crossing the whole cross-section of the ATLAS detector.
  Since the incident particles had a direction almost parallel to the beam axis, 
  they crossed many detector elements and were used for synchronization of the 
  individual TRT readout units (see Section~\ref{sec:timing}).  
  For reasons of detector safety, during this period the Pixel Detector
  and SCT barrel were switched off and the SCT endcaps were operated at a
  reduced bias voltage of 20~V instead of 150~V, with the readout threshold increased 
  to 1.2~fC to reduce the data volume. 
 \item
  Combined ATLAS cosmic-ray run. Data were taken by the full ATLAS detector with 
  different magnetic field combinations: toroid and solenoid switched on and 
  off independently. 
  %For Inner Detector calibration the most relevant condition is the status of the 
  %solenoid: in the following, the term {\em solenoid~on} indicates the 
  %condition 
  %in which the solenoid field was switched on to its design value of 2~T, and 
  %{\em solenoid~off} refers to the condition when the solenoid was switched off. 
 \item
  Standalone ID cosmic-ray run. Only the Inner Detector took part in this run,
  which used a newly introduced Level-1 tracking trigger 
  (see Section~\ref{sec:trktriggers}). All data taken during this period were
  with the solenoid off. 
\end{itemize} 

Cosmic rays come predominantly from the vertical direction. They were therefore particularly 
useful for studying the barrel region of the detector, where they resemble 
particles from collisions.

In the time between the combined and standalone cosmic-ray data-taking periods, 
a complete tuning and calibration of the detectors was performed as detailed in 
Section~\ref{sec:commissioning}.

A summary of the numbers of reconstructed tracks in the 2008 cosmic-ray 
data-taking periods is shown in Table~\ref{tab:cosmicstat}.
Similar data-taking periods in 2009 have been used to confirm the performance
achieved in the 2008 commissioning period.

\begin{table}
\begin{center}
\begin{tabular}{lrr}
  \hline\hline
  Detector & Solenoid off & Solenoid on  \\ 
  \hline
  All      & 4~940~000 & 2~670~000 \\ 
  $\ge$1 SCT hit      & 1~150~000 &   880~000 \\ 
  $\ge$1 Pixel hit   &   230~000 &   190~000 \\ 
  \hline\hline
  \end{tabular}
\end{center}
\caption{Number of tracks collected during the 2008 cosmic-ray runs. Numbers
         are given for all reconstructed Inner Detector tracks, those having
         at least one SCT hit and those having at least one Pixel hit. }
\label{tab:cosmicstat}
\end{table}

\subsection{Operating conditions}
\label{sec:operation}
Most of the detector was operational during the cosmic-ray data-taking periods.
Loss of coverage was mainly due to issues with the recently-commissioned evaporative
cooling system and the optical links. 
The fractions of non-operational channels in each sub-detector are summarised in
Table~\ref{tab:deadfrac}. 

In the Pixel Detector three cooling loops, each serving 12 modules, showed apparent
leaks, two on the positive-$z$ endcap and one on the negative-$z$ endcap.
For safety, these loops were disabled in 2008, but were operated successfully in 2009, 
after the installation of a leak-monitoring system during the winter shutdown.
In the SCT, 36 modules in the negative-$z$ endcap were turned off because of problems 
in two cooling loops. One of these loops was repaired after the end of 2008 operation, 
resulting in the recovery of 23 modules. 

A major problem with the optical links for the SCT and Pixel detectors
was the failure of VCSEL arrays in the off-detector electronics. 
The loss of data for the SCT was reduced because of the redundancy system,
but the problem prevented the read-out of 35 pixel modules in the combined run.
These were recovered by replacing the defective VCSEL arrays with spare parts between 
the combined and standalone data-taking periods. 
The VCSEL failures are believed to be due to Electro Static Discharge (ESD) 
damage.
During the 2008--2009 shutdown all VCSEL arrays in the off-detector electronics were 
replaced with new components produced with much tighter ESD controls.
A very low rate of problems was observed in 2009.

Remaining inactive parts in the Pixel Detector and SCT were mainly due to failure
in high- or low-voltage connections.

%A further 24 SCT modules could not be operated because of low- or high-voltage 
%problems. In addition there were 27 dead chips (128 strips each) and about 
%10k individual bad strips;
%these problems were spread fairly randomly across the detector. 
%At the end of the 2008 run the active fraction of the SCT was 97.9\%. 
%
%Remaining inactive parts in the Pixel Detector were 28 modules, 
%the bigger fraction due to failure in the high-voltage connection, 
%while individual channel failures are at the level of 0.2\%. The optoboard repair
%resulted in an improvement of the Pixel Detector active fraction 
%from 94\%\ during the combined run to 96\%\ in the standalone data-taking. 
%
%The TRT was in a stable configuration during this time with a total of 2\% of the 
%channels inactive. 
In the TRT barrel   1.6\% of the straws were inactive due to 
mechanical problems in the detector which had been noted prior to installation and 0.7\% 
were inactive due to scattered electronics problems at the board and chip level after 
installation. In the endcaps about 1.6\% of the electronics channels were inactive,
largely due to high- and low-voltage power connection problems, while only 0.3\% of the straws 
had known mechanical problems. 
The mechanical defects were always straw cathodes that
had been deformed during module or wheel construction so that they would 
not reliably hold high-voltage, and in these
cases the anode wires were removed. These numbers 
remained essentially constant throughout the 2008 and 2009 data-taking periods.

\begin{table}
\begin{center}
\begin{tabular}{llcc}
  \hline\hline
  Detector &Reason & 2008 & 2009  \\ 
  \hline
  Pixel    &Cooling        &2.1\%          &0.0\%   \\
           &Optical links  &2.0\% -- 0.0\% &0.3\%   \\
           &Other          &1.9\%          &2.4\%   \\ \cline{2-4}
           &Total          &6.0\% -- 4.0\% &2.7\%   \\ 
  \hline
  SCT      &Cooling        &0.9\%          &0.3\%  \\ 
           &Optical links  &0.4\%          &0.0\%  \\
           &Other          &0.8\%          &0.7\%  \\ \cline{2-4}
           &Total          &2.1\%          &1.0\%  \\
  \hline
  TRT      &Total   &2.0\% &2.0\%  \\
  \hline\hline
  \end{tabular}
\end{center}
\caption{Fraction of non-operational channels for each sub-detector in the 2008 cosmic-ray run and at the beginning of LHC collisions in 2009. 
For the Pixel Detector 
         in 2008 the first numbers correspond to the earlier combined run, the second to the
         later standalone run.}
\label{tab:deadfrac}
\end{table}

The detector conditions were supervised and monitored by a Detector Control System~\cite{bib:DCSPaper}, which monitored high-voltage and low-voltage values, temperatures and 
other environmental parameters. In particular the applied bias voltage on the silicon detectors 
was used to compute the Lorentz angle~(Section~\ref{sec:Lorentz}) during track 
reconstruction, and the detector status was used to assess the data quality.

Monitoring software~\cite{bib:Monitoring} running within the ATLAS Athena 
framework~\cite{bib:AthenaCore} was used to analyse data 
and to reconstruct tracks as described in Section~\ref{sec:tracking}, 
both online during the 
physics run and during offline reconstruction. The light-weight online 
monitoring ran on a limited subset of data, while the offline monitoring 
provided more in-depth analysis over larger samples of data.

%During the 2008 run both the online and offline versions of the
%monitoring systems were thoroughly tested on all parts of the detector in operation.
%During the standalone data-taking period, 
%almost 3.3 million tracks were recorded by the monitoring packages with a track rate
%of 8--10 Hz, consisting mainly by TRT segments alone whereas, due to their smaller
%acceptance, the SCT and
%Pixel Detector segment rate was around 2.5 and 0.5 Hz respectively. The average number
%of Pixel, SCT and TRT hits per track was recorded by the monitoring to be on
%average 4, 15 and 35 for the respective subsystem. The noise levels recorded
%both online and offline were as expected and no significant noise correlations
%between the three detectors were observed.  The monitoring software was not 
%only used to supervise the collection of data, check track parameter
%distributions and determine the data quality but was also able to catch 
%problems and errors occurring during a run, such as synchronisation 
%failures and readout errors.

\subsection{Track reconstruction}
\label{sec:tracking}

% Editor Andreas.Wildauer@cern.ch
% references: A. Salzburger (Editor) et al, The new ATLAS Track Reconstruction (NEWT), ATLAS Public
% Note, ATL-SOFT-PUB-2007-007, 2007.
Data were reconstructed using ATLAS software in the Athena framework~\cite{bib:AthenaCore}. 
In a first step, groups of contiguous pixels (in the Pixel Detector) or strips (in the SCT) 
with a hit were grouped into clusters. Channels which were noisy, as determined from either 
online calibration data or offline monitoring, were rejected at this stage. The one-dimensional 
strip clusters from the two sides of an SCT module were combined into three-dimensional 
space-points using knowledge of the stereo angle and the radial (longitudinal) positions of the 
barrel (endcap) modules; 
in the case of pixel clusters, only the knowledge of the radial (longitudinal) position was 
necessary to construct a barrel (endcap) space-point. 
The construction of TRT drift circles, i.e. the radial distance of the particle trajectory to the 
wire in a tube, required knowledge of the time of the cosmic ray passing through, which was 
determined using the iterative procedure described in Section~\ref{sec:timing}. 
The three-dimensional space-points, in the Pixel Detector and SCT, and the drift circles, in the 
TRT, formed the input to the pattern-recognition algorithms.

The track reconstruction~\cite{bib:NewTrackingPubNote} started the pattern recognition by using space-points from the silicon detectors.
In cosmic-ray data, these track candidates were allowed to span the central beam-axis region, and no cut was
placed on the transverse impact parameter $d_0$. These silicon-only tracks were extended in both directions into
the TRT, and refitted using all associated space-points from the silicon and TRT detectors.
As shown in Table~\ref{tab:cosmicstat}, a significant fraction of tracks from cosmic rays do not pass through the silicon detectors,
and these were found by running a TRT stand-alone track-finding algorithm on the remaining measurements.
At all stages, the track fitting was performed using the global $\chi^2$ fitter described in~\cite{bib:GlobalChi2TrackFitter}.

\begin{figure}
\centering
\subfigure[Momentum distribution]{
\label{fig:d0_resolution_vs_PT}
\includegraphics[width=\columnwidth]{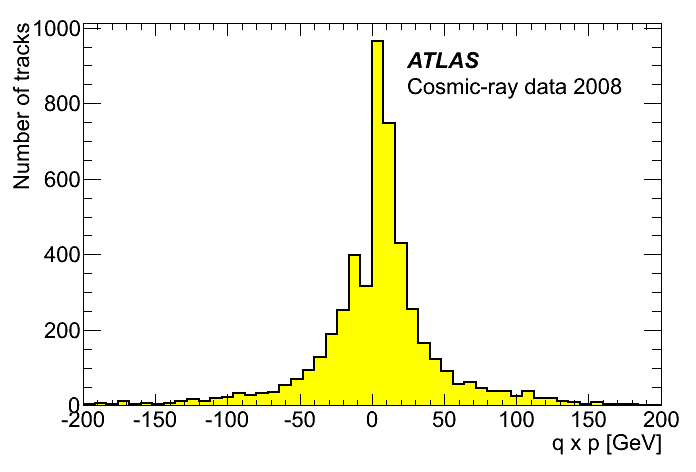}
}
\subfigure[$\phi_0$ distribution]{
\label{fig:z0_resolution_vs_PT}
\includegraphics[width=\columnwidth]{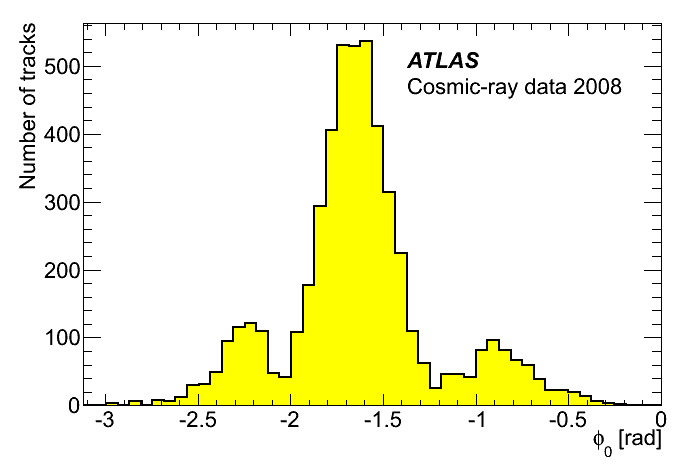}
}
\subfigure[$\theta$ distribution]{
\label{fig:phi0_resolution_vs_PT}
\includegraphics[width=\columnwidth]{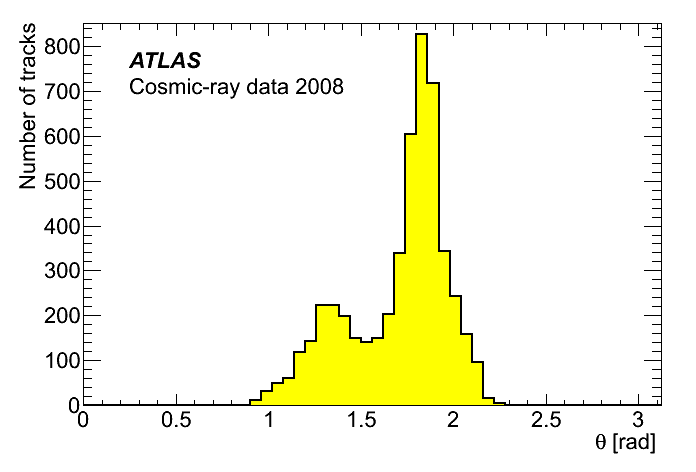}
}

\caption{Distribution of split-track parameters for a set of cosmic-ray data with solenoid on: 
(a) particle charge multiplied by momentum ($q\times p$), 
(b) azimuthal ($\phi_0$) and (c) polar ($\theta$) angles.}
\label{fig:IP_TrackParameters}
\end{figure}

To measure the resolution of the track parameters the cosmic-ray tracks which traverse the ATLAS 
detector from top to bottom were split into two halves. This was done by fitting two new tracks, 
each containing the hits in the upper or lower half of the detector only. These new tracks are 
referred to as split tracks. Figure~\ref{fig:IP_TrackParameters} shows the momentum and angular 
distributions of the split tracks as measured in data. The shapes of the $\phi_0$ and $\theta$ 
distributions reflect the fact that particles could enter the ATLAS cavern through the access 
shafts more easily than through the rock. The range of $\phi_0$ is always negative as the split tracks in both the upper and lower halves of the detector are reconstructed from top to bottom. The high $\mu^+/\mu^-$ asymmetry in the low momentum bins in \ref{fig:d0_resolution_vs_PT} is due to the toroid deflecting $\mu^-$ coming from the shafts away from the ID. The resolution results are presented in Section~\ref{sec:trackpar}.
% \begin{figure}
% \begin{tabular}{cc}
% \includegraphics[width=0.4\textwidth]{figures/CosmicsStandardPlots_realVSsim_generalNote.png}
% \end{tabular}
% \caption{Comparison between Monte Carlo simulation and real data of the transverse momentum distrubution of tracks reconstructed by the tracking algorithm. MORE PLOTS WILL BE ADDED HERE FOR THETA AND PHI.}
% \label{fig:TrackEffMC}
% \end{figure}

\subsection{Tracking triggers}
\label{sec:trktriggers}

The ATLAS trigger system has a three-level architecture: Level-1, Level-2 and
Event Filter. Level-2 and Event Filter together form the High Level Trigger 
(HLT)~\cite{bib:ATLASDetectorPaper}.

The trigger for cosmic-ray events was provided by the muon or calorimeter systems at Level-1. 
For the ID standalone data-taking, a Level-1 TRT trigger was added, based on a fast digital OR of 
 groups of approximately 200 TRT straws~\cite{bib:FastOr}. 
%For most of the data-taking period, the HLT was configured to run the trigger algorithms to add 
%information to the event used for streaming, but not to reject events. Running the HLT in active 
%mode, saving only events containing reconstructed tracks at the trigger level, was also 
%successfully tested. 

%One of the output data streams contains events with an ID track and is used for offline ID 
%alignment. It is therefore important to have an efficient track-based trigger that has as 
%uniform coverage as possible over the entire ID volume.

Three Inner Detector tracking algorithms were run at Level-2. One algorithm
%, TrigTRTSegFinder, 
was specifically designed for cosmic-ray running and used only barrel TRT information. It 
reconstructed tracks in a search window of up to about 45$^\circ$ to the vertical in azimuthal 
angle. The other two algorithms~\cite{bib:HLTTracking} were designed for collisions but were adapted for cosmic-ray 
running in order to exercise the algorithms online and also to complement the coverage of the TRT 
trigger. These algorithms started with track reconstruction in the silicon detectors and then 
extrapolated tracks to the TRT. As a consequence of being designed for collisions, the 
cosmic-particle trajectory was reconstructed as two tracks: one going upwards and the other 
downwards.  
The two algorithms used a common input consisting of space-points formed from clusters of hits 
in the pixel layers and from associated stereo-layer hits in the SCT. They shared common tools 
for track fitting and extrapolation to the TRT, but differed in the initial track-finding step:
\begin{itemize}
\item
SiTrack was based on a combinatorial method. It first looked for pairs of space-points in the inner 
layers consistent with beam-line constraints, then combined these pairs with space-points in other
layers to form triplets and finally merged triplets to form track candidates. In order to achieve
good efficiency in cosmic-ray data-taking, the beam-line constraints were relaxed compared with
those used for collision data. 
\item
IDSCAN used a three-stage histogramming method to first determine the $z$-coordinate (position 
along the beam) of the interaction point in collision events, and then look for track candidates 
consistent with this interaction point. For cosmic-ray data-taking a first step was introduced 
which shifted the space-points in the direction transverse to the beam-axis, so that the shifted 
points lay on a trajectory passing close to the nominal beam position.
\end{itemize}

\begin{figure}[htbp]
\centering
\includegraphics[width=\columnwidth]{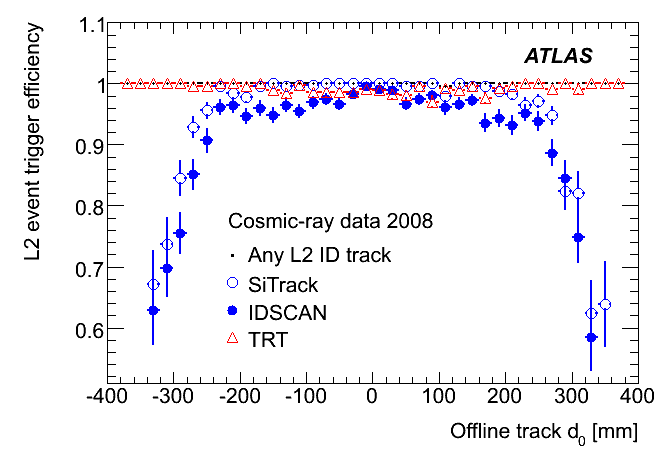}
\caption{Efficiency of Level-2 tracking algorithms in cosmic-ray events, as a function of $d_0$; the efficiency drop for the silicon based algorithms at about 300~mm corresponds to the acceptance of the first SCT barrel layer.}
\label{fig:L2Efficiency}
\end{figure}

The efficiency of the Level-2 ID cosmic-ray trigger was determined using events triggered by 
the Level-1 muon trigger and containing an offline ID track. In Fig.~\ref{fig:L2Efficiency} the 
efficiency is shown as a function of the transverse impact parameter of the offline track, $d_0$, 
for each of the three different algorithms as well as for the combined trigger. The efficiency was 
calculated for the sample of offline tracks with 3+3 space-points on the upper+lower track segments
in the silicon barrel. The track was also required to be within the TRT readout time window. The 
efficiency for IDSCAN and SiTrack falls off for tracks with $d_0$ approaching the radius of the 
first SCT layer (300~mm). The space-point shifting step that precedes IDSCAN fails for high 
curvature tracks, and this is reflected in a lower efficiency for IDSCAN. The combined efficiency is $(99.96\pm0.02)\%$.

\subsection{Simulation}
Cosmic-ray events were simulated by a sequence which
first generated single particles at the surface above ATLAS, then
filtered them for acceptance in the detector and finally ran the
standard detector simulation, digitisation and reconstruction.

The generator used the flux calculations in Ref.~\cite{bib:muonFlux1983}
and a standard cosmic-ray momentum spectrum~\cite{bib:PDG}.
Muons pointing to a sphere representing the inside of the experimental 
cavern were propagated through the rock, cavern structures and the detector
using simulation software based on GEANT4~\cite{bib:geant4a,bib:geant4b}.
To increase the acceptance, only events with at least one hit in a given volume 
inside the detector were submitted to the digitization algorithms and the event 
reconstruction. The digitisation was adapted to reproduce the timing properties
of cosmic-ray muons (see Section~\ref{sec:timing}), and tracks were reconstructed as described in 
Section~\ref{sec:tracking}.

\section{Detector timing}     
\label{sec:timing}

All sub-detectors use a common clock signal, with a 25~ns period corresponding
to the spacing of LHC bunch-crossings (BC).
This is either an ATLAS internal clock
or one provided by the LHC and synchronised to the bunch-crossing. 
A delay to this signal is then applied by each detector component in order 
to account for signal propagation times.  

A major difference between cosmic-ray running and detector operation with LHC collisions is that cosmic-ray events occur evenly distributed in the interval 
between two clock edges.
In order to properly treat cosmic-ray events, it is therefore necessary to measure for each event the time difference between the clock edge and the passage of the cosmic-ray particle. This time difference is then an input to the track reconstruction and analysis. The TRT timing determines the precision of this measurement,
because the granularity
of its leading-edge measurement is $3.125\,\rm{ns}$ ($1/8$ of a BC)
instead of one~BC as for the silicon detectors. It is therefore used as a reference. The broader readout window of the Pixel Detector helped in verifying the coarse selection of beam clock offsets for both the TRT and SCT, and in understanding the trigger time offsets for the various triggers used in cosmic-ray data-taking.

\subsection{TRT timing}

%The TRT measures the time of the leading-edge and trailing-edge of a signal 
%with the granularity of 3.125~ns. 

TRT timing requirements are set by the constraint that 
both the leading-edge and trailing-edge transitions of a signal 
must be within the $75$~ns (three BC) readout window.
About $50\,\rm{ns}$ are required to cover the range of electron drift 
times at the full $2\,\rm{T}$ magnetic field.  Propagation time differences
 within a front-end board are about $5\,\rm{ns}$ and, combined with  small cabling
and time-of-flight effects, imply that a time offset bigger than 
$10\,\rm{ns}$ would result in acceptance losses.  
The readout timing was initially synchronized across the 
detector using measured cable lengths, which gave a spread of $\pm 5\,\rm{ns}$ in the barrel,
and within one bunch-crossing in the endcaps.

\begin{figure}[htbp]
\begin{center}
% \begin{tabular}{cc}
%Subfigure (a)
\subfigure[Time offsets $T_0$ for TRT barrel A ]{
\label{fig:TRT_timing_beamSplash}
\includegraphics[width=\columnwidth]{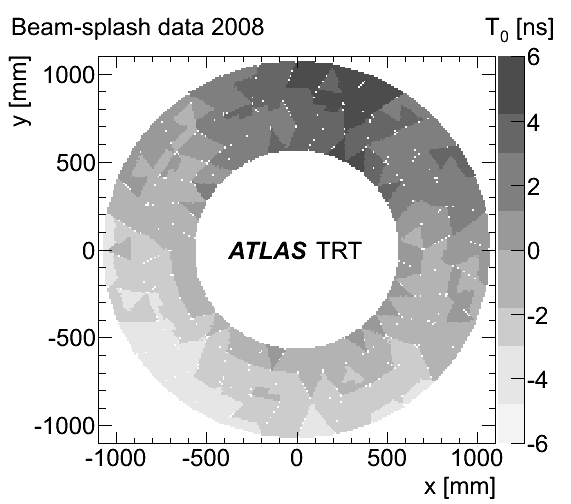}
}
%Subfigure (b)
\subfigure[$T_{\rm TRT}$ difference]{
\label{fig:TRT_event_phase}
\includegraphics[width=\columnwidth]{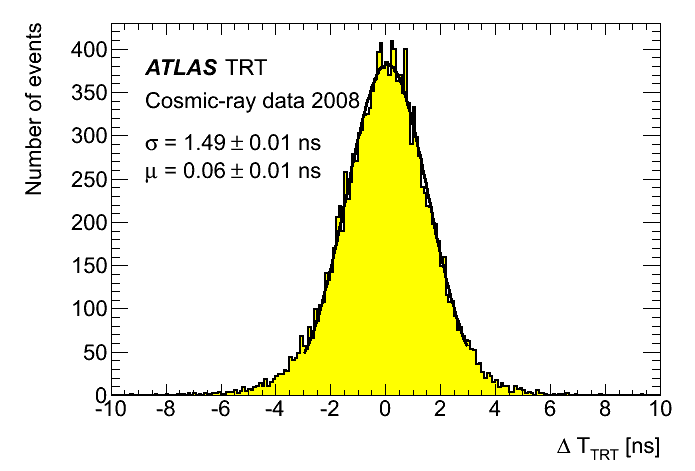}
} 
% \end{tabular}
\end{center}
\caption{\subref{fig:TRT_timing_beamSplash}~Validation of TRT $T_0$ hardware settings in TRT barrel A with September 2008 beam-splash data. 
\subref{fig:TRT_event_phase}~Difference between the $T_{\rm TRT}$ value obtained from the upper and lower parts of a split track for a sample of cosmic-ray tracks.
\label{fig:TRT_timing}}
\end{figure}

In the barrel region, the time offset $T_0$ for each 
Trigger, Timing and Control unit~\cite{bib:TRTElectronics} 
was improved using cosmic-ray tracks, and the corresponding 
corrections were applied to the hardware settings.
These offsets were validated using the LHC beam-splash events.
In these events many particles passed through the 
detector at the same time. Almost every TRT straw was hit multiple times and, 
apart from time-of-flight effects, different parts of the detector were 
hit simultaneously. 
Figure~\ref{fig:TRT_timing_beamSplash} shows $T_0$ settings 
which were estimated with a single beam-splash event. Since the readout 
timing before beam-splash events had already been adjusted using 
cosmic-ray events, the systematic effect due to time-of-flight in 
cosmic-ray data can clearly be seen.  
Apart from this, the measured time is
uniform, with variations of about $1\,\rm{ns}$. These settings were monitored in the subsequent running periods and they have remained stable.

In the endcap regions very
few cosmic-ray events had been collected by September 2008. 
The initial correction was derived from beam-splash data. 
This adjustment was validated using cosmic-ray data and, after subtracting the 
time-of-flight,
the measured $T_0$ constants in the endcap showed an accuracy of 1.3~ns.
%After subtracting the systematic time-of-flight effect 
%for cosmic rays 
%the measured $T_0$ constants in the endcap had a spread of only 
%$1.3\,\rm{ns}$.
%, mainly due to the uncertainty from the cosmic-ray results. 

In the cosmic-ray
run the TRT time measurement was used to determine the time, $T_{\rm TRT}$, 
of a cosmic ray passing through the ID. This was determined by
the average of measured TRT leading-edge times for all hits 
on a track, corrected for electron drift time and offline $T_0$ 
calibration constants (see Section~\ref{sec:calibrations}). 
Since the estimated electron drift time depends on the track trajectory,
the track was first fit
using only the position of the centre of each hit wire, without using the 
drift-time information. These track parameters were then used to estimate 
$T_{\rm TRT}$ and this estimate was used to correct the 
position of TRT hits and to repeat the track fit.

The accuracy of this $T_{\rm TRT}$ measurement procedure was studied by 
splitting the cosmic-ray track into upper and lower parts and fitting 
$T_{\rm TRT}$ separately for each. The time difference between the two 
segments is shown in Fig.~\ref{fig:TRT_event_phase}.
The resolution is estimated as the 
spread of this difference, divided by two. This factor assumes a statistical 
error only, and is a combination of a $\sqrt{2}$ due to both upper and 
lower $T_{\rm TRT}$ uncertainties contributing to the spread, and another
factor of $\sqrt{2}$ because split tracks have half the number of hits. 
%The procedure was checked with cosmic-ray Monte Carlo simulated events, 
%where the measured $T_{\rm TRT}$ can also be compared with the true 
%(simulated) value. 
The accuracy of $T_{\rm TRT}$ for barrel tracks in
the 2008 cosmic-ray data was shown to be better than 
$1\,\rm{ns}$.
%\footnote{\it 
%Recent study by Brett shows that the  $T_{\rm TRT}$ accuracy is actually 
%about twice as good once you exclude outliers in time residual
%(as it should have been, based on single hit time resolution). 
%Hopefully he can get final results for data in time and we can then 
%update this part accordingly.}

%This TRT time measurement provides the time reference which is used to synchronize the rest of ID.

%{\it Shift the scale in Fig. (a) such that the average is at 0. Divide the x-scale in Fig. (b) by 2  such that the plot directly shows full track resolution. 
%Add MC cross-check  in Fig (b), i.e. same distribution as is shown in data and measured - truth distribution. }

\subsection{Pixel Detector timing}

\label{sec:PixelTiming}

The Pixel Detector front-end electronics can read out up
to 16~consecutive BC for each trigger~\cite{bib:FEI3chip}. 
Each recorded hit includes the number of the BC in which
it occurred.

At luminosities higher than $10^{32}$~cm$^{-2}$s$^{-1}$, the expected 
occupancy will only permit read-out of a single BC per trigger. 
In cosmic-ray data-taking the low trigger rate allows a broader
time window. In the 2008 commissioning run, eight~BC 
were read out per trigger.

\begin{figure}%![ht]
     \centering
     %\subfigure[Pixel timing relative to $T_{\rm TRT}$]{
     \subfigure[Timing of pixel clusters]{
          \label{fig:allClusterL1A:vsTRT}
          \includegraphics[width=\columnwidth]{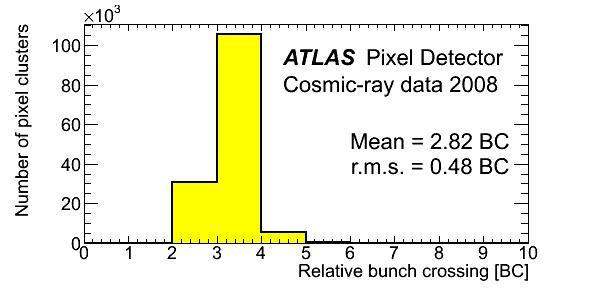}}  
     %\subfigure[Average Pixel timing per-module relative to $T_{\rm TRT}$]{
     \subfigure[Average pixel module timing]{
          \label{fig:moduleClusterL1A:vsTRT}
          \includegraphics[width=\columnwidth]{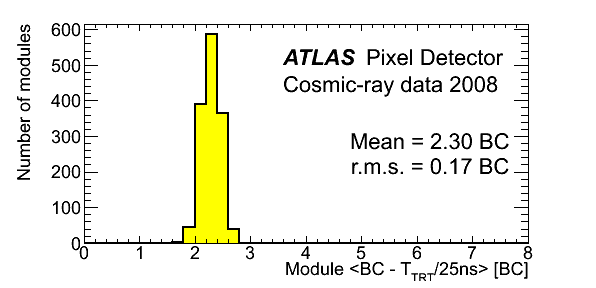}}  
     \caption{ Pixel Detector BC distributions for individual clusters on track \subref{fig:allClusterL1A:vsTRT} and per-module average BC relative to the $T_{\rm TRT}$ in units of 25~ns \subref{fig:moduleClusterL1A:vsTRT}. 
The dispersion in \subref{fig:allClusterL1A:vsTRT} is due to timewalk and event
time spread, while in \subref{fig:moduleClusterL1A:vsTRT} is the module-to-module synchronization. 
}
     \label{fig:allClusterL1A}
    \end{figure}

The BC distribution for hits from cosmic-ray muons is shown in Fig.~\ref{fig:allClusterL1A:vsTRT}. The spread is due to the convolution of the front-end electronics timewalk, 
which results in low pulse-height hits being assigned to a late BC,
and to the uniform time distribution of cosmic rays. 

The distribution of hits among bunch crossings can be used to  
improve the detector timing relative to the corrections
computed from   measured signal delays   in   cables and read-out electronics.

Module-to-module synchronization in the barrel was assessed averaging 
the BC, corrected for $T_{\rm TRT}$, of clusters 
with a pulse height greater than 15\,000~$e$. The subtraction of $T_{\rm TRT}$ reduces
the spread due to the event time and the requirement on pulse height 
removes the timewalk effect.
The measured values are shown in 
Fig.~\ref{fig:moduleClusterL1A:vsTRT} and indicate a time variation
of 0.17~BC, equivalent to 4.25~ns without any specific module-to-module tuning.
This is sufficient to obtain full efficiency in the readout window used for
detector commissioning. To reduce the spread and extend the tuning to the
endcap region, the higher statistics from collision events will be
needed.

%------------------------------------------------------------------------

\subsection{SCT timing} 

The readout of the SCT needs to be synchronized with the bunch-crossing time to ensure 
that the signal is sampled at the peak of the charge-response curve. 
In cosmic-ray data-taking, a strip is read out if the signal is above threshold in 
any one of three 25~ns time-bins centred on the triggered bunch-crossing.

%The latency of the Level-1 Accept signal can be tuned by two methods. Firstly
%the transmission of the clock-and-control stream from the BOC to the modules
%can be tuned in steps of 25~ns LHC clock cycles and/or steps of $\sim$280~ps.
%Secondly, the Timing Interface Module (TIM) can be used to delay signals relevant
%to triggering in steps of 25ns LHC clock cycles. The BOC transmission delays were 
%used to compensate for fibre-length variations between the BOC and the modules
%prior to the cosmic-ray runs. During cosmic-ray data-taking, the SCT was
%timed in by gradually changing the TIM delays until a peak in occupancy 
%associated with tracks was observed. No attempt was made to refine this timing 
%using the BOC transmisson delays, and no corrections for time-of-flight were 
%applied.

Prior to cosmic-ray data-taking, the timing of each module was adjusted to
compensate for differences in the lengths of the optical fibres used for
data transmission to and from the modules. During data-taking, the overall
timing of the SCT was adjusted in steps of 25~ns until a peak in occupancy 
associated with tracks was observed. No attempt was made to refine this timing 
using finer adjustments, and no corrections for time-of-flight were applied.

\begin{figure}[htbp]
  \begin{center}
  %\subfigure[ ]{\label{fig:SCTtimingfit}
      %\includegraphics[width=\columnwidth,viewport=0 0 567 368,clip]{figures/SCTtimingfit.pdf} 
      \includegraphics[width=\columnwidth]{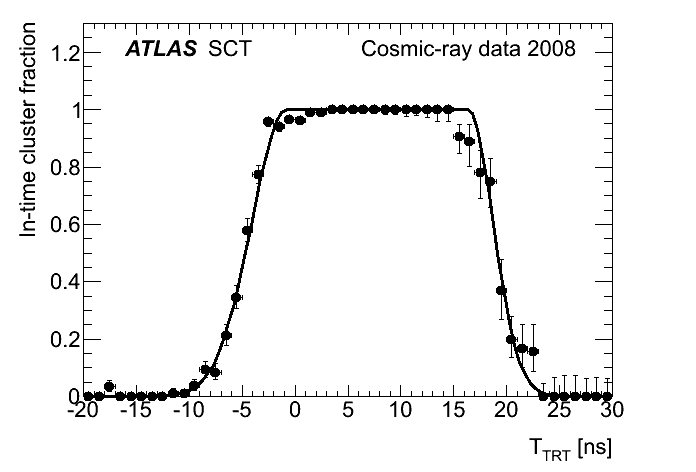} 
  %}
  %\subfigure[ ]{\label{fig:peaktime}
  %\includegraphics[width=\columnwidth,viewport=0 0 567 368,clip]{figures/peaktime1D.pdf}
  %} 
  \end{center}
  \caption{Fraction of in-time clusters on track as a function of $T_{\rm TRT}$ 
   for SCT barrel modules. The curve shows a fit to a step function convolved 
   with two Gaussian functions. The peak time of the response curve is assumed to
   be at the centre of the step function.
   %\subref{fig:peaktime} The fitted peak of the timing curve for each harness in the SCT barrel.
  }
\label{fig:scttiming}
\end{figure}

The degree of synchronisation of the SCT was studied using the cosmic-ray
timing derived from the TRT.
Figure~\ref{fig:scttiming} shows the fraction of in-time clusters on a
track as a function of $T_{\rm TRT}$ for barrel modules. The clusters were
required to contain at least two strips, all from the same BC, to reduce the effect of variations in the 
charge-collection time. The distribution has a flat top with
a width of about 25~ns and can be fitted to a step function convolved with two Gaussian functions. The peak time of the charge response corresponds to the mid-point of the step function. Separate fits have been
performed for the SCT barrel modules served by a single optical-fibre 
`harness' (each harness serves six modules on a barrel at the same azimuthal 
angle). Most of 
the barrel harnesses are well synchronised: the r.m.s.\ width of the distribution 
is 1.8~ns.

\subsection{BCM timing}

\begin{figure}
\includegraphics[width=\columnwidth]{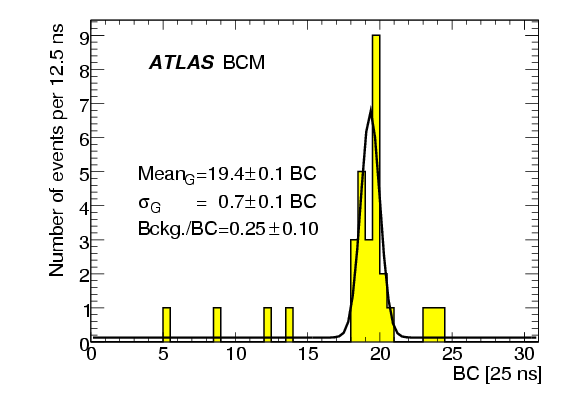}
\caption{Timing distribution of BCM events triggered by the TRT Fast-Or. The 
data are fitted with a Gaussian over a flat background.}
\label{f:BCMtiming}
\end{figure}

Even though the BCM acceptance for cosmic rays is very 
limited, during the November 2008 operation, a total of 131 events had muons passing through 
this detector. These allowed the relative timing between the BCM signal and the 
trigger to be measured. From the timing distributions, an offset of $19.5\pm 0.4$~BC was observed for triggers based on the muon system
and of $19.4\pm 0.1$~BC for the events triggered by the TRT Fast-Or, as shown in Fig.~\ref{f:BCMtiming}. These 
observed time offsets agree well with the expectation of 19~BC from the 
estimation of propagation time along cables and optical fibres.

\section{Sub-detector calibration}
\label{sec:commissioning}
%%%Calibration Text%%%%

\begin{figure*}[ht]
\begin{center}
\begin{tabular}{cc}
\includegraphics[width=0.45\textwidth]{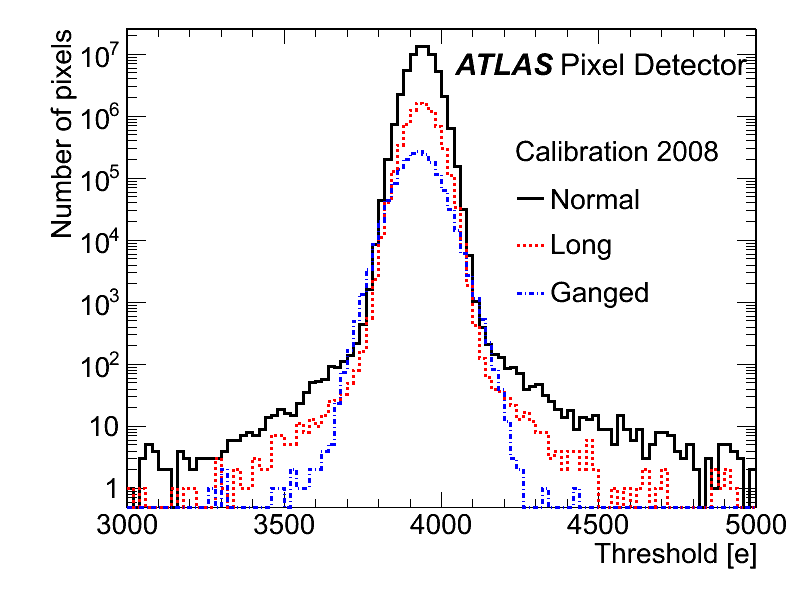} &
\includegraphics[width=0.45\textwidth]{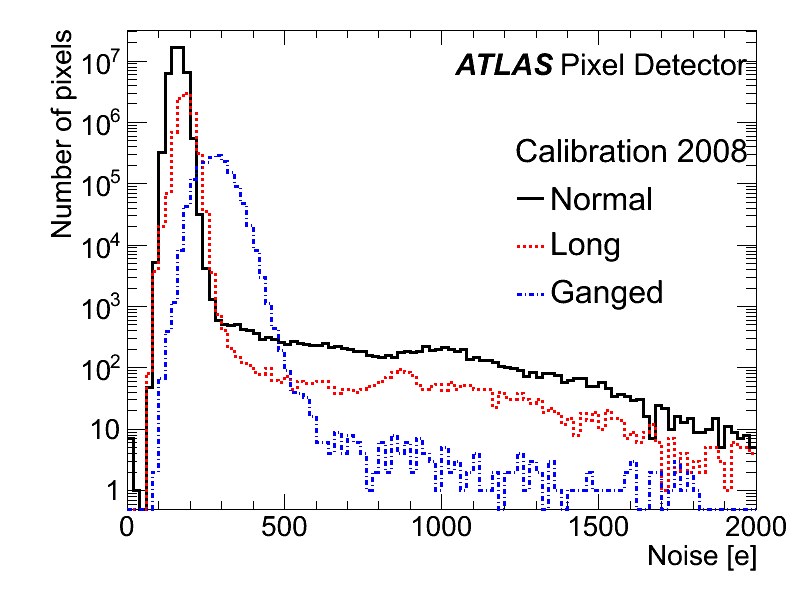} \\
(a) Pixel Detector thresholds & (b) Pixel Detector noise 
\end{tabular}
\end{center}
\caption{Pixel Detector threshold~(a) and noise~(b) distributions,
as obtained from
in-situ calibrations based on charge injection.}
\label{fig:PixelCalib}
\end{figure*}

To be prepared for data-taking, each sub-detector performs a set 
of calibrations necessary to provide a uniform response, to map defective
channels and to ensure an acceptable noise rate. Offline calibrations
are then obtained during normal data-taking. They consist of
additional noise suppression and, for the Pixel Detector and TRT, 
corrections to the position measurement of reconstructed tracks.

During collision data-taking, it is planned that offline calibrations will be performed
on a subset of the data and   the bulk processing of most data will start only after 
these calibrations have been validated. This model could not be applied 
during the 2008 data-taking, since the rate of events with tracks,
especially in the 
silicon detectors, is many orders of magnitude lower than  
in LHC collisions. Therefore offline calibration 
results were only used in the reprocessing at the end of the data-taking period.

\subsection{Pixel Detector calibration}
\label{sec:Pixelcalibration}

The calibration of the Pixel Detector consists in tuning   the 
optical communication links and adjusting the front-end electronics to provide uniform thresholds and response to injected charge. Suppression of noisy channels 
is also done at this time. Data for these calibrations are acquired in special runs. A detailed description of the hardware calibration is available 
in~\cite{bib:PixelCosmicPaper}. 
The quality of the calibration is then verified using measurements of noise rate, charge collection and timing in normal ATLAS runs. The cluster reconstruction algorithm, which uses the pulse height to improve the accuracy of the
position measurement is also calibrated.

The optical data-links contain arrays of 8 or 16 VCSEL 
devices~\cite{bib:PixelOptoElectronics,bib:PixelOpticalLinks}. 
The bias voltage which controls optical power can only be 
adjusted for the data-link as a whole. Due to the spread 
in the device characteristics, the optical power 
for a setting is not uniform and a scan of the bias voltage 
  is performed to determine a suitable value for all devices 
in the data-link.
A bit-error rate of $<2.7\times 10^{-8}$ 
with a confidence level of 99\%\ was measured for the
two bandwidth configurations,  40 and 80~Mbits/s, which will be used for 
operation up to a luminosity of $10^{33}$~cm$^{-2}$s$^{-1}$.
At higher luminosity, the innermost layer will be operated at a readout speed of
160~Mbits/s, by using two 80~Mbits/s channels for each module.
%~\cite{bib:PixelOptolinkNote}.
 
Threshold calibration of the front-end electronics is performed by injecting 
known amplitude signals into the input of the electronics chain. The fraction 
of observed hits as a function of the injected charge is fitted with an 
error function, providing the threshold, defined as the 50\%\ efficiency point, 
and the electronic noise. 
An 8-bit DAC is used to adjust the threshold to the target value.  
The distributions of threshold and noise for the whole
detector are shown in Fig.~\ref{fig:PixelCalib}. At the 
nominal working point, corresponding to a 4000~$e$ threshold, 
a uniformity of 40~$e$ r.m.s. is achieved after tuning.   
In these conditions the average noise level is 160~$e$
for most pixels, and slightly higher for pixels of 600~$\mu$m size 
(long pixels) or for pairs of interconnected pixels (ganged pixels), which are
used to cover the otherwise dead area between front-end 
chips~\cite{bib:PixelSensors}. The tails in Fig.~\ref{fig:PixelCalib}
correspond to $4\times 10^{-5}$ of channels differing by more than 
250~$e$ from the nominal threshold and $1.3\times 10^{-4}$ of channels with 
noise greater than 600~$e$, which may give high noise occupancy 
during operation.

Due to the finite electronics rise-time, low-amplitude pulses may be 
assigned to a BC later than the one in which the signal is 
generated~\cite{bib:FEI3chip}. Therefore the
 {\em in-time} threshold is also measured. This is 
the minimal signal for which the hit 
is located in the same BC as the particle crossing. For the reference 
4000~$e$ threshold, the {\em in-time} threshold is 5400~$e$, with a r.m.s. spread of 240~$e$.  

Due to the high threshold-to-noise ratio, 
random noise occupancy, i.e. the probability for a channel to give
a noise hit per BC, is extremely low. Dedicated standalone runs with random
triggers are used to find and mask the small fraction of channels that show
an anomalous occupancy, greater than $10^{-6}$~hits/BC. 
Random triggers during normal data-taking runs are used for 
monitoring additional noisy channels
which are not used in reconstruction if they have an occupancy 
greater than $10^{-5}$~hits/BC.

The actual fraction of noisy pixels was below 
$2.2\times 10^{-4}$ for all the 2008 data-taking. After masking these 
channels, the noise occupancy was 
$\sim10^{-10}$~hits/BC, 
corresponding to less then one noise hit per event in the Pixel Detector.

\begin{figure}
\begin{center}
\includegraphics[width=\columnwidth]{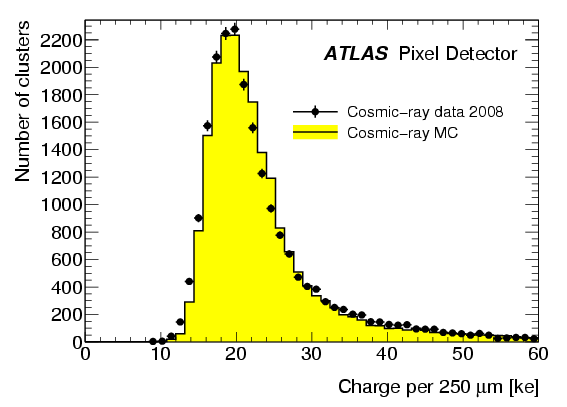} 
\end{center}
\caption{Spectrum of charge release by cosmic-ray muons in the Pixel Detector,
as obtanied from the Time-over-Threshold measurement.}
\label{fig:PixToT}
\end{figure}

The pulse height is measured using the Time-over-Threshold (ToT) method. 
The relationship between amplitude and 
ToT is calibrated with charge injection and the resulting
calibration curve is used to reconstruct the 
energy deposited in the detector by charged particles. 
The absolute scale of the ToT calibration can be estimated by comparing
the observed spectrum of collected charge with the 
expectation obtained by combining the theoretical model of energy loss in 
silicon~\cite{bib:Bichsel}, the average energy needed to 
create an electron-hole pair, $W=3.68\pm 0.02$~eV/pair~\cite{bib:SiW2},
and the 
effect of losses of collected charge due to the finite threshold of pixels
(Fig.~\ref{fig:PixToT}). For this study two methods were used. The first
selected two-pixel clusters on tracks with incident 
angle $\alpha < 25\degr$: for these clusters the losses due to threshold
effects are negligible and the most probable value could be directly compared
to theoretical predictions. The second compared the pulse height of one-pixel 
and two-pixel clusters in data and Monte Carlo as a function of $\alpha$
in the range 
$\alpha < 30\degr$. Both methods agreed, providing a calibration factor
for the charge scale of 
$0.986\pm 0.002\stat\pm 0.030\syst$, consistent with 
unity. The largest systematic uncertainties are 2.4\%\ from the spread 
of the measured values of $W$~\cite{bib:SiW2,bib:SiW1,bib:SiW3,bib:SiW4}
and 2\%\ from the theoretical modelling of energy loss in 
silicon.

%\begin{figure}[htbp]
%\begin{center}
%\includegraphics[width=\columnwidth,viewport=0 0 567 380,clip]{figures/Pixelnoiseocc_91900.pdf} 
%\caption{Average pixel occupancy per bunch crossing from noise hits in 
%         cosmic-ray data-taking for the 3 barrel layers and 2 endcap Pixel 
%         detectors, after masking noisy channel.
%          }
%\label{fig:Pixelnoise}
%\end{center}
%\end{figure}

Pulse-height measurements improve the accuracy of the 
position measurement, in both the local $x$ and $y$ coordinates,
for clusters consisting of more than one pixel.  
The charge-sharing ratios, $\Omega_x$ and $\Omega_y$, between the signals 
collected on the first and last row or column in the cluster 
$$\Omega_x =\frac{Q_{\rm last\ row}}{Q_{\rm last\ row}+Q_{\rm first\ row}}
\qquad \Omega_y =\frac{Q_{\rm last\ column}}{Q_{\rm last\ column}+Q_{\rm first\ column}}$$ 
are used to correct the geometrical centre-of-cluster positions $(x_c,y_c)$ 
with a linear function
\begin{equation}
\left( x_c, y_c \right)  \ra 
 \left[ x_c + \Delta_x \left( \Omega_x - \frac{1}{2} \right) ,
y_c + \Delta_y \left( \Omega_y - \frac{1}{2} \right) \right],
\label{eq:chargesharing}
\end{equation}
with weights, $\Delta_x$ and $\Delta_y$, depending on
the particle incident angle and cluster size~\cite{bib:PixelLorentzAngle}.

\begin{figure}
\includegraphics[width=\columnwidth]{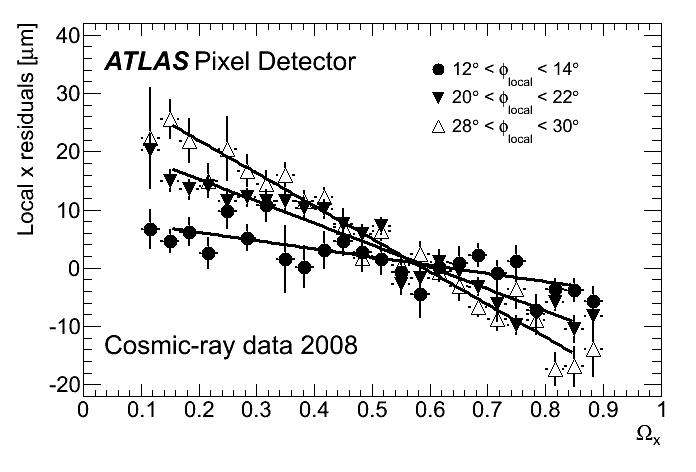}
\caption{Residual between track extrapolation and the centre-of-cluster 
position in the Pixel Detector for two-pixel clusters in the local~$x$ direction and
different incident angles. The measured slopes are used to improve the position resolution
with respect to the purely binary readout according to Eq.~(\ref{eq:chargesharing}).}
\label{fig:Pixel_omega}
\end{figure}

Cosmic rays with transverse momenta $p_{\rm T}>5$~\GeV\ provided 
a calibration of $\Delta_x$ for two- and three-pixel clusters and 
$\phi_{\rm local}<45^{\circ}$ (Fig.~\ref{fig:Pixel_omega}),
a range much wider than expected for particles from proton-proton collisions. 
Along the
beam direction, the limited range of cosmic-ray polar angles 
(Fig.~\ref{fig:phi0_resolution_vs_PT}) 
only allowed the $\Delta_y$ calibration for two-pixel clusters up 
to $|\eta|<1$; 
collisions are needed to cover the full acceptance 
in pseudorapidity.
This calibrated position-reconstruction
 algorithm is expected to provide a measurement accuracy of 6~$\mu$m 
in the transverse plane for two-pixel clusters.

%The multiple bunch-crossing readout window together with the measurement
%of the particle crossing time  allows 
%for a direct observation of the front-end electronic timewalk, 
%by measuring the average BC at which signals of a fixed pulse height are 
%collected~\cite{bib:PSD6}.
%During cosmic-ray data-taking the crossing time is provided 
%by the TRT time measurement $T_{\rm TRT}$ (Section~\ref{sec:timing}).
%Figure~\ref{fig:PixelTimewalk} shows that low-charge pixels 
%indeed are assigned a bunch-crossing later than high-charge pixels. The 
%pulse height for which the delay reaches 1 BC is consistent with the 
%5400~e {\em in-time} threshold measurement.

%\begin{figure}
%\includegraphics[width=\columnwidth]{figures/PixelTimewalk.png}
%\includegraphics[width=\columnwidth]{figures/pixel_timing_figure_2.png}
%\caption{Average hit bunch-crossing as a function of pixel signal. 
%Low-charge hits are detected late because of the higher time needed
%to cross the discriminator threshold.}
%\label{fig:PixelTimewalk}
%\end{figure}

\begin{figure*}[ht]
\begin{center}
\begin{tabular}{cc}
\includegraphics[width=0.45\textwidth]{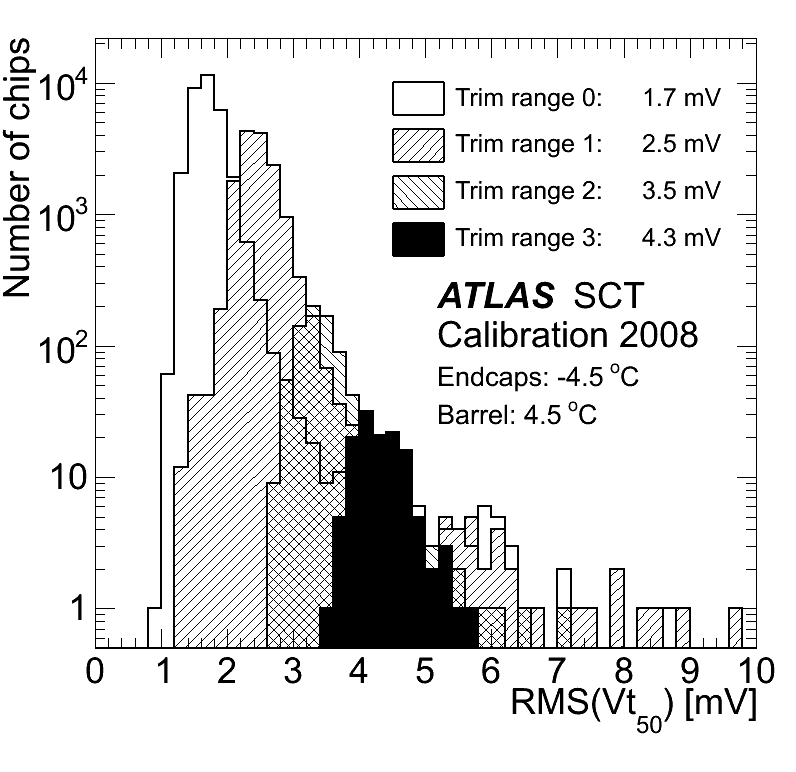} &
\includegraphics[width=0.45\textwidth]{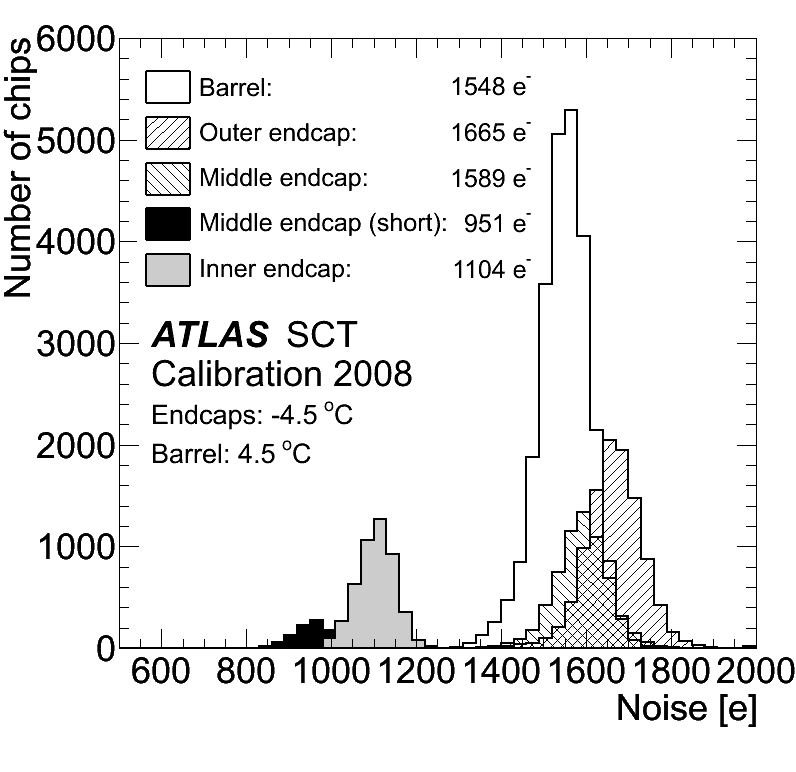} \\
(a) SCT threshold dispersion & (b) SCT noise  
\end{tabular}
\end{center}
\caption{SCT threshold dispersion and noise from calibrations at 2~fC 
threshold based on charge injection. (a) Distribution of the r.m.s. spread
of the threshold $V_{\rm t 50}$ for each chip. The average values for each trim 
range are given.
(b) Distribution of the input noise values for each chip as obtained in
response curve tests. The average values for each detector region are given.
The average SCT sensor temperatures for barrel and endcap modules as estimated 
from the operation conditions are also given.}
\label{fig:SCTCalib}
\end{figure*}

\subsection{SCT calibration}
\label{sec:SCTcalibration}
Good front-end calibration is essential to the operation of the SCT because of
the binary readout employed. The channel thresholds must be set to provide good 
efficiency ($>$99\%) and uniformity of response while keeping the noise occupancy 
below $5\times 10^{-4}$~hits/BC. 
The calibration procedure is described in~\cite{bib:SCTDAQ} and it follows
a sequence similar to the one described for the Pixel Detector. Calibration 
runs are performed with the SCT data-acquisition system in a standalone mode, and the data 
analysed online. As a first step the parameters of the optical data 
links~\cite{bib:SCTOpticalLinks} are tuned to ensure reliable communication
to and from the modules.

Threshold calibration is performed by injecting known charges into the front-end
of each readout channel and measuring the occupancy as a function of threshold.
For each input charge the dependence is parameterized using a complementary 
error function. The threshold at which the occupancy is 50\% ($V_{\rm t 50}$)
corresponds to the median of the injected charge while the sigma gives the 
noise after amplification. Channel gains are extracted from the dependence of
$V_{\rm t 50}$ on the input charge, and are used to set the discriminator thresholds.
Channel-to-channel variations are compensated using a 4-bit DAC (TrimDAC).
The TrimDAC steps can themselves be set to one of four different values
to allow uniformity of response to be maintained when uncorrected
channel-to-channel variations increase after irradiation.
The achieved uniformity of response is
shown in Fig.~\ref{fig:SCTCalib}(a), which shows the distribution of the
r.m.s. spread of $V_{\rm t 50}$ values on a chip. Distributions are shown
separately for chips in each TrimDAC range; most of the chips are configured
in the finest setting, with a small spread. After irradiation it is expected
that coarser settings will become necessary. The uniformity at the 
nominal threshold of 1fC, corresponding to a signal of 54--58 mV, is $\sim$4\%.
The corresponding noise level, shown in Fig.~\ref{fig:SCTCalib}(b), is 
between 900 and 1700 $e$, depending on the strip length.

\begin{figure}[ht]
\begin{center}
\includegraphics[width=\columnwidth]{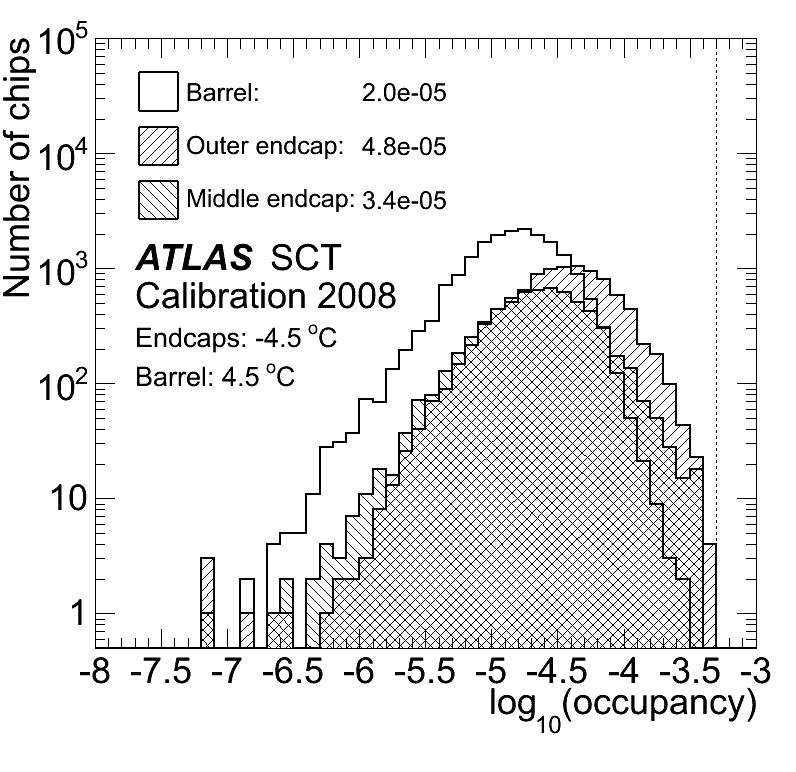} 
\caption{The SCT noise occupancy per channel measured in calibration mode 
         at 1~fC threshold for
         barrel and endcap modules in 2008 data. The dotted line is 
         the specification value of $5\times10^{-4}$. A fraction of $~0.2\%$ 
         of strips with occupancy 
         above specification are excluded.  The average noise occupancies and 
         operational temperatures are shown.
         }
\label{fig:SCTnoise}
\end{center}
\end{figure}

Threshold scans with no injected charge are used to measure the noise 
occupancy and strips with occupancy greater than $5\times10^{-4}$~hits/BC
are disabled.  Figure~\ref{fig:SCTnoise} shows 
the occupancy values measured in calibration mode after removing 
the $\sim 0.2\%$ of noisy strips.
Normal data-taking runs are used for the 
identification of noisy channels
which escape detection during the calibration runs. Strips with 
an occupancy above $5\times10^{-3}$~hits/BC are subsequently removed during reconstruction. The number of such strips never exceeds $0.1\%$ of the  channels. 
The noise occupancy in 
cosmic-ray data was calculated as the number of hits per event not associated to
a track, per channel and BC. This rate was found to be of order of $10^{-5}$,
in good agreement with the calibration-mode data.

\subsection{TRT calibration}
\label{sec:calibrations}

As for the other sub-systems, the first step in calibrating the TRT  
is to adjust the data-links to provide reliable communication.  
There are separate steps for adjusting, on one hand, the  phasing of the 
clock and the trigger and control lines  and, 
on the other hand,  the phasing of the data lines from the front end into the  optical links going to the TRT RODs.
Noise data are then acquired 
 in special calibration runs and are used for the high-uniformity tuning 
of detector thresholds.  

\label{sec:thcalibration}

The effective gain and inherent noise of the front-end chips were  measured during production by 
injecting each channel with
known amplitude signals at multiple threshold settings. At the board, module and detector level, thresholds were set to give a noise occupancy corresponding to the desired threshold in fC. The uniformity of the random noise occupancy (or rate) for different detector elements at the same effective threshold gives a measure of element-to-element matching.

\begin{figure}
\begin{center}
\includegraphics[width=\columnwidth]{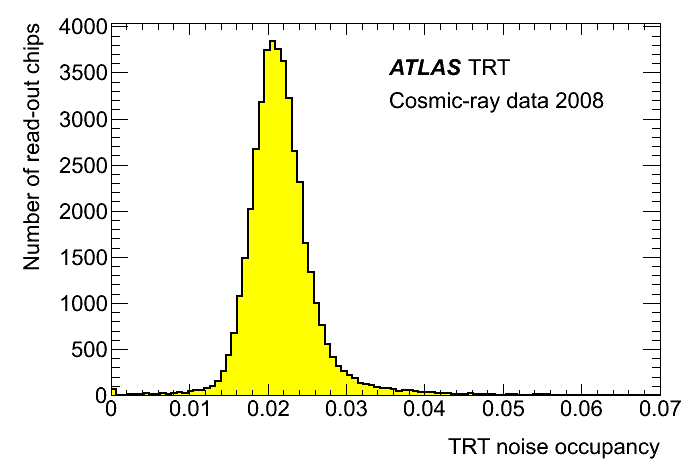}
\end{center}
\caption{TRT low threshold noise occupancy for 2008 cosmic-ray data averaged over each group of eight straws.}
\label{fig:TRTnoise}
\end{figure}

The TRT low (tracking) threshold is set to about 2~fC, corresponding to  250~eV of 
deposited ionization energy. This setting gives an average noise 
%rate of about 300 kHz or an 
occupancy of about 2\% for the three bunch-crossings sampled by each trigger.
This calibration process achieves a uniform response to particles
across the detector, correcting, for example, for the effect on the physical 
thresholds of ground offsets in the low voltage levels supplied to the front-end
electronics. Figure~\ref{fig:TRTnoise} shows the TRT low threshold noise 
occupancy in 2008 cosmic-ray data. The occupancy is 
uniform with a r.m.s. spread across the detector of 0.5\%.  
The $\sim$2\% permanently dead straws and the handful with 100\% occupancy are 
discarded.

\begin{figure}
\begin{center}
\includegraphics[width=\columnwidth]{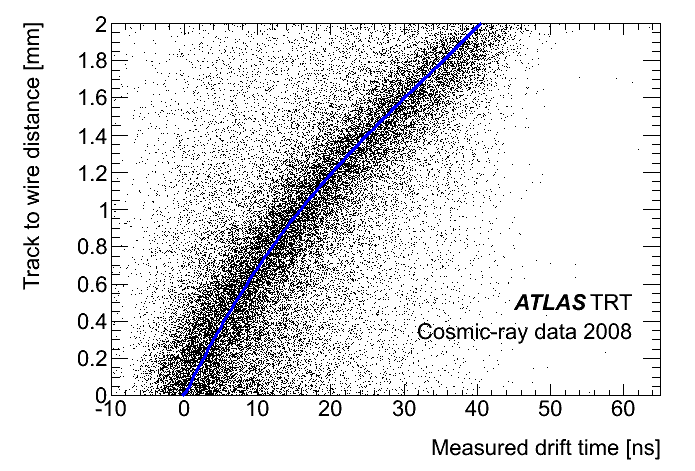}
\end{center}
\caption{Measured time-distance ($R-T$) relationship for the TRT barrel 
with solenoid field on.}
\label{fig:TRT_RT}
\end{figure}

Normal data-taking runs are used for the identification of noisy channels and measurement of random noise. 
These runs are also used to compute parameters needed to optimize the determination 
of the particle crossing point. The parameters consist of the 
$T_0$ for each 16-straw time-measuring chip
and the  global time-distance 
relationship, $R-T$, shown in Fig.~\ref{fig:TRT_RT}. The $R-T$ 
relationship is obtained by fitting a 
third-order polynomial  to the distance of the  reconstructed track from 
the centre of the straw as a function of the time of the leading-edge, corrected by $T_{\rm TRT}$.

%%%dummy line added to try to force svn to commit this file%%%%%%
%
%this is more extra junk to try to make the file different?????
%
%
%

\section{Alignment}
\label{sec:alignment}
\begin{table*}[ht]
\begin{center}
\begin{tabular}{clrr}
\hline \hline
Level & Brief description & Structures & Degrees of freedom \\
\hline
 0  & \multicolumn{1}{r}{Total:} & 7 & 41 \\
    & Whole Pixel detector & 1 & 6 \\
    & SCT barrel and 2 endcaps & 3 & 18 \\
    & TRT barrel (except $T_z$) and 2 endcaps & 3 & 17 \\
\hline
 1  & \multicolumn{1}{r}{Total:} & 14 & 84 \\ 
    & Pixel barrel layers split into upper and lower halves plus 2 endcaps & 6+2 & 48 \\
      & SCT barrel split into 4 layers plus 2 endcaps & 4+2 & 24 \\
\hline
 2  & \multicolumn{1}{r}{Total:} &  & 2472 \\ 
    & Pixel barrel layers split into staves plus 2 endcaps & 112+2 & 684 \\
    & SCT barrel layers split into staves plus 2 endcaps & 176+2 & 1068 \\
    & TRT barrel modules (except $T_z$) & 96 & 480 \\
    & TRT endcap wheels (only $T_x$, $T_y$ and $R_z$) & $40\times 2$ & 240 \\
\hline
 3  & \multicolumn{1}{r}{Total:} & 3568 & 7136 \\ 
    & Pixel barrel modules (only $T_x$ and $R_z$) & 1456 & 2912 \\
    & SCT barrel modules (only $T_x$ and $R_z$)   & 2112 & 4224 \\
\hline \hline
\end{tabular}
\end{center} 
\caption{Alignment levels used with cosmic-ray data for the Inner Detector 
subsystems. Naming, brief description, 
         number of structures and the total number of degrees of freedom to be aligned at each level are
         given. The six degrees of freedom per structure
         in Eq.~(\ref{eq:alparam}) 
         are used, 
         unless otherwise indicated.
         \label{tab:idalign_levels}}
\end{table*}

%\label{sec:alignment}
The accuracy with which particle tracks can be reconstructed is limited by how precisely the positions and 
orientations of the ID sensor modules and wires are known.
The requirement on the alignment quality is that the resolution of track 
parameters is to be degraded by no more than 20\% with respect 
to the intrinsic resolution~\cite{bib:IDTDR}. The silicon pixel and 
strip modules must be aligned
with a precision of respectively 7~$\mu$m and 12~$\mu$m in the sensitive 
$R\phi$~direction.
In the $z$ ($R$ for the endcap) direction of silicon modules and for the 
TRT, the  alignment precision is required to be of several 
tens of micrometres.  
In addition, the alignment should have minimal systematic 
effects which could bias the track-parameter determination.

The alignment is specified by a set of constants, six for each individual module or assembly structure 
(barrel layer, endcap disk, etc.) corresponding to the six degrees-of-freedom of a rigid body: three translations
$T_x$, $T_y$ and $T_z$ with respect to the nominal position and three rotations $R_x$, $R_y$ and $R_z$ with respect 
to the nominal axis orientations. 

Track-based alignment algorithms were used to determine alignment constants using the 
cosmic-ray data collected in 2008. The algorithms use the tracking residual distributions of the modules;
a residual is defined as the distance between the position of the measurement and the intersection 
of the fitted track with that module.
% If 
%a particular module is displaced in the $R\phi$~direction its $R\phi$~residual distribution is expected to be 
%biased accordingly. The magnitude of this bias will be proportional to the size of the displacement.
The alignment constants can be determined via a minimisation of the following 
$\rm{\chi^{2}}$~function:

\begin{equation}
\chi^{2} = \sum_{\mathrm{tracks}} \mathbf{r}^{T}V^{-1}\mathbf{r}
\label{eq:alignment_chi2}
\end{equation}

\noindent
where the sum is over all tracks in a given event sample, $\mathbf{r}$ is the vector of residuals for a given track 
and $V$ is the covariance matrix of those residuals. In general, $\mathbf{r}$ is a function of both the track 
parameters, 
\begin{equation}
\boldsymbol{\tau}=\left(d_0,z_0,\phi_0,\theta,q/p\right),
\end{equation}
and of the alignment constants, 
\begin{equation}
\mathbf{a}=\left(T_x,T_y,T_z,R_x,R_y,R_z\right),
\label{eq:alparam}
\end{equation}
of those modules with hits contributing to the track
fit. The alignment was determined using the Global $\rm{\chi^{2}}$ algorithm~\cite{bib:GLX2}. In this algorithm 
the $\rm{\chi^{2}}$ given by Eq.~\ref{eq:alignment_chi2} was simultaneously minimised with respect to $\boldsymbol{\tau}$ 
and $\mathbf{a}$ to determine the alignment constants. 

The results 
were cross-checked using two alternative algorithms, which gave consistent results. In the Local $\rm{\chi^{2}}$ 
algorithm~\cite{bib:LocalX2_1,bib:LocalX2_2} the minimisation was done 
only with respect to $\mathbf{a}$.  
In the Robust algorithm\cite{bib:RobustAlignment}, used only for silicon 
detectors, 
the alignment corrections were calculated directly from the size 
of the residual bias. In all cases, an iterative procedure was used.

The 7.6 million tracks reconstructed in the Inner Detector
during the 2008 cosmic-ray data-taking period were used to perform a 
preliminary alignment of the tracking system which significantly improved
the tracking performance.
 
%Silicon modules are strongly affected by the distribution of cosmic-ray 
%tracks. 
Because cosmic rays come from above and not from the centre of the ATLAS detector, more hits were recorded 
in silicon modules in the top and bottom quadrants of the barrel than the side quadrants or the endcaps.
In addition, the large incidence angles in the side and endcap modules result in poor-resolution large or fragmented 
clusters. This limits the precision to which these regions of the Pixel Detector and SCT can be aligned. 
Due to its structure and larger acceptance, the TRT is less 
sensitive to this anisotropy and its alignment precision was more uniform.

\subsection{Global alignment}

The alignment proceeds in stages from larger structures
to the individual module level, as detailed in Table~\ref{tab:idalign_levels}. 
At each stage more degrees of freedom are introduced, but the expected sizes 
of the corrections are smaller.

In the first step, the Level~0 alignment, the SCT barrel and two endcaps are 
aligned relative to the entire Pixel Detector, followed by the TRT alignment
with respect to the silicon detectors. 
%This step required cosmic-ray tracks reconstructed simultaneously by the silicon detectors and the TRT. 
In aligning the TRT barrel, 
only 5 degrees of freedom are used; the $T_z$
is not considered because the TRT barrel modules are almost 1~m long and 
do not measure the $z$ coordinate.
 
Cosmic-ray simulation studies with a misaligned geometry showed that, 
using solenoid-on 
tracks for the silicon detectors' Level~0 alignment,
may lead to corrections being underestimated. 
The presence of a misalignment between the sub-detectors could lead to a bias in reconstructed track
momentum, with part of the misalignment being absorbed into the curvature. Therefore these alignment corrections 
were derived using only solenoid-off data. 
The simulation tests also showed that the solenoid-off data were able to estimate the Level~0 misalignments with 
a precision better than  100~$\mu$m. This precision is limited by misalignments of the internal structures and 
by multiple Coulomb scattering effects.

For the TRT instead, both a solenoid-on and a solenoid-off sets of tracks 
were used. The results were compared and found consistent within the 
uncertainties.

Shifts from the nominal positions of up to 2~mm were observed, with rotations $R_z$ of several mrad, as shown in 
Table~\ref{tab:level1results}; the rotations $R_x$ and $R_y$ were all consistent with zero.

\begin{table}[hbt]
\begin{center}
\begin{tabular}{lcccc}
\hline \hline
Structure & $T_x$ [mm] & $T_y$ [mm] & $T_z$ [mm] & $R_z$ [mrad] \\
\hline
SCT barrel    &  \pho0.9 & \pho0.6 & \pho0.5 &   --1.8 \\
SCT endcap A  &    --1.8 & \pho0.5 & \pho0.0 &   --1.3 \\
SCT endcap C  &    --0.4 & \pho0.6 & \pho1.0 &   --1.3 \\
TRT barrel    &  \pho0.2 &   --0.1 &     N/A & \pho0.0 \\
TRT endcap A  &    --1.5 & \pho0.2 &   --3.4 &   --7.0 \\
TRT endcap C  &    --1.0 & \pho1.7 & \pho2.1 & \pho6.4 \\
\hline \hline
\end{tabular}
\end{center} 
\caption{Level 0 alignment parameters, translations ($T_x$, $T_y$ and $T_z$) and rotation ($R_z$ only), of the SCT and TRT  barrel, 
         endcap A (positive $z$) and endcap C (negative $z$). The statistical errors were much smaller than the last digit.}
         \label{tab:level1results} 
\end{table}

\subsection{Local alignment of the Pixel Detector and SCT}

After the initial alignment of the detector components as a whole, the 
subsequent alignment levels consider smaller structures.

Due to the low statistics
the endcaps were aligned globally, but no attempt was made to align individual disks or modules.
The initial geometry for the alignment was based on the nominal position of the modules.

The first stage in the internal alignment of the Pixel Detector and SCT (Level~1) was the alignment of the pixel 
half-shell barrel layers, the full SCT barrel layers and the four endcap structures (two for each of the Pixel 
Detector and the SCT). 
The SCT barrel layers were considered to be rigid cylinders, whilst the pixel half-shells were considered rigid 
half-cylinders. 
For all the structures, the full set of 6 degrees of freedom was considered in the alignment. 
This level was aligned combining both solenoid-on and solenoid-off cosmic-ray data. 
The computed alignment corrections were of the order of hundreds of micrometres in all $T_x$, $T_y$ and 
$T_z$, with in particular a rotation of the first pixel upper half shell of almost 2~mrad with respect to the 
other layers.

The next step was the alignment of the Pixel Detector and SCT stave-by-stave (Level~2). 
The pixel staves are real structures, composed of 13 modules in the same $\phi$ position, which were assembled and surveyed. 
The SCT was not assembled in staves but the modules were individually mounted on the support cylinder. 
Nevertheless, for alignment purposes the SCT barrel was also split into rows of 12 modules. 
The staves were considered a rigid body and all 6 degrees of freedom were used. 
The alignment corrections for the translations of the staves were of the order of tens of micrometres.  

Once the staves were aligned the alignment at module-to-module level (Level~3) 
was performed. 
The positions of pixel modules mounted within the staves were surveyed just after assembly~\cite{bib:PixelSurvey}.  
This survey information was used as a starting point for the internal alignment of the pixel modules, but not to constrain the alignment corrections, because 
the deformation of staves after the survey was expected to be significantly 
larger than survey errors. This step was performed in the local coordinate 
system described in Section~\ref{sec:ID} for individual silicon modules.

The number of hits per module 
was much smaller than for the larger structures, and thus the statistical precision of the alignment becomes
a significant consideration. Therefore the 
number of 
degrees of freedom was reduced to just two per module, $T_x$ and $R_z$. 
These two parameters were chosen because they were appropriate to describe the lateral bending along the pixel staves, the largest deformation observed in the residuals, with an amplitude 
reaching 500~$\mu$m for the worst case.

%\begin{figure}[htb]
%\begin{center}
%\subfigure[Scheme of the pxiel stave bow]{
% \includegraphics[width= 0.45\textwidth]{figures/NegBow_TxRz.png}
%}
%\subfigure[alignment constants of the pixel stave bow]{
% \includegraphics[width= 0.45\textwidth]{figures/PixelPositiveBowXY.pdf}
%}
%\caption{Left: schematical view of the bowing of the pixel staves. The bowing implies a translation in the local X reference ($T_x$) plus a within plane rotation ($R_z$) which are correlated. Right: alignment corrections obtained for a given pixel stave. Each module (given in the x-axis by his global Z coordinate along the stave) has two measurements: the $T_x$ and its $R_z$. The translation really shows the bow and the rotation of the module confirms its shape. }
%\label{fig:idalign_pix_stave_bow}
%\end{center}
%\end{figure}

\begin{figure*}[ht]
\begin{center}
\subfigure[Pixel local-$x$ barrel residuals]{
\includegraphics[width=0.9\columnwidth]{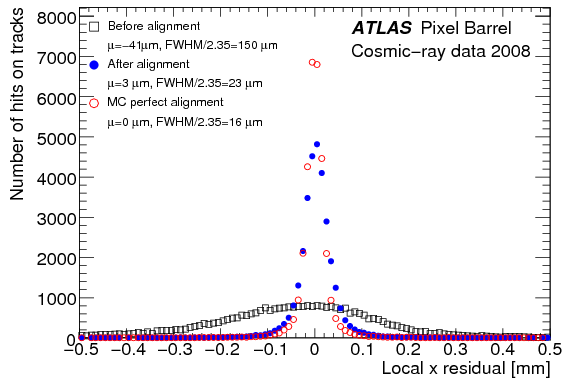}
\label{fig:idalign_pix_resx}
}
\subfigure[Pixel local-$y$ barrel residuals]{
\includegraphics[width=0.9\columnwidth]{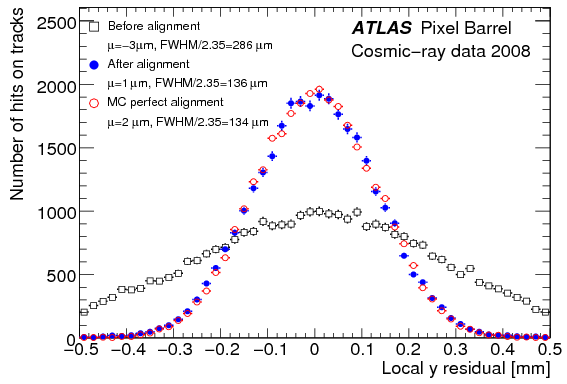}
\label{fig:idalign_pix_resy}
} 
\subfigure[SCT barrel residuals]{
\includegraphics[width=0.9\columnwidth]{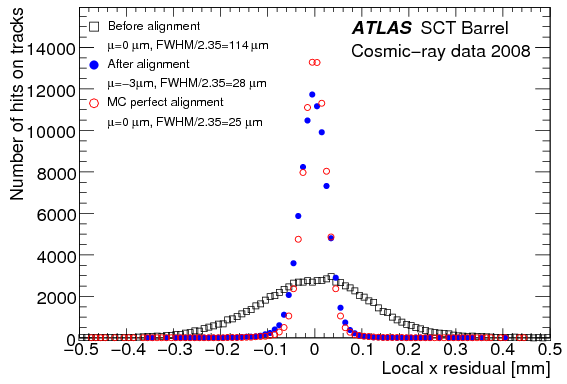}
\label{fig:idalign_sct_res}
}
\subfigure[TRT barrel residuals]{
\includegraphics[width=0.9\columnwidth]{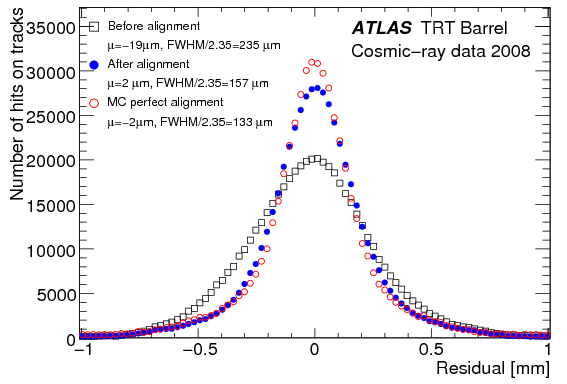}
\label{fig:idalign_trt_res}
}
\caption{Residual distributions in the local reference frame for hits in barrel regions for all ID sub-detectors. 
The plots show the results for 2008 cosmic-ray tracks before and after alignment and a comparison with a 
         perfectly aligned cosmic-ray Monte Carlo simulation. Tracks are selected requiring $\pT>2$~GeV.}
\label{fig:idalign_res}
\end{center}
\end{figure*}

Pixel Detector and SCT residual distributions before and after the alignment procedure are shown in 
Fig.~\ref{fig:idalign_res} 
for tracks with $\pT>2$~GeV and $|d_0|<50$~mm.
These are compared to distributions obtained using a perfectly-aligned Monte Carlo simulation of cosmic rays. 
Before alignment the residual distributions are very wide compared to the Monte Carlo simulation 
and also biased. After alignment their widths were substantially reduced and the means are consistent with zero 
to within a few micrometres. 

The residuals cannot be used to quote the point resolution, 
because their errors include a contribution from 
extrapolation uncertainties larger than the point resolution. This
contribution also depends on the track momentum and silicon layer, 
resulting in strongly non-Gaussian distributions.
By comparing the width of the aligned residual distributions to the simulation, and 
assuming that the only contribution to the increased width is from misalignments, the size of the remaining 
module-level misalignments is estimated to be approximately 
$20\,\mu{\rm m}$.

\subsection{Local alignment of the TRT}

The second step of the TRT barrel alignment internally aligned the 96 individual 
TRT barrel modules (three layers of 32 $\phi$-sectors each). 
Although the straw anodes inside the barrel modules are physically separated at $z$=0, no such 
distinction exists at the module level. As  for the Level~0 barrel alignment, 
only five degrees of freedom were used, $T_z$ being non-measurable.
The internal alignment was determined separately for different periods of cosmic-ray 
data taking, which could either be solenoid~on or solenoid~off. 
This internal alignment used TRT stand-alone tracks, giving high statistics because of the larger 
acceptance of the TRT volume. 
The size of the translation alignment corrections was of the order of 
200-300~$\mu$m with respect 
to the nominal position of the modules. 

In each endcap, the 40~wheels were aligned in three degrees of freedom: 
$T_x$, $T_y$, and $R_z$. The corrections for the translations were of the
order of 100~$\mu$m and the rotations were tenths of a milliradian. 

Figure~\ref{fig:idalign_trt_res} shows the residual distribution for 
tracks with $\pT>2\GeV$ in the barrel modules, both before and after 
alignment. The distributions are compared to those obtained using a perfectly
aligned cosmic-ray Monte Carlo simulation. Again the 
width and bias of the residual distribution were improved after alignment.

\subsection{Summary and perspectives}

The cosmic-ray alignment significantly improved the track reconstruction 
and the track-parameter resolutions, presented in 
Section~\ref{sec:trackpar}. The achieved level of precision, 
about 20~$\mu$m, ensures that track reconstruction efficiency with early 
LHC data will not be significantly affected by 
residual misalignments.

Local alignment with cosmic rays is statistically limited by the small 
acceptance of individual detector modules, especially in the endcap region. 
Therefore it was not possible to perform a Level~3 alignment in the endcaps.
In addition, a reduced set of degrees of freedom was used in the barrel 
region. That not all possible misalignments can be recovered using only 
cosmic-ray data partially
explains
why the nominal Monte Carlo resolution has not yet been achieved.

In order to reach the design granularity, a high statistics sample of tracks 
from proton-proton collisions is needed. When this has been collected,
all 1744 and 4088 Pixel Detector and SCT modules will be aligned
with the full set of degrees of freedom~(\ref{eq:alparam}). Individual 
TRT wires  will also be aligned with the two more sensitive degrees of freedom: 
the translation along the $\phi$ direction and 
the rotation about the $R$ or $z$ directions
in the barrel and endcap regions, respectively.

\section{Detector performance}
\label{sec:performance}
\subsection{Intrinsic detector efficiency}
The intrinsic detector efficiency measures the probability of a hit being
registered in an operational detector element when a charged particle traverses
the sensitive part of the element. Both a high intrinsic efficiency and a 
low non-operational fraction are essential to ensure good-quality tracking.

The intrinsic efficiencies of the Pixel and SCT detectors are measured 
by extrapolating well-reconstructed tracks through the detector and
counting the numbers of hits (clusters) on the track and `holes' where
a hit would be expected but is not found. The track extrapolation uses 
the full track fit described in Section~\ref{sec:tracking} to compute
the intersections of the track with all modules along its trajectory.
If a module (module side for the SCT) does not have a cluster associated 
to the track and the intersection point is more than 3$\sigma$ from the 
edge of the sensitive area the absence is called a hole. 
The efficiency, $\varepsilon$, is defined 
as the ratio of the number of clusters found to the number expected:
\begin{eqnarray}
  \varepsilon = \frac{N_{\rm clusters}}{N_{\rm clusters}+N_{\rm holes}}
  \label{equ:eff}
\end{eqnarray}
where $N_{\rm clusters}$ is the number of clusters found and $N_{\rm holes}$
is the number of holes.

%Only well-reconstructed tracks are used to measure the efficiency.
Pixel efficiencies are determined using tracks with at 
least 30 TRT hits (40 for the data with solenoid off), at least 12 SCT hits 
and $\sin\alpha < 0.7$. 
%where $\alpha$ is the angle between the track and the normal to the sensor. 
There must be only one track passing these cuts in the 
event. Tracks used to measure the SCT efficiency must have at least 30 TRT 
hits or 7 SCT hits, a hit both before and after the module side under 
investigation and $|\phi_{\rm local}| < 40^{\circ}$.
%, where $\phi_{\rm local}$ 
%is the angle between the track and the normal to the sensor in the plane 
%defined by the normal to the sensor and the axis in the plane of the sensor 
%perpendicular to the strip direction. 
A run-dependent cut on $T_{\rm TRT}$
is applied to ensure good timing. The angular cuts are applied because
the tracking algorithm does not function as well at high incidence angle; charge 
sharing among many channels combined with the readout threshold may result in 
multiple clusters and reduced apparent efficiency.  

The track extrapolation does not predict holes near the sensor edges or 
ambiguously mapped pixels, so these areas are excluded from the efficiency 
calculation. For the Pixel detector, clusters or holes within 0.6~mm of ganged 
pixels in the $\phi$ direction, or within 1.0~mm of the sensor edge in the $\phi$ 
or $z$ direction, are excluded. Similarly, for the SCT the intersection of the 
track with the sensor is required to be at least 2~mm from the edge in $\phi$ and 
at least 3~mm in $z$.    
%To reduce the effect of misalignments, any Pixel cluster not 
%already associated to a track but within 10~mm of an intersection is 
%included in the number of clusters in the efficiency calculation, and
%removed from the number of holes. The size, charge and time distributions of 
%these clusters are very consistent with those from clusters associated 
%to a track, hence it is likely that these result from track reconstruction 
%inefficiencies rather than noise. 
%In the case of the SCT, unassociated clusters within 2~mm of an 
%intersection point are included in the number of clusters, as in 
%previous studies~\cite{bib:comb_tests}. 
%The inclusion of unassociated clusters increases the measured efficiency of the
%Pixel barrel by $\sim$0.4\% and of the SCT barrel by $\sim$~0.2\%.
To reduce the bias due to the track fitting and pattern recognition criteria,
which are affected by residual misalignments, clusters not already associated to 
a track but close to an intersection are included in $N_{\rm clusters}$ in Eq.~(\ref{equ:eff}) 
and removed from $N_{\rm holes}$. 
%The size, charge and 
%time distributions of such pixel clusters are very consistent with those from 
%clusters associated to a track,
Due to the low noise occupancy (Section~\ref{sec:commissioning}), it is likely that these result from 
track reconstruction inefficiencies rather than noise.
The inclusion of these clusters improves the efficiency by 0.04\%
in the Pixel barrel and 0.2\% in the SCT barrel. Varying the distance  
for inclusion of non-associated clusters between 2~mm and 10~mm changes the efficiencies 
by at most 0.002\% and 0.004\% for Pixel Detector and SCT respectively, and is included 
in the systematic uncertainties.

Non-functioning detector elements (Section~\ref{sec:operation}) are not included 
in the calculation of the 
intrinsic efficiency. In the SCT, complete module sides and chips are excluded;
these amount to $\sim$2\% of the detector. The measured inefficiency contains a 
contribution from isolated dead strips for which no correction is applied.
For the Pixel detector, non-operational modules and front-end chips amount to 
4--6\% of the detector.

\begin{figure}[htbp]
\begin{center}
\subfigure[Pixel and SCT barrel efficiencies]{\label{fig:pixscteff}
\includegraphics[width=\columnwidth]{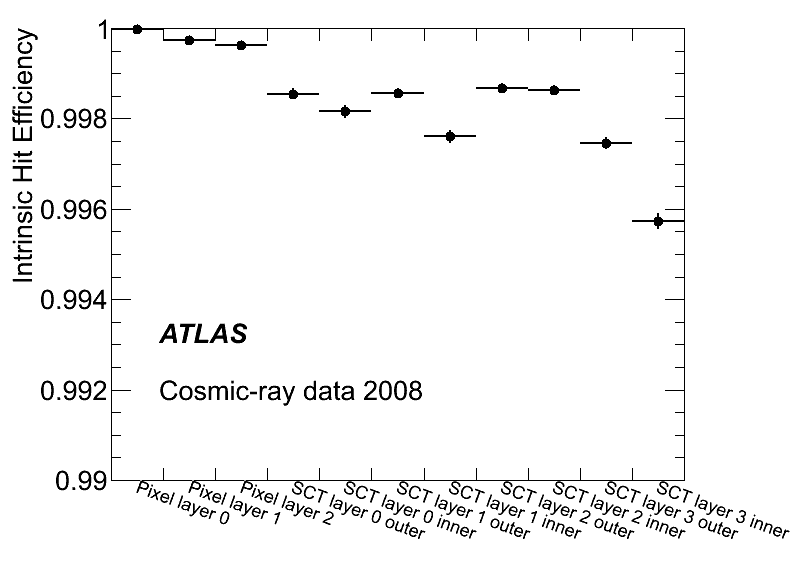} }
\subfigure[TRT efficiency]{\label{fig:trteff}
\includegraphics[width=\columnwidth]{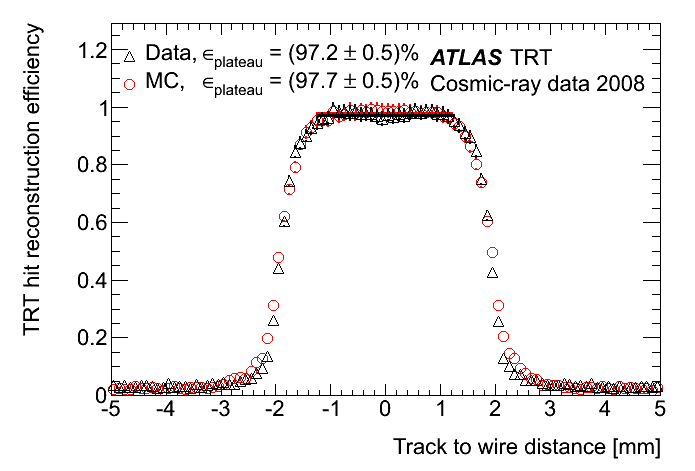}}
\caption{\subref{fig:pixscteff} Intrinsic efficiency of each Pixel Detector and SCT barrel layer.
         \subref{fig:trteff} TRT efficiency as a function of distance from the wire.}
\label{fig:efficiency}
\end{center}
\end{figure}

The measured efficiency of each barrel layer is shown for the Pixels and SCT in 
Fig.~\ref{fig:efficiency}(a) for data
taken with solenoid on. Efficiencies measured with solenoid off are typically
$\sim$0.2\% lower, indicating some residual inefficiencies arising from track 
reconstruction when the particle momentum is unknown.  
The overall efficiency of the Pixel barrel is 
$(99.974\pm0.004(\rm stat.)\pm0.003(\rm syst.))\%$ and of the SCT barrel is 
$(99.78\pm0.01(\rm stat.)\pm0.01(\rm syst.))\%$; the systematic error in each
case is determined by varying the track selection criteria. 
Of the remaining 0.026\% pixel inefficiency, $(0.017\pm0.004)\%$ is the contribution 
due to known defective channels observed during detector construction.

The efficiency of the TRT is determined in a similar manner to that of
the silicon detectors, excluding the 2\% non-functioning channels. Tracks are extrapolated through the TRT in a series of 
steps. To reduce tracking biases, at each point all straws in a region containing 
up to the third nearest neighbour are considered. The efficiency is determined by 
dividing the number of hit straws by the total number of straws within the region. 
The efficiency depends on the path length of a track inside a straw, and is therefore 
determined as a function of the distance of a track from the wire. 
Tracks are required to have at least 20 TRT hits, at least 6 SCT hits,
$T_{\rm TRT}$ between 5~ns and 25~ns and an angle to the vertical of
less than $15^{\circ}$. The efficiency of the TRT barrel, for data with 
solenoid on, is shown in Fig.~\ref{fig:efficiency}(b). The overall efficiency
over the plateau region is $(97.2\pm0.5)\%$.

%Figure

\subsection{Lorentz angle measurement}

\label{sec:Lorentz}

The charge carriers in the silicon detectors are subject to the electric 
field $\mathbf{E}$, generated by the bias voltage and oriented normal to the 
module plane, and the solenoid magnetic field $\mathbf{B}$. 
In the endcaps the fields are nearly parallel and the charge carriers drift
directly towards the electrodes. In the barrel modules these fields are perpendicular 
and the charge carriers drift at the Lorentz angle, $\theta_{\rm L}$, with respect to 
the normal to the sensor plane. 
%This is called the Lorentz angle and its value is given by:
%\begin{equation}
%\tan\theta_L = \mu_H B = \gamma_H \mu_d B 
%\end{equation}
%where $\mu_H$ is the Hall mobility, the product of the charge carrier mobility in silicon 
%$\mu_d$ and the Hall factor $\gamma_H$ which is of the order unity.  
The Lorentz angle depends on the charge carrier mobility, which in turn depends on 
the bias voltage, the thickness of the depleted region and the temperature~\cite{bib:Jacoboni}. 
%For a certain depleted region of thickness $t$, 
%the drift along the Lorentz angle results in an 
%average shift of the collected charge by a distance $(t/2)\tan\theta_L$,
For fully-depleted modules, the average shift in collected charge  
is approximately 30~$\mu$m for the Pixel Detector and 
10~$\mu$m for the SCT, in both cases not negligible with respect to the
detector resolution and alignment precision. 
%It is therefore necessary to measure this effect in data.
Measurements of the Lorentz angle for the ATLAS sensors have already
been performed in test beams~\cite{bib:PixelLorentzAngle,bib:SCTLorentzAngle}, 
but in conditions different from the actual operation in ATLAS. 

The Lorentz angle is measured from the dependence of the cluster size on the 
incident angle of the particle. When the incident angle equals the Lorentz 
angle, all the charge carriers generated by the particle drift along 
the particle direction and, apart from charge diffusion, are collected
at the same point on the sensor surface, giving a minimum cluster size.  

\begin{figure}
\subfigure[Pixel Detector mean cluster width]{\label{fig:pixlorentz}
\includegraphics[width=\columnwidth]{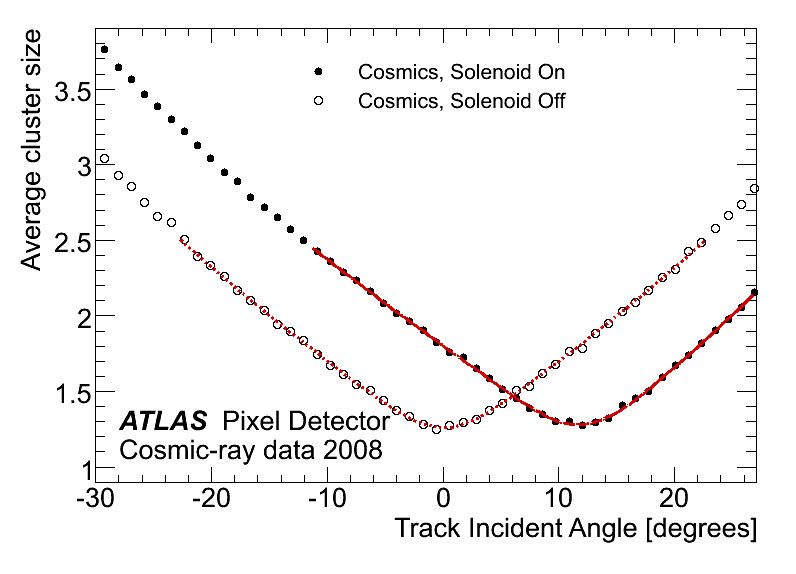}}
\subfigure[SCT mean cluster width]{\label{fig:sctlorentz}
\includegraphics[width=\columnwidth]{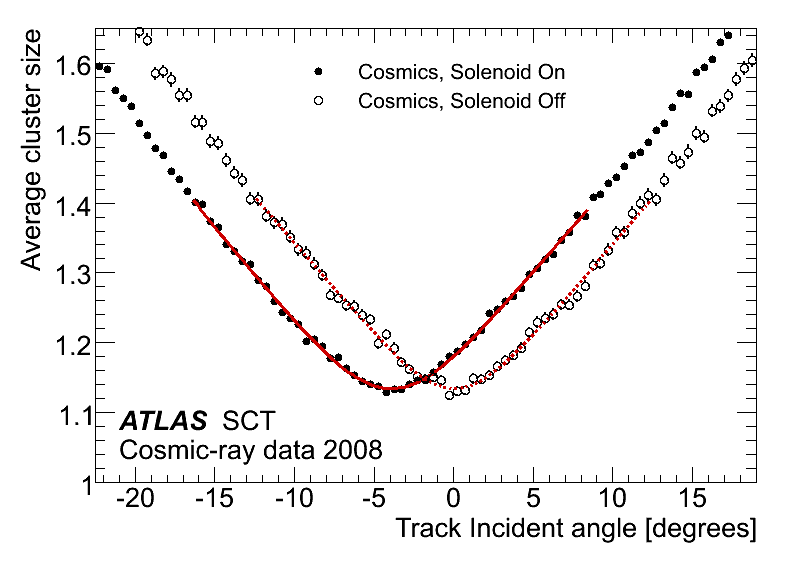}}
\caption{Cluster-size dependence on the particle incident angle for the Pixel Detector~\subref{fig:pixlorentz} and the SCT~\subref{fig:sctlorentz}. The displacement of the minimum for the data with 
solenoid on is a measurement of the Lorentz angle $\theta_L$.}
\label{fig:LorentzAngle}
\end{figure}

The dependence of the cluster size on the incident angle 
%in the $R\phi$ plane,
$\phi_{\rm local}$ is shown for 
the Pixel Detector and SCT in Fig.~\ref{fig:LorentzAngle}. 
Data are fitted using the convolution of the function:
\begin{equation}
f(\phi_{\rm local}) = a \left| \tan\phi_{\rm local} - \tan\theta_L \right| + b
\label{eq:LorentzGeo}
\end{equation}
with a Gaussian distribution. Fit parameters are the Lorentz angle $\theta_L$, the shape parameters $a$, $b$ and the width of the Gaussian. For the Pixel Detector an improvement of the fit quality was observed by replacing the second term in~(\ref{eq:LorentzGeo}) by 
$b/\sqrt{\cos\phi_{\rm local}}$, which is a phenomenological attempt to 
describe the bigger relative weight of diffusion effects for tracks at high incident angle.

The measured values are $11.77^\circ\pm 0.03^\circ$ and $-3.93^\circ\pm 0.03^\circ$ 
for the Pixel Detector and SCT respectively, where the errors are statistical 
only. The values differ by a factor of three due to the different mobility of the charge carriers
which provide the dominant signal: electrons in the Pixel Detector, holes in the SCT.

As a cross check for systematic effects, the same measurement was performed
for data with no magnetic field, giving values of  
$0.09^\circ\pm 0.03^\circ$ and $0.05^\circ\pm 0.05^\circ$ for the Pixel Detector and 
SCT respectively. Since for the Pixel Detector the disagreement with 
respect to the expected null value is statistically significant, it is used
as a component of the systematic uncertainty. The other dominant source of systematic uncertainty is the fit range, which has been estimated to give
a contribution of $0.07^\circ$ for the Pixel Detector and $0.10^\circ$ for the SCT. 
%The resulting measurements of the Lorentz angle in a 2~T magnetic field, are:
%\begin{eqnarray*}
%\theta_L \left(\mathrm {electrons}\right) 
%& = & 11.77^\circ\pm 0.03^\circ \stat \pm 0.11^\circ \syst,
%\\
%\theta_L \left(\mathrm{holes}\right) 
%& = & -3.93^\circ\pm 0.03^\circ \stat \pm 0.10^\circ \syst.
%\end{eqnarray*}
The measured values of the Lorentz angle in the 2~T magnetic field are shown in
Table~\ref{tab:la} where they are compared with the expectation from the model 
in~\cite{bib:Jacoboni}. The measurements are compatible with the model predictions 
within the uncertainties on the predictions arising from the values of charge-carrier 
mobilities.

Since Pixel Detector modules operated with different temperature ranges in 2008 and 2009, 
it was possible to measure the dependence of the Lorentz angle on the silicon 
temperature. The resulting dependence 
\begin{eqnarray}
{\rm d}\theta_L/{\rm d}T=(-0.042\pm 0.003)^{\circ}{\rm /}{\rm K}
\end{eqnarray}
is in agreement with the model expectation of $-0.042^\circ / {\rm K}$. 

\begin{table}
\newcommand\T{\rule{0pt}{2.6ex}}
\newcommand\B{\rule[-1.2ex]{0pt}{0pt}}

\begin{center}
\begin{tabular}{l@{\ }lcr@{$\pm$}c@{$\pm$}lr@{$\pm$}l}
  \hline\hline 
  \multicolumn{2}{l}{Detector} \B \T &$T$ [\degr C]
                               &\multicolumn{3}{c}{Measured $\theta_{\rm L} [^\circ]$}
                               &\multicolumn{2}{c}{Model $\theta_{\rm L} [^\circ]$}  \\ 
  \hline

  Pixel \T &(electrons) &--3   & 11.77 &0.03 &$^{0.13}_{0.23}$ &12.89  &1.55 \\
  SCT   \B &(holes)     &\phantom{--}5   &--3.93 &0.03 &0.10             &--3.69 &0.26 \\ 
  \hline\hline
  \end{tabular}
\end{center}
\caption{Measured values of the Lorentz angle in 2~T magnetic field at the average operational temperature in 2008, compared with model expectations~\protect \cite{bib:Jacoboni}.
         For the
         measurements, the first error is statistical and the second systematic. The
         error on the model prediction arises from uncertainties in the charge-carrier mobility. }
\label{tab:la}
\end{table}

\subsection{Track parameter resolution}
\label{sec:trackpar}
% Editor Andreas.Wildauer@cern.ch

The expected resolution of the perigee parameters $d_0$, $z_0$, $\phi_0$,
$\theta$ and $q/p$
of a particle emerging from proton-proton collisions in the LHC can be predicted using reconstructed 
and split tracks from cosmic-ray data. Since particles coming from cosmic-ray showers mostly traverse the detector
from top to bottom, the resolutions can only be derived for the ATLAS barrel detectors.

In order to select tracks with good quality, the split tracks are each required to have at least
2, 6 and 25 hits in the barrel of the Pixel, SCT and TRT detectors respectively, and a transverse momentum
of more than 1~GeV. The $|d_0|$ impact parameter has to be less than 40~mm to guarantee
that the split tracks originate in the interaction region inside the beam pipe.

The perigee parameters $T_{\rm up}$ and $T_{\rm down}$, where $T$ is any of the
five parameters, of each split-track pair
are compared to each other to extract the overall track parameter resolutions. 
Since both tracks come from the same particle, their difference $\Delta\tau = T_{\tau,\rm up}
- T_{\tau,\rm down}$ for each perigee parameter $\tau$ must have a variance $\sigma^2(\Delta\tau)$ which is two times the variance $\sigma^2(T_\tau)$
of the parameters of each track. The resolution of the track parameter $\tau$ is 
therefore given by the root mean square of the $\Delta\tau$ distribution
divided by $\sqrt{2}$. This method has been used to study the resolution of the perigee parameters 
of Inner Detector tracks. The variances were calculated excluding the outermost 0.3\% of
events in each distribution. 

The measured resolution is compared to the Monte Carlo 
expectation for a perfectly-aligned detector. The difference in performance
is attributed to the remaining misalignment after the procedure in Section~\ref{sec:alignment}.  
In addition, the refit of the split-track pair can be
restricted to a subset of measurements in the Inner Detector. This has been done
to study the perigee parameter resolutions of silicon-only tracks (Pixel and SCT)
and compare them to resolutions of the same tracks which have been fitted using
the full Inner Detector.

A summary of the measured track-parameter resolutions for $\pT>30$~\GeV, 
where the multiple-scattering contribution can be neglected, is given in 
Table~\ref{tab:TrackParametersResolution}.

%%%%%%%%%%%%%%%%%%%%%%%%%%%%
%%%%%%%%%%%%%%%%%%%%%%%%%%%%
%%%%%%%%%%%%%%%%%%%%%%%%%%%%

\paragraph{Impact parameter resolution}
\begin{figure*}[p]
\begin{tabular}{cc}
\includegraphics[width=0.48\textwidth]{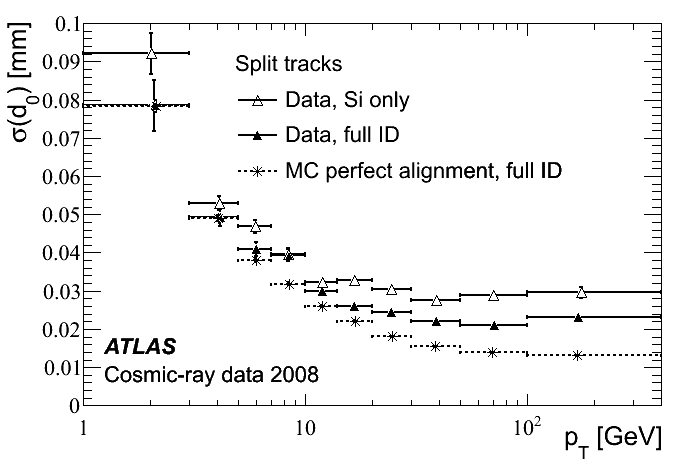} &
\includegraphics[width=0.48\textwidth]{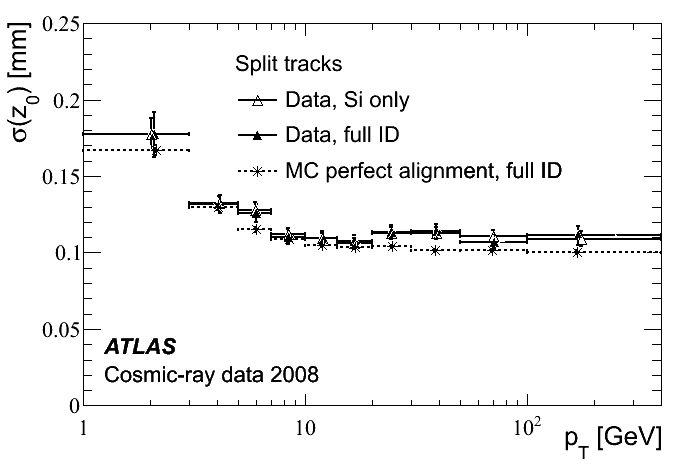}
\end{tabular}
\caption{Impact parameter resolution determined from data for the track
impact parameters as a function of transverse momentum. Resolutions of full
ID (solid triangles) and silicon-only (open triangles) tracks 
are compared to those from full tracks in MC simulation (stars).}
\label{fig:deltaIPvspT}
\end{figure*}%
Figure~\ref{fig:deltaIPvspT} shows the transverse and longitudinal impact parameter
resolutions as determined from the data using the track-splitting method.
They are displayed as a function of transverse momentum. At low momenta
the resolution is governed by multiple scattering in the beam pipe and
first pixel layers. For higher momenta, above about 10~\GeV, the impact 
parameter resolutions rapidly approach an asymptotic limit which is given 
by the intrinsic detector resolution and residual misalignments.% 

\begin{figure*}[p]
\begin{tabular}{cc}
\includegraphics[width=0.48\textwidth]{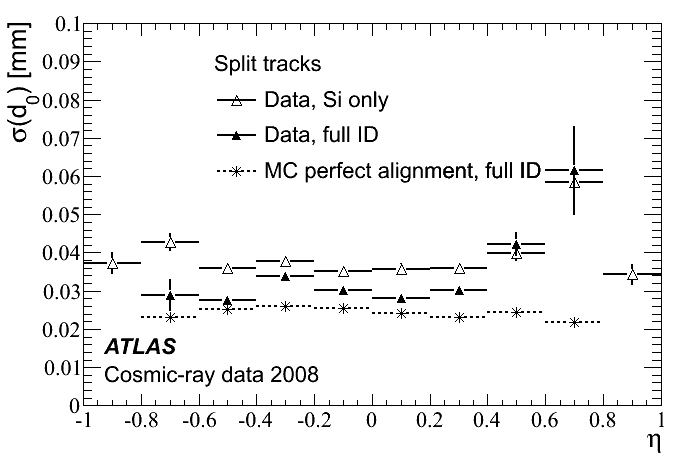} &
\includegraphics[width=0.48\textwidth]{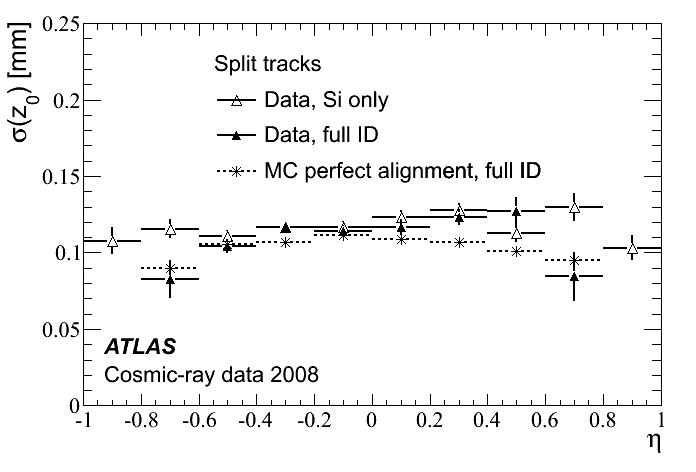}
\end{tabular}
\caption{Impact parameter resolution determined from data for tracks with 
$\pT>1$~\GeV,  as a
function of pseudorapidity $\eta$. The resolutions are shown for full ID tracks (solid triangles), silicon-only tracks (open triangles) and simulated full ID tracks (stars).}
\label{fig:deltaIPvsEta}
\end{figure*}%

Resolutions as a function of $\eta$ are constant and symmetric around $\eta=0$,
as shown in Fig.~\ref{fig:deltaIPvsEta}. Both Figs.~\ref{fig:deltaIPvspT} and
\ref{fig:deltaIPvsEta} compare the resolution obtained for Inner Detector tracks
with that from a fit to solely the silicon part. The $d_0$ resolution is
slightly more precise for full tracks, as the TRT measurements add to the
momentum resolution and thus to the precision of the track extrapolation to
the perigee point.

\begin{figure*}[p]
\begin{center}\begin{tabular}{cc}
\includegraphics[width=0.48\textwidth]{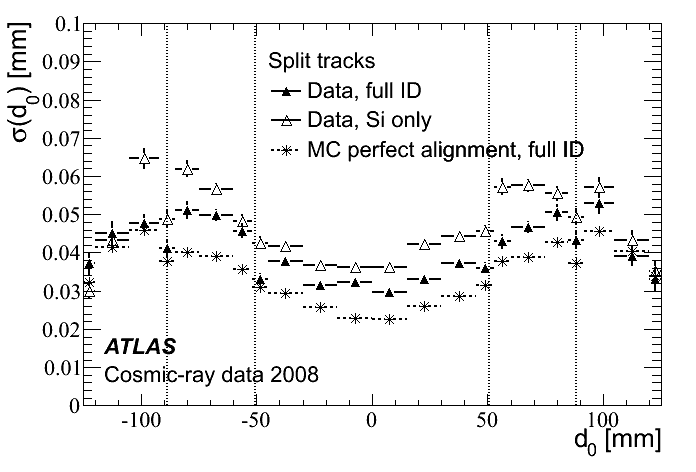} &
\includegraphics[width=0.48\textwidth]{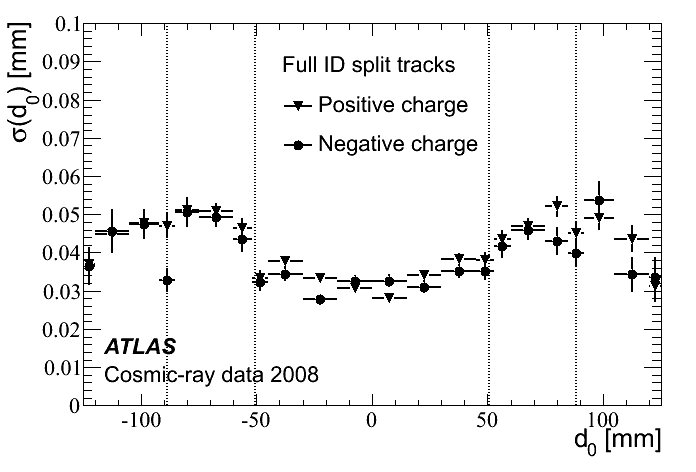}
% \fbox{\parbox{4cm}{\begin{minipage}{3.5cm}
%   Show charge dependence here?
%   \end{minipage}}}
\end{tabular}\end{center}
\caption{Transverse impact parameter resolution as a function of transverse
impact parameter for tracks with $\pT>1$~GeV. As for the previous figures, the left plot compares resolutions of full ID tracks, silicon-only tracks and simulated full ID tracks.
In the right plot resolutions are compared for full Inner Detector tracks with positive (circles) and negative charge (squares). The vertical lines indicate the positions of the pixel barrel layers.}
\label{fig:deltaIPvsd0}
\end{figure*}

The $d_0$ resolution has also been studied as a function of $d_0$ on
a sample without the cut on $|d_0|$. The results are presented in
Fig.~\ref{fig:deltaIPvsd0} and show a worsening in resolution towards
larger $|d_0|$, which corresponds to tracks crossing pixel layers at
high incident angle. Pixel clusters from such tracks are wider and
possibly fragmented due to a geometrically reduced charge deposition
per pixel. This effect degrades the resolution, as does the smaller
number of pixel layers crossed. The resolution of full ID tracks at $d_0$ values near to the radii of
pixel layers (about 50, 90 and 120 mm) improves
because of the reduction in the extrapolation length between the
closest measurement and the perigee of the track.

A dependence
on the charge of the reconstructed tracks has also been investigated as shown
in Fig.~\ref{fig:deltaIPvsd0} (right plot). Small differences appear in some bins,
but do not allow for a conclusive result.
A dependence of the resolutions on $z_0$ and $\phi_0$ has been checked
as well, and none was found. This means that the impact parameter
resolutions follow the symmetries in the barrel part of the Inner Detector.

%
% \begin{figure}[htbp]
% \begin{tabular}{cc}
% \includegraphics[width=0.48\textwidth]{figures/mean_delta_d0_vs_pT.eps} &
% \includegraphics[width=0.48\textwidth]{figures/mean_delta_d0_vs_d0.eps}
% \end{tabular}
% \caption{Mean of the $\Delta{d_0}$ distribution as function of $p_T$ (left)
%          and $d_0$ (right). The mean value for full ID tracks (full triangles)
%          is compared against that from silicon-only tracks (open triangles)
%          and that from MC (stars).}
% \label{fig:meand0}
% \end{figure}%
%
% The mean of the $\Delta{d_0}$ distribution shows a difference from 0 of around
% $10\,\mu\mathrm{m}$ on average. It is most pronounced for high momentum tracks and shows a
% dependence on the $d_0$ parameter itself, as can be seen in~\ref{fig:meand0}.
% This effect can be attributed to the presence of a residual systematic
% misalignment in the data. The mean of the $\Delta{z_0}$ distribution is
% compatible with 0 and does not show a structure.

\begin{figure*}[p]
\begin{tabular}{cc}
\includegraphics[width=0.48\textwidth]{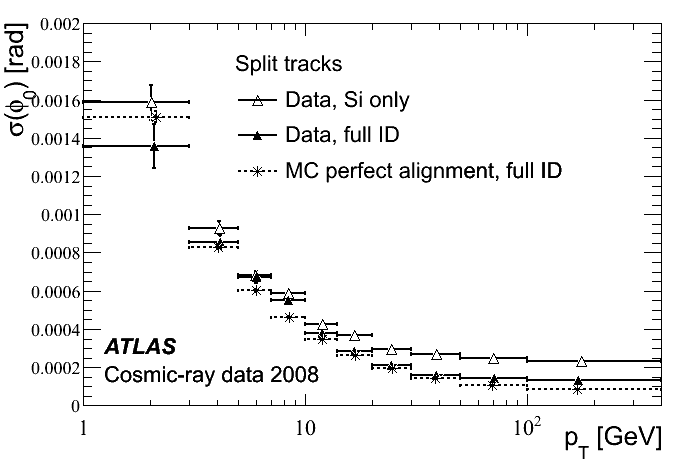} &
\includegraphics[width=0.48\textwidth]{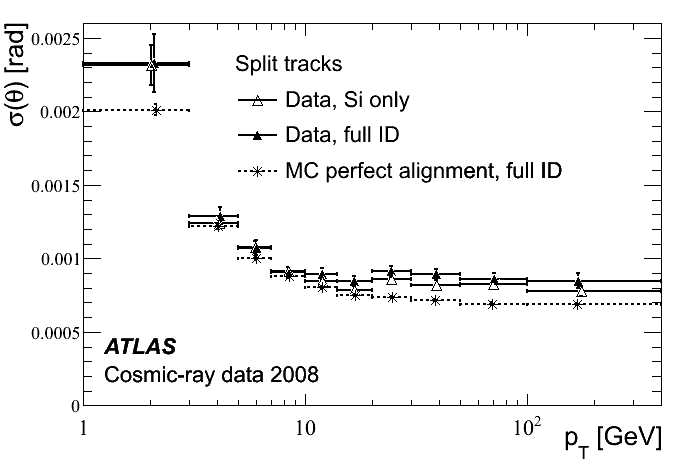}
\end{tabular}
\caption{Angular resolution determined from data as a
function of transverse momentum. The resolutions are shown for full ID tracks (solid triangles), silicon-only tracks (open triangles) and simulated full ID tracks (stars).}
\label{fig:deltaANvspT}
\end{figure*}

\begin{figure*}[p]
\begin{center}
\begin{tabular}{cc}
\includegraphics[width=0.48\textwidth]{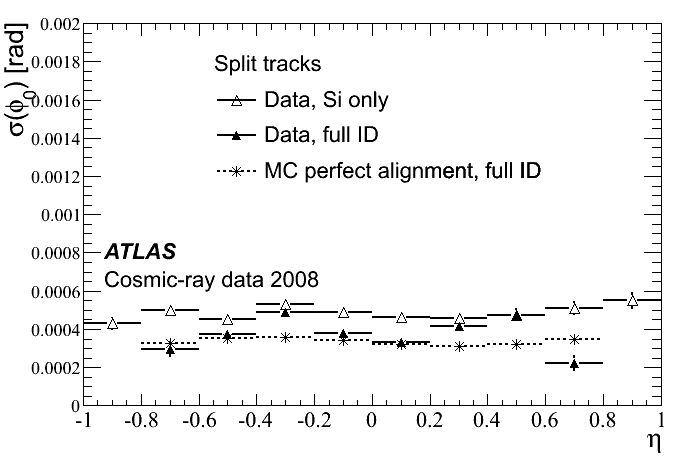} &
\includegraphics[width=0.48\textwidth]{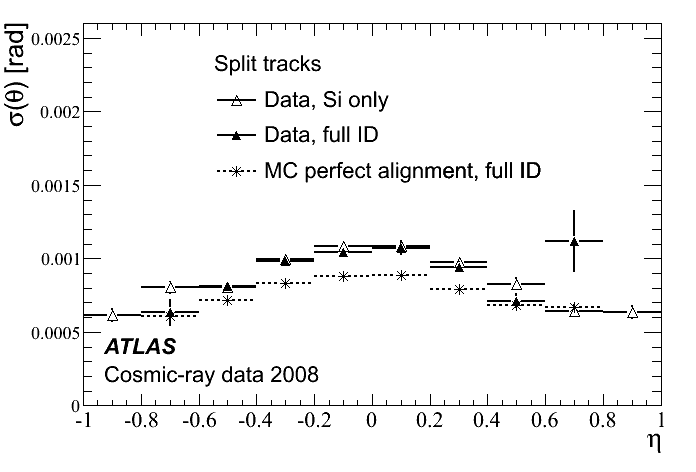} \\
% (a) Cosmic Data & (b) Cosmic MC
\end{tabular}
\end{center}
\caption{Angular resolution determined from data for tracks with $\pT>1$~GeV as a
function of pseudorapidity $\eta$. The resolutions are shown for full ID tracks (solid triangles), silicon-only tracks (open triangles) and simulated full
ID tracks (stars).}
\label{fig:deltaANvsEta}
\end{figure*}

\paragraph{Angular resolution}
A precise and reliable reconstruction of the track direction contributes
to the knowledge of the momentum vector and thus is vital for finding
decay vertices and matching with signals from other detectors.
A precision on the track angles
below 1\,mrad is achieved, as shown in Figs.~\ref{fig:deltaANvspT} and
\ref{fig:deltaANvsEta}.%

The angular resolutions have been found to be independent of other track
parameters, except for an expected small worsening at $|d_0|>50\,\mathrm{mm}$.
%The mean of the $\Delta\phi_0$ and $\Delta\theta$ distributions
%are compatible with 0.
%
%\begin{figure}[htbp]
%\begin{center}
%\includegraphics[width=0.48\textwidth]{figures/mean_delta_theta_vs_phi0.png}
%\end{center}
%\caption{Mean of $\Delta\theta$ as a function of angular parameter $\phi_0$.
%         The mean value for full ID tracks (full triangles)
%         is compared against that from silicon-only tracks (open triangles)
%         and that from MC (stars).}
%\label{fig:meantheta}
%\end{figure}%
%

\paragraph{Momentum resolution}
% \begin{itemize}
% \item $p_T$ resolution as a function of  $p_T$, $\eta$,  $d_0$ and $z_0$.
% \end{itemize}

\begin{figure*}[p]
\begin{tabular}{cc}
\includegraphics[width=0.48\textwidth]{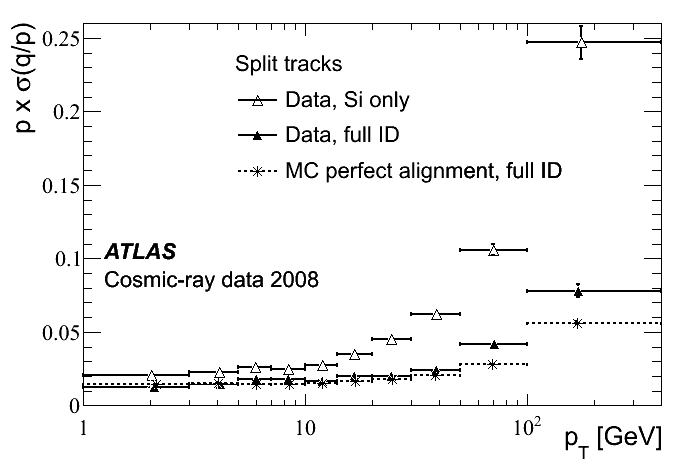} &
\includegraphics[width=0.48\textwidth]{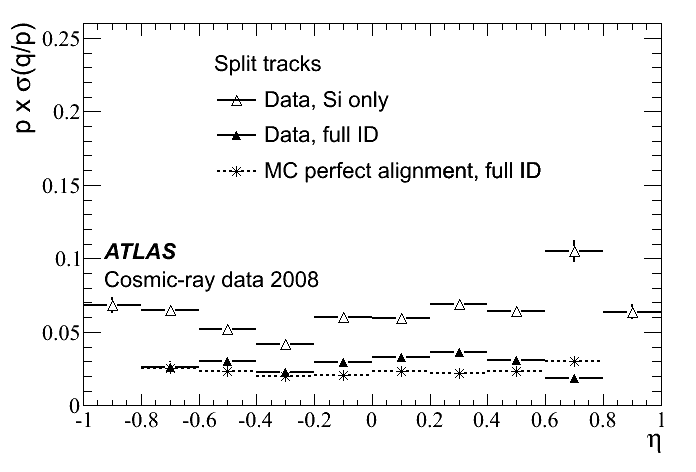}
\end{tabular}
\caption{Momentum resolutions determined from data as a
function of transverse momentum and $\eta$. The resolutions are shown for full ID tracks (solid triangles), silicon-only tracks (open triangles) and simulated full ID tracks (stars).}
\label{fig:deltaQoP}
\end{figure*}

A precise momentum determination of high-energy particles is a key 
ingredient for any physics analysis. In Fig.~\ref{fig:deltaQoP} the 
relative momentum resolution $p \times  \sigma(q/p)$ is shown as a 
function of $p_T$ (left plot) and $\eta$ (right plot). 
While the resolution is flat in $\eta$, it shows the expected degradation at higher transverse 
momenta. In this region, the contribution of the TRT to the momentum 
resolution becomes clearly visible. 

\begin{table}
\centering
\begin{tabular}{lcc}
\hline\hline
Parameter & \multicolumn{2}{c}{Asymptotic resolution} \\
          & Cosmic-ray data 2008 & Monte Carlo \\ 
\hline
$d_0$ [$\mu$m]         & $22.1\pm 0.9$& $14.3\pm 0.2$ \\ 
$z_0$ [$\mu$m]         & $112 \pm 4  $& $101 \pm 1  $ \\ 
$\phi_0$ [mrad]     & $0.147\pm 0.006$& $0.115\pm 0.001$ \\ 
$\theta$ [mrad]      & $0.88 \pm 0.03$ & $0.794\pm 0.006$ \\ 
$q/p$ [\GeV$^{-1}$]  & $(4.83\pm 0.16)\times 10^{-4}$ 
                     & $(3.28\pm 0.03)\times 10^{-4}$\\ 
\hline\hline
\end{tabular}
\caption{Track parameter resolution for tracks with 
$\pT>30$~\GeV\ 
in cosmic-ray data and simulation.}
\label{tab:TrackParametersResolution}

\end{table}

\subsection{Energy-loss measurement}
\label{sec:dEdx}
The average specific energy loss of charged particle d$E$/d$x$ is described by the 
Bethe-Bloch function~\cite{bib:PDG}. The specific
energy loss, sensitive to the particle speed $\beta=v/c$, can be combined 
with the momentum measurement to provide particle identification. Because of 
the energy loss tails (see Fig.~\ref{fig:PixToT}) a truncated mean can 
be used to reduce the variance of the estimation.

Split tracks from cosmic-ray muons have been used to measure the resolution 
on d$E$/d$x$ of the Pixel Detector. Tracks are required to have 
a transverse momentum $p_{\rm T}>0.5$~GeV and relative momentum resolution
$\sigma(p_{\rm T})/p_{\rm T}<20\%$. In addition a cut on the 
distance of closest approach to the beam axis, $|d_0|<10$~mm, is made
in order to select tracks similar to the ones generated by LHC collisions. 

The specific energy loss in a Pixel Detector module is derived from the
cluster charge, $Q$, taking into account the average energy needed to 
create an electron-hole pair $W$ 
(Section~\ref{sec:Pixelcalibration}) and the path in silicon $d/\cos\alpha$ 
where $d$ is the detector sensitive thickness (250~$\mu$m):
\begin{equation}
\frac{{\rm d}E}{{\rm d}x} = \frac{Q}{e} \frac{W\cos\alpha}{d}.
\label{eq:dedx}
\end{equation}
At high incident angle particles cross several pixel cells; 
the signal released in some of them 
may be below threshold and the energy loss underestimated. 
%This is especially relevant for high incident angles 
%in the $R-\phi$ plane, $\phi_{\rm local}$, 
%where the number of crossed pixels
%is high and the path length in the silicon approaches its minimum value, 
%corresponding to the 50~$\mu$m pitch.
To reduce this effect, only clusters with $\cos\alpha>0.6$ and 
$|\phi_{\rm local}|<0.5$~rad
are used. The correct association of clusters to the reconstructed track is 
ensured by requiring position residuals to be less than 300~$\mu$m in the
local~$x$ coordinate and less than 900~$\mu$m in local~$y$.  

\begin{figure}[t]
\begin{center}
\includegraphics[width=\columnwidth]{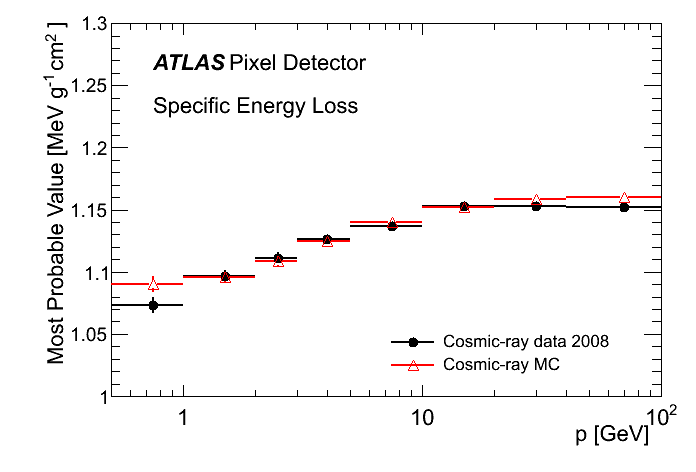}
\end{center}
\caption{Most probable value of the specific energy loss
d$E/$d$x$ in the Pixel Detector as a function of muon momentum in
the relativistic rise region. Monte Carlo points are scaled according 
to the absolute charge calibration determined 
in Section~\ref{sec:Pixelcalibration}.}
\label{fig:PixeldEdx}
\end{figure}

Figure~\ref{fig:PixeldEdx} shows the most probable d$E/$d$x$ value of 
individual clusters in the barrel region as a function of the track
momentum. 
The relativistic rise and its 
saturation due to the density effect are clearly visible and there is a 
good agreement between the $7.2\pm 0.4\%$ rise observed in data from 
0.5~GeV to 20~GeV in $p_{\rm T}$, and the $7.5\pm 0.4\%$ estimated from the 
simulation. For tracks with at least three 
clusters, a global d$E/$d$x$ estimation is made by averaging all the 
individual measurements after the exclusion of the cluster with the 
maximum $Q\cos\alpha$. This procedure has been verified to produce an almost
Gaussian estimator on the relativistic plateau, $p_{\rm T}>20$~GeV, with a
resolution of 15\%. This would allow a limited 
particle identification capability,
with a $2\sigma$ separation between $K$ and $\pi$ for $p<500$~MeV.

\subsection{Transition radiation measurement}
\label{sec:TR}
%%% Transition Radiation Measurement
The large spread of momenta of the cosmic rays recorded has allowed 
a validation of the transition-radiation performance of
the TRT by measuring the percentage of high-threshold hits on tracks at different 
momenta. 
The probability of producing a transition radiation photon at each material 
boundary is dependent upon the Lorentz gamma factor of the particle.
Since the threshold for producing transition radiation is $E/m \sim 1000$,
in LHC collision events transition radiation is essentially limited to 
electrons. However, the mean 
\pt\ of recorded cosmic-ray muons was 60~\GeV\ with a significant 
tail to almost 1~\TeV\ (see Fig.~\ref{fig:d0_resolution_vs_PT}). The high-momentum
muons produce enough transition-radiation photons to allow an initial calibration of the TRT as a transition radiation detector. 

The transition radiation study used 20\,000 nearly-vertical tracks in the barrel TRT. 
The tracks were required to have at least four SCT hits and at least 20 
TRT hits, a fit $\chi^2$ / Ndof $<$ 10.0, $\sigma (\pt) / \pt$ $<$ 3.0 and 
0.5 $< \pt <$ 1000~\GeV. The track angle to the vertical, measured using hits in 
the SCT, was restricted to be less than $15^\circ$. Tracks were assigned to 
(logarithmic) momentum bins, and the high-threshold hit probability calculated as 
a simple ratio in each bin. 

\begin{figure}[t]
	\centering
	{\includegraphics[width=\columnwidth]{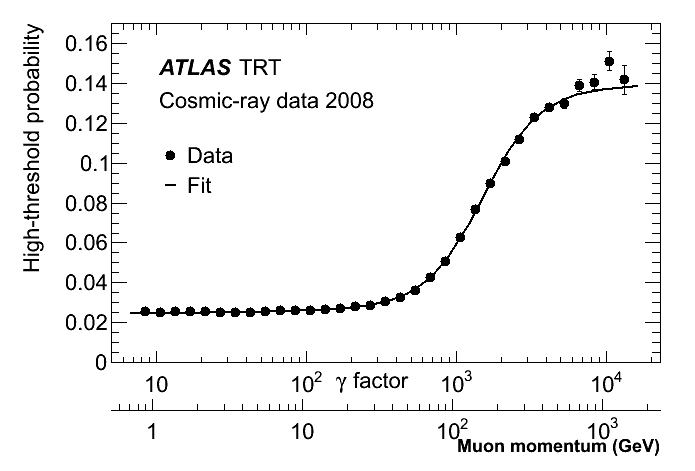}}
	\caption{High-threshold hit probability as a function of muon Lorentz $\gamma$ factor for selected tracks in the October 2008 cosmic-ray data. The line shows a sigmoid fit to the data. }
% the dashed line corresponds to test-beam measurements.}
	\label{TRturnon}
\end{figure}

Figure~\ref{TRturnon} shows the probability of seeing a high-threshold hit on a 
muon track in the TRT barrel as a function of the Lorentz gamma factor of the
particle; the probability is averaged over positively and negatively charged muons.
The fitted curve shown in Fig.~\ref{TRturnon} is consistent with the result 
obtained in the 2004 ATLAS combined test beam run and confirms the design of the TRT electron identification capabilities.

\section{Conclusions}
\label{sec:conclusions}

The final installation of the ATLAS Inner Detector in August 2008 was followed by a period
of commissioning and calibration. During this period the detector took data with high efficiency 
with both LHC single beams and cosmic rays. These data allowed full tests of trigger, data-acquisition 
and monitoring systems, and of offline track reconstruction.  
Some problems with the newly-installed evaporative cooling system and the optical links of the 
silicon detectors were exposed. These were addressed before data-taking with LHC beams in 2009, 
when more than 98\% of the detector was operational.

Detector gains were calibrated and thresholds adjusted to give good uniformity of response.
The components of the detector were timed-in with a precision of 1--2~ns.
Many detector performance properties were measured.
The average noise occupancies were $\sim 10^{-10}$ hit/channel/BC for the Pixel Detector
and $\sim 3 \times 10^{-5}$ hit/channel/BC for the SCT, well within specifications.
The intrinsic efficiencies of the silicon detectors were measured to be close to 100\%
and of the TRT to be 97.2$\pm$0.5\%.
The Lorentz angle in the silicon detectors in the 2~T magnetic field was found
to be consistent with model expectations. 
Energy loss in the Pixel Detector and transition radiation were measured 
and found to be in agreement with expectations from test beams. 

A new Level-1 track trigger based on a fast OR of TRT signals was 
commissioned.
The Level-2 trigger tracking-algorithms were modified for cosmic rays,
resulting in a trigger efficiency of 99.6$\pm$0.02\% for tracks reconstructed offline.
The cosmic-ray data were used to perform an initial detector alignment.
The resolution of track parameters was measured by comparing two segments of a cosmic-ray track.
After detector alignment, the impact parameter resolutions for high-momentum tracks were found to 
be $22.1\pm 0.9$~$\mu$m and $112\pm 4$~$\mu$m in the transverse and longitudinal directions, 
respectively. In this asymptotic limit, the relative momentum resolution was measured to be 
$\sigma_p/p=(4.83\pm 0.16)\times 10^{-4}\ \mathrm{GeV}^{-1} \times\pT$.

The observed performance on this early data showed the ATLAS Inner Detector to be fully 
operational and providing high-quality tracking before the first LHC collisions.

\section{Acknowledgements}

We are greatly indebted to all CERN's departments and to the LHC project for their immense efforts not only in building the LHC, but also for their direct contributions to the construction and installation of the ATLAS detector and its infrastructure. We acknowledge equally warmly all our technical colleagues in the collaborating Institutions without whom the ATLAS detector could not have been built. Furthermore we are grateful to all the funding agencies which supported generously the construction and the commissioning of the ATLAS detector and also provided the computing infrastructure.

The ATLAS detector design and construction has taken about fifteen years, and our thoughts are with all our colleagues who sadly could not see its final realisation.

We acknowledge the support of ANPCyT, Argentina; Yerevan Physics Institute, Armenia; ARC and DEST, Australia; Bundesministerium f\"ur Wissenschaft und Forschung, Austria; National Academy of Sciences of Azerbaijan; State Committee on Science \& Technologies of the Republic of Belarus; CNPq and FINEP, Brazil; NSERC, NRC, and CFI, Canada; CERN; CONICYT, Chile; NSFC, China; COLCIENCIAS, Colombia; Ministry of Education, Youth and Sports of the Czech Republic, Ministry of Industry and Trade of the Czech Republic, and Committee for Collaboration of the Czech Republic with CERN; Danish Natural Science Research Council and the Lundbeck Foundation; European Commission, through the ARTEMIS Research Training Network; IN2P3-CNRS and CEA-DSM/IRFU, France; Georgian Academy of Sciences; BMBF, DFG, HGF and MPG, Germany; Ministry of Education and Religion, through the EPEAEK program PYTHAGORAS II and GSRT, Greece; ISF, MINERVA, GIF, DIP, and Benoziyo Center, Israel; INFN, Italy; MEXT, Japan; CNRST, Morocco; FOM and NWO, Netherlands; The Research Council of Norway; Ministry of Science and Higher Education, Poland; GRICES and FCT, Portugal; Ministry of Education and Research, Romania; Ministry of Education and Science of the Russian Federation and State Atomic Energy Corporation ROSATOM; JINR; Ministry of Science, Serbia; Department of International Science and Technology Cooperation, Ministry of Education of the Slovak Republic; Slovenian Research Agency, Ministry of Higher Education, Science and Technology, Slovenia; Ministerio de Educaci\'{o}n y Ciencia, Spain; The Swedish Research Council, The Knut and Alice Wallenberg Foundation, Sweden; State Secretariat for Education and Science, Swiss National Science Foundation, and Cantons of Bern and Geneva, Switzerland; National Science Council, Taiwan; TAEK, Turkey; The Science and Technology Facilities Council and The Leverhulme Trust, United Kingdom; DOE and NSF, United States of America.

%
%%%%%%%%%%%%%%%%%%%%%%%%%%%%%%%%%%%%%%%%%%%%%%%%%%%%%%%%%%%%%%%%%%%%%%%%%%%%%%%
% Bibliography
%%%%%%%%%%%%%%%%%%%%%%%%%%%%%%%%%%%%%%%%%%%%%%%%%%%%%%%%%%%%%%%%%%%%%%%%%%%%%%
%
% Style file to use with mcite.
% Use atlasstyle with just cite.
\bibliographystyle{atlasstylem}
\bibliography{idpaper}

\providecommand{\href}[2]{#2}\begingroup\raggedright\begin{thebibliography}{10}

\bibitem{bib:ATLASDetectorPaper}
{G. Aad et al.}, {\em {The ATLAS Experiment at the CERN Large Hadron
  Collider}\/},  \href{http://dx.doi.org/10.1088/1748-0221/3/08/S08003}{JINST
  {\bf 3} (2008)  S08003}.

\bibitem{bib:LHCPaper}
{L. Evans and P. Bryant eds.}, {\em {LHC Machine}\/},
  \href{http://dx.doi.org/10.1088/1748-0221/3/08/S08001}{JINST {\bf 3} (2008)
  S08001}.

\bibitem{bib:PixelDetectorPaper}
{G. Aad et al.}, {\em {ATLAS pixel detector electronics and sensors}\/},
  \href{http://dx.doi.org/10.1088/1748-0221/3/08/P07007}{JINST {\bf 3} (2008)
  P07007}.

\bibitem{bib:FEI3chip}
{I. Peric et al.}, {\em {The FEI3 readout chip for the ATLAS pixel
  detector}\/},  \href{http://dx.doi.org/10.1016/j.nima.2006.05.032}{\NIM {\bf
  A 565} (2006)  178--187}.

\bibitem{bib:SCTBarrelModules}
{A. Abdesselam et al.}, {\em {The barrel modules of the ATLAS semiconductor
  tracker}\/},  \href{http://dx.doi.org/10.1016/j.nima.2006.08.036}{\NIM {\bf A
  568} (2006)  642--671}.

\bibitem{bib:SCTEndcapModules}
{A. Abdesselam et al.}, {\em {The ATLAS semiconductor tracker end-cap
  module}\/},  \href{http://dx.doi.org/10.1016/j.nima.2007.02.019}{\NIM {\bf A
  575} (2007)  353--389}.

\bibitem{bib:SCTsensors}
{A. Ahmad et al.}, {\em {The Silicon microstrip sensors of the ATLAS
  semiconductor tracker}\/},
  \href{http://dx.doi.org/10.1016/j.nima.2007.04.157}{\NIM {\bf A 578} (2007)
  98--118}.

\bibitem{bib:SCTABCD3TA}
{F. Campabadal et al.}, {\em {Design and performance of the ABCD3TA ASIC for
  readout of silicon strip detectors in the ATLAS semiconductor tracker}\/},
  \href{http://dx.doi.org/10.1016/j.nima.2005.07.002}{\NIM {\bf A 552} (2005)
  292--328}.

\bibitem{bib:TRTBarrel}
{E. Abat et al.}, {\em {The ATLAS TRT Barrel Detector}\/},  JINST {\bf 3}
  (2008)  P02014.

\bibitem{bib:TRTEndcap}
{E. Abat et al.}, {\em {The ATLAS TRT end-cap detectors}\/},  JINST {\bf 3}
  (2008)  P10003.

\bibitem{bib:TRTElectronics}
{E. Abat et al.}, {\em {The ATLAS TRT electronics}\/},
  \href{http://dx.doi.org/10.1088/1748-0221/3/08/S08003}{JINST {\bf 3} (2008)
  P06007}.

\bibitem{bib:BCMPaper}
{V. Cindro et al.}, {\em {The ATLAS Beam Conditions Monitor}\/},
  \href{http://dx.doi.org/10.1088/1748-0221/3/02/P02004}{JINST {\bf 3} (2008)
  P02004}.

\bibitem{bib:SCTOpticalLinks}
{A. Abdesselam et al.}, {\em {The optical links of the ATLAS SemiConductor
  tracker}\/},  \href{http://dx.doi.org/10.1088/1748-0221/2/09/P09003}{JINST
  {\bf 2} (2007)  P09003}.

\bibitem{bib:PixelOpticalLinks}
{K. K. Gan et al.}, {\em {Optical Link of the ATLAS Pixel Detector}\/},  \NIM
  {\bf A 570} (2007)  292--294.

\bibitem{bib:GOL}
{P. Moreira et al.}, {\em {G-Link and Gigabit Ethernet compliant serializer for
  LHC data transmission}\/},  IEEE Nuclear Science Symposium {\bf 2} (2000)
  96--99.

\bibitem{bib:SirOD}
{M.L. Chu et al.}, {\em {The off-detector opto-electronics for the optical
  links of the ATLAS semiconductor tracker and pixel detector}\/},
  \href{http://dx.doi.org/10.1016/j.nima.2004.04.228}{\NIM {\bf A 530} (2004)
  293--310}.

\bibitem{bib:TRTROD}
{P. Lichard et al.}, {\em {Evolution of the TRT backend and the new TRT-TTC
  board}\/},  in {\em {Proceedings of the 2005 LECC, Heidelberg}}, p.~253.
\newblock CERN-LHCC-2005-038, CERN, Geneva, 2005.

\bibitem{bib:SCTDAQ}
{A. Abdesselam et al.}, {\em {The Data acquisition and calibration system for
  the ATLAS semiconductor tracker}\/},
  \href{http://dx.doi.org/10.1088/1748-0221/3/01/P01003}{JINST {\bf 3} (2008)
  P01003}.

\bibitem{bib:CoolingPaper}
{D. Attree et al.}, {\em {The evaporative cooling system for the ATLAS inner
  detector}\/},  \href{http://dx.doi.org/10.1088/1748-0221/3/07/P07003}{JINST
  {\bf 3} (2008)  P07003}.

\bibitem{bib:DCSPaper}
{A. Barriuso Poy et al.}, {\em {The detector control system of the ATLAS
  experiment}\/},  \href{http://dx.doi.org/10.1088/1748-0221/3/05/P05006}{JINST
  {\bf 3} (2008)  P05006}.

\bibitem{bib:Monitoring}
{M. White}, {\em {Data Quality Monitor for the ATLAS Inner Detector}\/},  in
  {\em {$17^{\rm th}$ International Workshop on Vertex detectors}}.
\newblock {Proceedings of Science}, 2008.
\newblock {PoS (Vertex 2008) 044}.

\bibitem{bib:AthenaCore}
{ATLAS Collaboration}, {\em {The Athena Framework}\/},  in {\em {ATLAS
  Computing Technical Design Report}}, p.~27.
\newblock CERN-LHCC-2005-022, CERN, Geneva, 2005.

\bibitem{bib:NewTrackingPubNote}
{T. Cornelissen et al.}, {\em {Concepts, Design and Implementation of the ATLAS
  New Tracking (NEWT)}\/},  ATLAS Note ATL-SOFT-PUB-2007-007.

\bibitem{bib:GlobalChi2TrackFitter}
{T. Cornelissen et al.}, {\em {The global $\chi^2$ track fitter in ATLAS}\/},
  \href{http://dx.doi.org/10.1088/1742-6596/119/3/032013}{{J. Phys.: Conf.
  Ser.} {\bf 119} (2008)  032013}.

\bibitem{bib:FastOr}
{A. Fratina et al.}, {\em {The TRT Fast-OR Trigger}\/},  ATLAS Note
  ATL-INDET-PUB-2009-002.

\bibitem{bib:HLTTracking}
{G. Aad et al.}, {\em {HLT Track Reconstruction Performance}\/},  in {\em
  {Expected performance of the ATLAS experiment : detector, trigger and
  physics}}, p.~565.
\newblock {CERN-OPEN-2008-020, CERN, Geneva}, 2009.

\bibitem{bib:muonFlux1983}
{A. Dar}, {\em {Atm. Neutrinos, Astrophysical Neutrons, and Proton-Decay
  Experiments}\/},  Phys. Rev. Lett. {\bf 51} (1983)  227.

\bibitem{bib:PDG}
{C. Amsler et al.}, {\em The review of particle physics\/},  \PL {\bf B 667}
  (2008)  1.

\bibitem{bib:geant4a}
{S. Agostinelli et al.}, {\em {Geant4 -- a simulation Toolkit}\/},
  \href{http://dx.doi.org/10.1016/S0168-9002(03)01368-8}{\NIM {\bf A 506}
  (2003)  250--303}.

\bibitem{bib:geant4b}
{J. Allison et al.}, {\em {Geant4 developments and applications}\/},
  \href{http://dx.doi.org/10.1109/TNS.2006.869826}{IEEE Trans. Nucl. Sci. {\bf
  Vol. 53, Issue 1} (2006)  270--278}.

\bibitem{bib:PixelCosmicPaper}
{Atlas Collaboration}, {\em {Commissioning of the ATLAS Pixel Detector}\/},
  {to be submitted to \EPJ {} C}  .

\bibitem{bib:PixelOptoElectronics}
{K. E. Arms et al.}, {\em {ATLAS pixel opto-electronics}\/},  \NIM {\bf A 554}
  (2005)  458--468.

\bibitem{bib:PixelSensors}
{M. S. Alam et al.}, {\em {The ATLAS silicon pixel sensors}\/},  \NIM {\bf A
  456} (2001)  217--232.

\bibitem{bib:Bichsel}
{H. Bichsel}, {\em {Straggling in thin silicon detectors}\/},  Rev. Mod. Phys.
  {\bf 60} (1988)  663--669.

\bibitem{bib:SiW2}
{R. D. Ryan}, {\em {Precision Measurements of the Ionization Energy and Its
  Temperature Variation in High Purity Silicon Radiation Detectors}\/},  IEEE
  Trans. Nucl. Sci. {\bf Vol. 20, Issue 1} (1973)  473--480.

\bibitem{bib:SiW1}
{P. Christmas}, {\em {Average Energy Required to Produce an Ion Pair}\/},
  Tech. Rep. Report 31, ICRU, 1979.

\bibitem{bib:SiW3}
{R.H. Pehl et al.}, {\em {Accurate determination of the ionization energy in
  semiconductor devices}\/},  \NIM {\bf 59} (1968)  45--55.

\bibitem{bib:SiW4}
{F. Scholze et al.}, {\em {Determination of the electron-hole pair creation
  energy for semiconductors from the spectral responsivity of photodiodes}\/},
  \NIM {\bf A 439} (2000)  208--215.

\bibitem{bib:PixelLorentzAngle}
{I. Gorelov et al.}, {\em {A measurement of Lorentz angle and spatial
  resolution of radiation hard silicon pixel sensors}\/},
  \href{http://dx.doi.org/10.1016/S0168-9002(01)01413-9}{\NIM {\bf A 481}
  (2002)  204--221}.

\bibitem{bib:IDTDR}
{ATLAS Collaboration}, {\em {Alignment requirements}\/},  in {\em {ATLAS Inner
  Detector Technical Design Report, vol. I}}, p.~215.
\newblock CERN-LHCC-1997-016, CERN, Geneva, 1997.

\bibitem{bib:GLX2}
{P. Br\"ukman, A. Hicheur and S.J. Haywood}, {\em {Global $\chi^2$ approach to
  the Alignment of the ATLAS Silicon Tracking Detectors}\/},  ATLAS Note
  ATL-INDET-PUB-2005-002.

\bibitem{bib:LocalX2_1}
{R. H\"artel}, {\em {Iterative local $\chi^2$ alignment approach for the ATLAS
  SCT detector}\/},  Master's thesis, MPI Munich, 2005.

\bibitem{bib:LocalX2_2}
{T. G\"ottfert}, {\em {Iterative local $\chi^2$ alignment algorithm for the
  ATLAS Pixel detector}\/},  Master's thesis, {Universit\"at W\"urzburg and MPI
  Munich}, 2006.

\bibitem{bib:RobustAlignment}
{F. Heinemann}, {\em {Track Based Alignment of the ATLAS Silicon Detectors with
  the Robust Alignment Algorithm}\/},  ATLAS Note ATL-INDET-PUB-2007-011.

\bibitem{bib:PixelSurvey}
{A. Andreazza, V. Kostyukhin and R.J. Madaras}, {\em {Survey of the ATLAS Pixel
  Detector Components}\/},  ATLAS Note ATL-INDET-PUB-2008-012.

\bibitem{bib:Jacoboni}
{C. Jacoboni et al.}, {\em {A review of some charge transport properties of
  silicon}\/},  \href{http://dx.doi.org/10.1016/0038-1101(77)90054-5}{Solid
  State Electronics {\bf 20} (1977)  77--89}.

\bibitem{bib:SCTLorentzAngle}
{F. Campabadal et al.}, {\em {Beam Tests of ATLAS SCT Silicon Strip Detector
  Modules}\/},  \href{http://dx.doi.org/10.1016/j.nima.2004.08.133}{\NIM {\bf A
  538} (2005)  384--407}.

\end{thebibliography}\endgroup

\end{document}